\documentclass[12pt,BCOR12mm,DIV12,headsepline,chapterprefix,bibtotoc,pointlessnumbers]{scrbook}

\usepackage[T1]{fontenc}
\usepackage[dvips]{graphicx}

\usepackage[british,ngerman]{babel}

\usepackage[latin1]{inputenc}
\usepackage[margin=10pt,font=small,labelfont=bf,tableposition=top]{caption}
\usepackage{exscale}
\usepackage[english]{minitoc}
\usepackage[tight]{subfigure}
\usepackage{scrpage2}
\usepackage{amssymb}
\usepackage{wasysym}

\begin{document}
\selectlanguage{british}

\frontmatter
\begin{titlepage}
\vspace*{-1cm}\raggedleft{\usekomafont{title} FAU-PI4-DISS-06-002}

\vspace*{1cm}
  \begin{center}
    {\usekomafont{title}\LARGE Detection of ultra high energy
      neutrinos \\[1mm]
      with an underwater very large volume \\[1mm]
      array of acoustic sensors: \\[1mm]
      A simulation study \\}

    \selectlanguage{ngerman}
    \vfill Den Naturwissenschaftlichen Fakult"aten \\
    der Friedrich-Alexander-Universit"at Erlangen-N"urnberg \\
    zur \\
    Erlangung des Doktorgrades

    \vfill vorgelegt von \\
    Timo Karg \\
    aus N"urnberg
  \end{center}
\end{titlepage}

\selectlanguage{ngerman}

Als Dissertation genehmigt von den Naturwissenschaftlichen
Fakult"aten der \\
Universit"at Erlangen-N"urnberg.

\vfill
\begin{tabular}{p{6cm}l}
  Tag der m"undlichen Pr"ufung: & 25.~April 2006\tabularnewline[3mm]
  Vorsitzender der \tabularnewline
  Promotionskommision: & Prof.~Dr.~D.-P.~H"ader
  \tabularnewline[3mm]
  Erstberichterstatter: & Prof.~Dr.~G.~Anton \tabularnewline[3mm]
  Zweitberichterstatter: & Prof.~Dr.~K.-H.~Kampert
\end{tabular}

\selectlanguage{british}

\dominitoc
\tableofcontents
\listoffigures


\mainmatter
\chapter{Introduction}
\label{chap:introduction}

In astroparticle physics the processes powering the most energetic
objects in our universe are studied as well as particle interactions
at energies not accessible at accelerator laboratories.  Hard x-rays,
TeV gamma rays, electrons, hadrons, nuclei, and neutrinos emitted from
single stars, but also, for example, from active galactic nuclei and
gamma-ray bursts can be observed by experiments on the Earth, by
high-altitude balloon experiments, or by satellites.

Until the beginning of the twentieth century, astronomy, and with it
our knowledge of the universe, was limited to the observation of
visible light.

The field of particle astrophysics was born in 1912, when Viktor Hess
undertook several balloon flights up to an altitude of 5200\,m to
measure the assumed decrease of the ionising radiation known to exist
on the Earth's surface, which was believed to originate in the decay
of radioactive nuclei in the Earth's crust. What he found was a
completely unexpected {\em increase} of the flux with altitude
\cite{Hess:1912}.  In the same publication Hess already suggested,
that the radiation must be of extra-terrestrial origin. He was further
able to rule out a solar origin, because he did not measure any
intensity variations, neither during a day night cycle, nor during a
solar eclipse.

\begin{figure}[ht]
  \centering
  \includegraphics{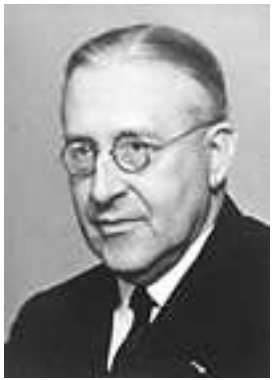}
  \caption[Viktor F. Hess]{Viktor F. Hess (1883 -- 1964).}
  \label{fig:hess}
\end{figure}

Soon people started to study the composition of cosmic rays, and it
was found, that they provided excellent means to study particle
physics at the highest energies available at this time. For his
discovery Hess was awarded the Nobel prize in 1936 together with Carl
D. Anderson, who discovered the positron in the cosmic radiation.

\medskip Another important discovery on the way to neutrino astronomy
was, of course, the first detection of the (electron anti-)neutrino
(which had already been predicted in 1930 by Wolfgang Pauli) in 1956
in the ``Project Poltergeist'' \cite{Cowan:1992xc, Reines:1956}, for
which Frederick Reines was awarded the Nobel prize in 1995. To measure
the inverse beta decay, his group designed a detector situated near
the core of the Savannah River nuclear reactor. It consisted of a
water target sandwiched between 4200 litres of liquid scintillator
read out by 330 photomultiplier tubes.

\begin{figure}[ht]
  \centering
  \includegraphics{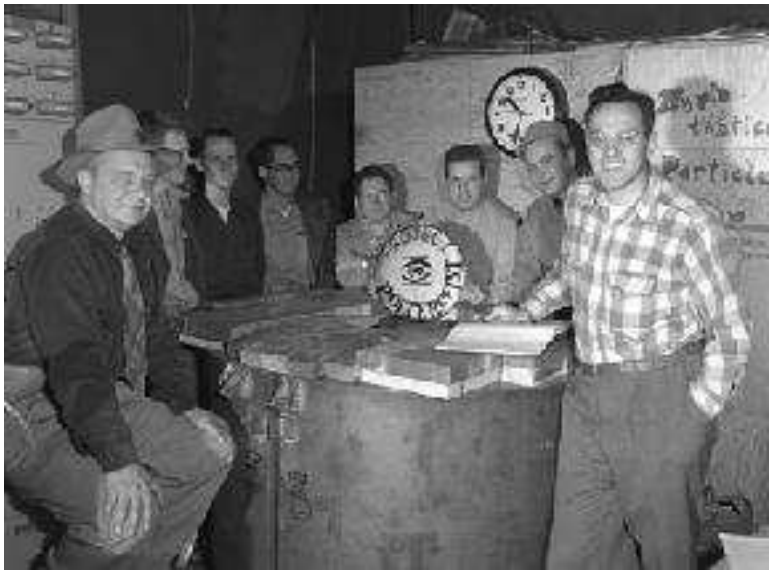}
  \caption[Project Poltergeist]{Frederick Reines' Project Poltergeist
    group with the ``Herr Auge'' detector, a smaller predecessor of
    the Savannah River detector.}
  \label{fig:poltergeist}
\end{figure}

Soon after, also the muon neutrino was discovered, and the
possibility of observing extraterrestrial neutrinos was discussed.

\medskip Doing astronomy with neutrinos is especially appealing
because of their unique properties. Neutrinos are electrically neutral
particles, so they are not deflected in the electromagnetic fields
present nearly everywhere in the universe. If one can determine the
direction of an observed neutrino, the direction will always point
back to the source, which is essential for imaging astronomy.
Neutrinos have a very small total cross section, so they will
(practically) not be absorbed on their way through the interstellar
medium. They allow a direct view into cosmological objects, whereas
optical astronomy is always confined to the observation of the surface
of the source.

The small cross section also poses the great challenge of neutrino
astronomy: Enormous target masses are required to observe at least a
few neutrino interactions. The flux of cosmological neutrinos is
believed to decrease steeply with energy. For energies up to a few
hundred TeV, gigaton (1\,km$^3$ of water or ice) detectors are built,
which detect the \v{C}erenkov light emitted by muons or cascades
produced in neutrino interactions. To measure neutrinos at even higher
energies, detectors with an observed target mass in the teraton range
(1000\,km$^3$ of water or ice) will be required. There are several
different experimental approaches to build such a detector, which want
to use a variety of different target media ranging from water and ice,
over the Earth's atmosphere, to the moon.

\begin{figure}[ht]
  \centering
  \includegraphics[width=0.9\textwidth]{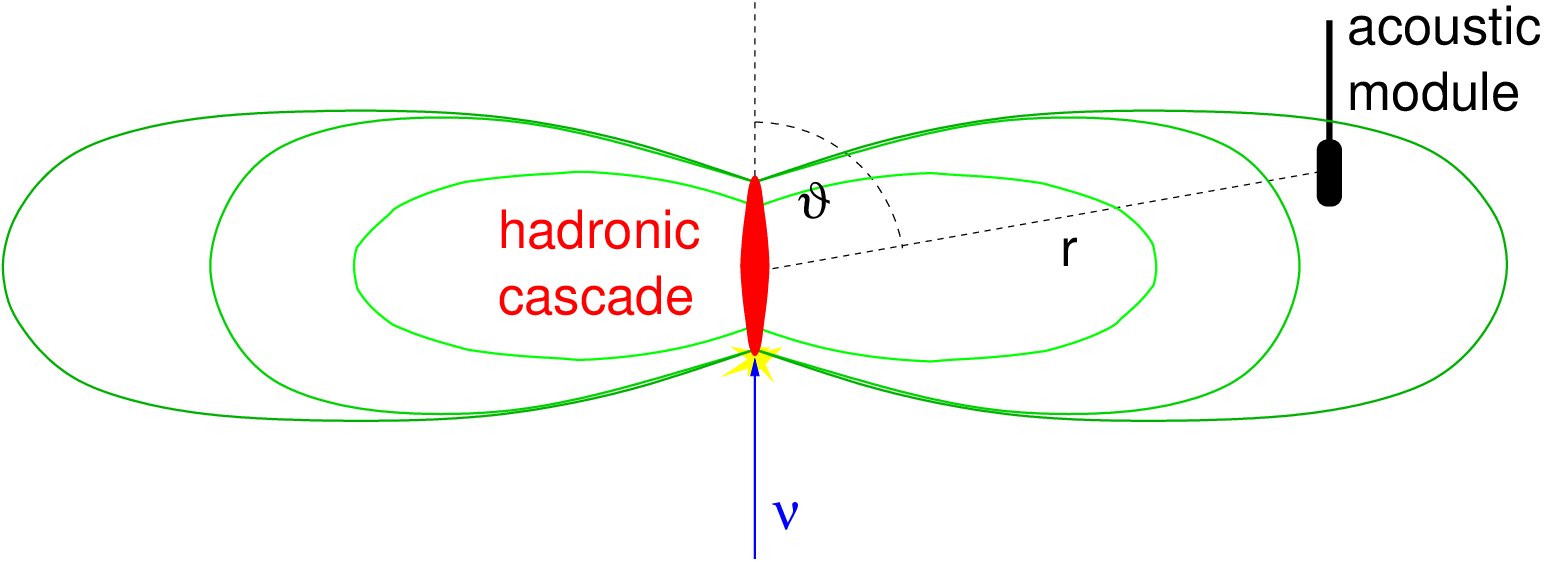}
  \caption[Principle of acoustic neutrino detection]{Principle of
    acoustic neutrino detection: Measurement of bipolar pressure
    pulses, which are emitted perpendicular to a neutrino induced
    particle cascade developing in a fluid.}
  \label{fig:schematic_acoustics}
\end{figure}

In this thesis neutrino detection utilising the thermoacoustic sound
generation mechanism in fluids is discussed: A neutrino induced
hadronic cascade heats a narrow region of the medium, leading to a
rapid expansion, which propagates perpendicular to the cascade axis as
a bipolar sonic pulse through the fluid
(cf.~Fig.~\ref{fig:schematic_acoustics}). If this pulse of a few ten
microseconds length can be recorded at different positions, the
direction and energy of the neutrino can be inferred. We will show
that a detector consisting of 1000\,km$^3$ of sea water instrumented
sparsely with acoustic sensors would allow to detect neutrinos with
energies above some EeV\footnote{1\,EeV = 10$^{18}$\,eV.}.

In chapter~\ref{chap:sources} theoretical models of various
cosmological sources are presented, which are expected to produce
ultra high energy (UHE) neutrinos. Chapter~\ref{chap:experiments}
discusses different existing experimental techniques for the detection
of ultra high energy neutrinos exemplified by existing or planned
experiments.  In chapter~\ref{chap:thermoacoustic_model} the
thermoacoustic sound generation model is introduced, and measurements
for its verification are described. Afterwards, the sound generation
of UHE neutrinos in water is analysed in
chapter~\ref{chap:tam_uhe_neutrinos}, followed by a discussion of the
propagation of acoustic signals in sea water and their detection
(Chap.~\ref{chap:acoustics}). Finally, we will present in
chapter~\ref{chap:detectors} a simulation study of an underwater
acoustic neutrino telescope, and will derive its sensitivity to a
diffuse flux of neutrinos.


\chapter{Sources of ultra high energy neutrinos}
\label{chap:sources}
\minitoc

\bigskip In this chapter several theoretical models are discussed,
which predict the emission of ultra high energy neutrinos.

\section{Active Galactic Nuclei}
\label{sec:agn}

The model of the Active Galactic Nucleus (AGN) was developed to
describe a whole range of cosmological objects. These include Seyfert
galaxies\footnote{Galaxies with very broad emission lines.}, blazars,
and quasars \cite{Longair:1994}. AGNs are spiral galaxies with a super
massive (10$^7$ -- 10$^9$ solar masses) black hole at the centre,
which accretes matter from its host galaxy. Particles can be
accelerated to ultra high energies by Fermi acceleration in jets
perpendicular to the accretion disc.

\begin{figure}[ht]
  \centering
  \includegraphics[width=0.4\textwidth]{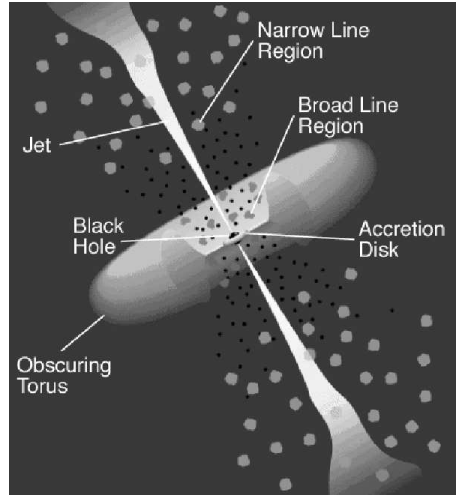}
  \caption[Schematic of an Active Galaxy]{Unified model of an Active
    Galactic Nucleus: The characteristic features observed depend on
    the direction to the observer.}
  \label{fig:agn}
\end{figure}

AGNs are characterised by a very high energy output from a relatively
small volume. A schematic of an Active Galactic Nucleus is shown in
Fig.~\ref{fig:agn}. Depending on the direction under which an AGN is
observed, its spectrum will show various prominent features. If the
accretion disc is observed edge-on all visible light is usually
absorbed in molecular clouds in and around the disc, and a strong
radio source with no optical counterpart is measured. In the case
where AGNs are observed in the direction of the jet, a very luminous
object with high variability, a {\em blazar}, is seen. The light from
blazars shows a high degree of polarisation, which indicates a
synchrotron production mechanism in the magnetic fields of the jet.

AGNs are believed to be a source of ultra high energy neutrinos, since
the accelerated protons can interact with hadrons or gamma-rays
producing charged pions which then decay into leptons and neutrinos,
e.g.:

\begin{displaymath}
  \begin{array}{ccccccccccccc}
    p & + & \gamma & \rightarrow & \Delta^+ (1232) & \rightarrow &
    \pi^+ & + & n \\
    & & & & & & \rightarrow & \mu^+ & + & \nu_\mu \\
    & & & & & & & \rightarrow & e^+ & + & \bar{\nu}_\mu & + & \nu_e
  \end{array}
\end{displaymath}

The $\Delta^+ (1232)$ can also decay into a proton and a neutral pion,
the latter one decaying into two high energy photons, which are a
possible source for the high energy (TeV) gamma-rays observed in the
cosmic radiation.

The first prediction for the neutrino flux from AGNs was derived in
1991 by Stecker et~al.~\cite{Stecker:1991vm, Stecker:2005hn} assuming
a hidden-core model where all hadronic energy is transformed into
neutrinos, and the AGN can thus only be observed in neutrinos.  They
also pointed out that the main production mechanism must be proton
photon interactions, because the column density of hadronic matter
surrounding the AGN is limited by the absence of strong x-ray
absorption lines in their spectra. Other models predict a neutrino
flux at higher energies, e.g.~\cite{Halzen:1997hw}, which could
already be ruled out by the AMANDA and Baikal
\cite{Wischnewski:2005rr} experiments. A summary of the neutrino
fluxes predicted by the source models described in this chapter is
shown in Fig.~\ref{fig:flux_models}.

\section{Gamma-ray bursts}
\label{sec:grb}

Gamma-ray bursts (GRBs) are sudden, very short (from milliseconds up
to a few seconds) and very intense flashes of gamma-rays, which were
discovered in the late 1960s by the Vela military satellites
monitoring the Nuclear Test Ban Treaty. It was found that GRBs are
accompanied by an afterglow in the x-ray, optical, and radio band.
GRBs are distributed isotropically over the sky, and spectral analysis
of the afterglow revealed that they have high redshifts.

\begin{figure}[ht]
  \centering
  \includegraphics[width=0.8\textwidth]{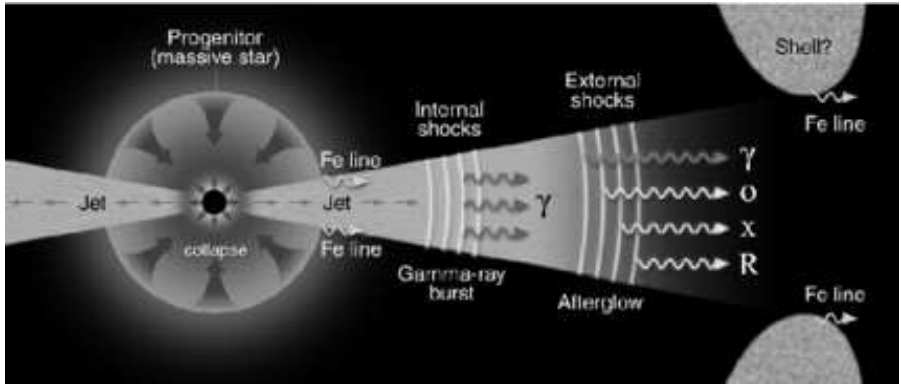}
  \caption[Fireball shock model of gamma-ray bursts]{The fireball
    shock model of gamma-ray bursts (from \cite{Meszaros:2001ig}).}
  \label{fig:grb}
\end{figure}

The large distances (high redshifts) require a model, where, under the
assumption of isotropic emission, photon energies of 10$^{52}$\,erg to
10$^{54}$\,erg are produced during the short duration of the
burst\footnote{For comparison: Luminosity of the sun $L_{\astrosun} =
  3.8 \cdot 10^{33} \, \mathrm{erg} / \mathrm{s}$.}. In addition,
causality and the short timescale limit the spatial extension of the
source to the order of tens to hundreds of kilometres. In the {\em
  fireball shock model} (e.g. \cite{Meszaros:2001ig}) the collapse of
a massive star induces a relativistically expanding $e^\pm$, $\gamma$
fireball with Lorentz factors $\Gamma \approx 100$
\cite{Waxman:1997ti}. The fireball kinetic energy is converted into
non-thermal particle and radiation energy in collisionless shocks
mediated by chaotic electric and magnetic fields in which the
electrons produce a synchrotron power-law radiation spectrum similar
to that observed, while inverse Compton scattering of these
synchrotron photons extends the spectrum into the GeV range. There are
also modifications to this model, which propose an energy release into
relativistic jets, reducing the total energy produced to 10$^{52}$ --
10$^{54}$ $(\Omega / 4 \, \pi)$\,erg, where $\Omega$ is the solid
angle of the jets (cf.~Fig.~\ref{fig:grb}).

In the same shock fronts where electrons are accelerated to highest
energies also protons should be accelerated. These ultra high energy
protons will be converted into neutrinos by the same mechanism as
described for AGNs (cf.~Sec.~\ref{sec:agn}).

The model predicts a neutrino flux with a power law energy spectrum
$\Phi_\nu \propto E_\nu^{-2}$, as expected for first order Fermi
acceleration \cite{Waxman:1997ti, Waxman:1998yy}:

\begin{displaymath}
  E_\nu^2 \, \Phi_{\nu,GRB} = 3.0 \cdot 10^{-9} \, \mathrm{GeV}
  \mathrm{cm}^{-2} \mathrm{s}^{-1} \mathrm{sr}^{-1} \qquad 10^5 \,
  \mathrm{GeV} < E_\nu < 10^7 \, \mathrm{GeV}
\end{displaymath}

For energies $E_\nu >$ 10$^7$\,GeV the neutrino spectral index rises
(i.e. the flux decreases faster), since these neutrinos are produced
by charged pions and muons with energies above the synchrotron
emission threshold. These pions and muons lose a considerable amount
of their energy by synchrotron emission before decaying, producing
lower energy neutrinos. The expected neutrino spectrum is shown in
Fig.~\ref{fig:flux_models}.

\section{The Waxman Bahcall upper limit}
\label{sec:waxman_bahcall}

The Waxman Bahcall (WB) limit \cite{Waxman:1998yy} is a model
independent upper bound to the flux of ultra high energy neutrinos
derived from the measured flux of UHE cosmic hadrons and the
assumption of optically thin sources, i.e.~sources of size not
much larger than the mean free path for proton photon interactions.

The assumption of optically thin sources is motivated by the fact,
that cosmic rays with energies up to 10$^{11}$\,GeV have been observed
at Earth, which disfavours the AGN hidden core model, where all
hadrons are absorbed within the source. It is applicable to all
sources where protons are Fermi accelerated to ultra high energies and
interact with ambient photons or protons, like AGNs or GRB fireballs.
Assuming that the {\em complete} energy observed in UHE cosmic rays
would be transformed into neutrinos one can derive a strict upper
limit on the flux of UHE neutrinos:

\begin{displaymath}
  E_\nu^2 \, \Phi_{\nu,WB} = 2.0 \cdot 10^{-8} \, \mathrm{GeV}
  \mathrm{cm}^{-2} \mathrm{s}^{-1} \mathrm{sr}^{-1}
\end{displaymath}

This limit overestimates the most likely neutrino flux by a factor of
approximately $5 / \tau$, for small optical depths $\tau$ (number of
proton photon interaction lengths) \cite{Bahcall:1999yr}, since in
proton photon interactions only 20\% of the proton energy is
transfered to the charged pion.

The flux of GRB neutrinos presented in Sec.~\ref{sec:grb} is
compatible with the WB bound. However, the AGN neutrino fluxes
discussed in Sec.~\ref{sec:agn} exceed the bound, but they were
derived under the assumption of optically thick sources. These models
can thus not be tested by cosmic ray observations on Earth, but only
by the detection of the corresponding neutrinos.

\section{GZK neutrinos}
\label{sec:gzk_neutrinos}

One of the most promising models predicting ultra high energy
neutrinos is the GZK effect. It was first described in 1966 by Greisen
\cite{Greisen:1966jv}, Zatsepin, and Kuzmin \cite{Zatsepin:1966jv} who
inferred that cosmic ray protons with energies $E_p \gtrsim 50$\,EeV
should interact with the 2.7\,K cosmic microwave background (CMB). In
this interaction mostly $\Delta^+ (1232)$ are produced, which partly
decay into $\pi^+$ and subsequently into neutrinos giving rise to a
flux of ultra high energy neutrinos if there are sources of UHE
protons with redshifts $z \gtrsim 1$ (which is approximately the
interaction length for proton photon interactions). It is widely
believed that for example AGNs and GRBs, which have been measured with
much higher redshifts, are the required proton sources. The
non-existence of the GZK cutoff would lead to the need of a strongly
modified cosmology, in the sense that the (yet unknown) sources of
ultra high energy cosmic rays must be very close to us, at redshifts
$z <$ 1.

\begin{figure}[ht]
  \centering
  \includegraphics[width=0.9\textwidth]{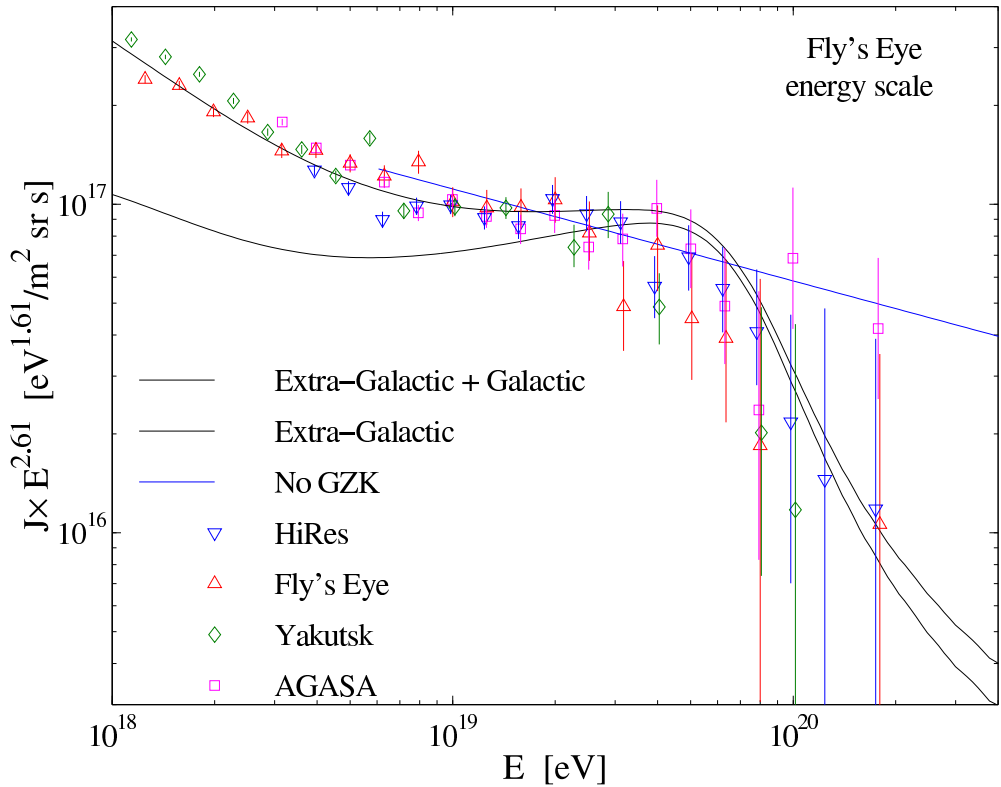}
  \caption[Cosmic ray spectrum at the ankle]{The spectrum of cosmic
    rays at the ankle measured by different experiments (from
    \cite{Bahcall:2002wi}). The Fly's Eye and HiRes data suggest the
    existence of the GZK cutoff (cosmic ray spectrum expected
    including the GZK cutoff is indicated by the black curves); AGASA
    on the other hand does not see a cutoff (blue curve).}
  \label{fig:cr_ankle}
\end{figure}

However, there is still disagreement between cosmic ray experiments
whether there is a GZK cutoff. Figure~\ref{fig:cr_ankle} shows the
energy spectrum of cosmic rays at the ``ankle'' measured by different
experiments. Clarification is expected from the results of the Pierre
Auger Observatory (cf.~Sec.~\ref{sec:eas}), which began to measure
cosmic rays at highest energies in 2004.

Detailed calculations \cite{Engel:2001hd} using different cosmic ray
source distributions, cosmic ray injection spectra, cosmological
evolution of the sources, and different cosmologies lead to a
considerable neutrino flux exceeding even the Waxman Bahcall bound
between 10$^8$\,GeV and 10$^{11}$\,GeV (see
Fig.~\ref{fig:flux_models}).  This is not in contradiction to the WB
upper limit, since via the GZK effect, the integrated proton flux
above the GZK threshold is scaled down in energy below this threshold
through neutrino production. Thus there is an accumulation of the
neutrino flux around the GZK threshold.

\section{Z-Burst neutrinos}
\label{sec:zburst_neutrinos}

Another possible production mechanism for ultra high energy neutrinos
are Z-bursts, which are particularly interesting, since they would
allow, if detected, the determination of the absolute neutrino mass.

Similarly to the cosmic microwave background the universe is
presumably filled with a background of relic neutrinos from the big
bang, which decoupled when the universe had cooled down to a
temperature of $kT \approx 2$\,MeV \cite{Bernstein:1998}. At present
time these neutrinos have a number density of 114\,cm$^{-3}$ per
massive neutrino family with a black body temperature of 1.9\,K.

\begin{figure}[ht]
  \centering
  \includegraphics[width=0.9\textwidth]{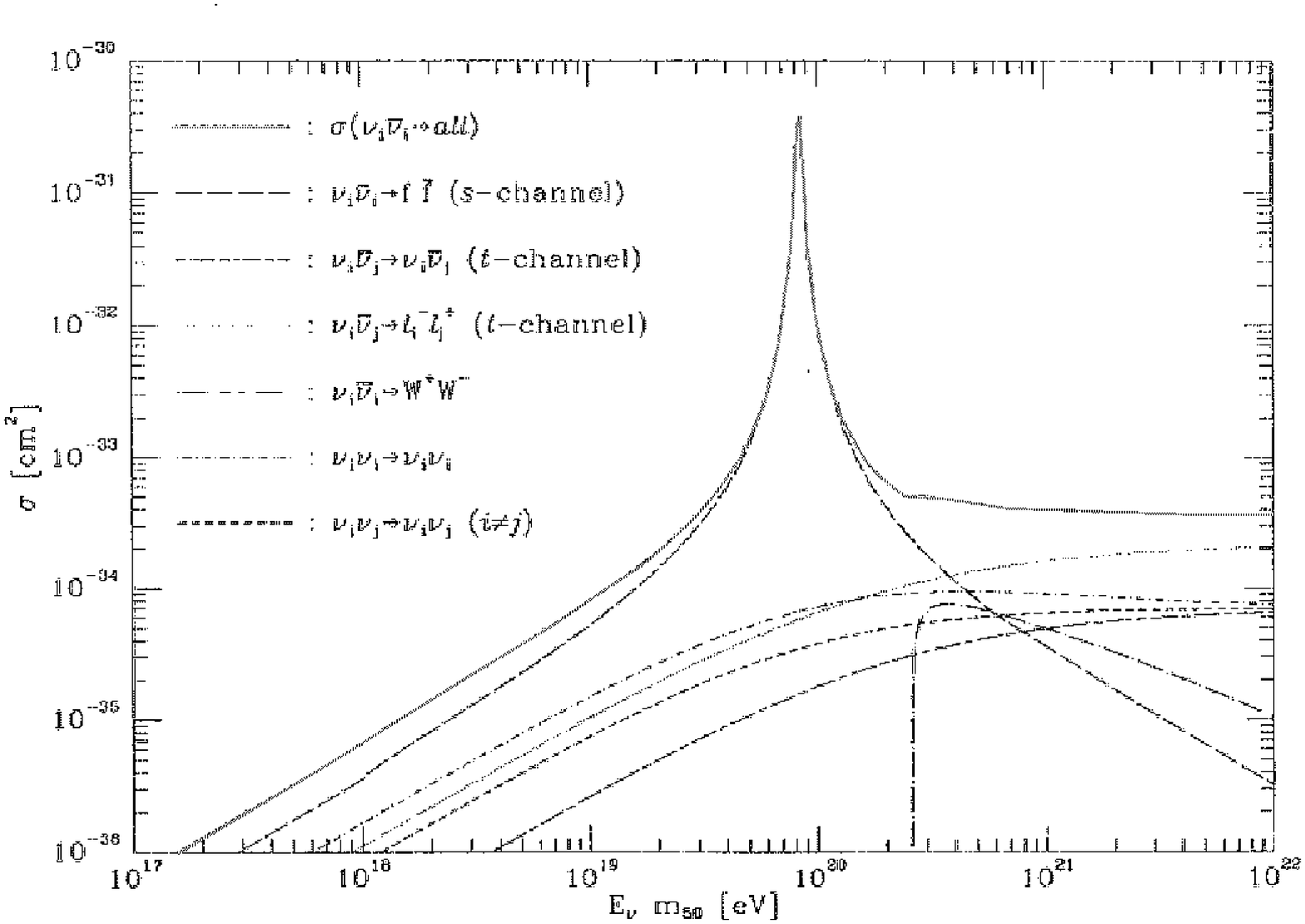}
  \caption[Z-burst cross section]{Cross section for the production of
    $Z$ bosons by ultra high energy neutrinos propagating through the
    1.9\,K black body neutrino background as a function of the
    neutrino energy times $m_{50} = m_\nu /$50\,eV (from
    \cite{Roulet:1992pz}).}
  \label{fig:z-burst_cross_section}
\end{figure}

The only cosmic ray process sensitive to these relic neutrinos is the
Z-burst scenario \cite{Weiler:1982qy}. An ultra high energy neutrino
emitted by a cosmic accelerator and propagating through the universe
can interact with a relic neutrino producing a $Z$ boson. The $Z$
decays in 70\% of all cases into hadronic jets, which produce
neutrinos with a somewhat lower energy, but with a great multiplicity.
This leads to a very high neutrino flux around the $Z$ resonance.
Figure~\ref{fig:z-burst_cross_section} shows the cross sections for
ultra high energy neutrinos propagating through the black body
neutrino background. For lower neutrino masses the resonance is
shifted to higher energies.

In Fig.~\ref{fig:flux_models} two different predictions for the
Z-burst neutrino flux are shown.

\section{Topological Defects}
\label{sec:topological_defects}

The last model to be presented in this section is different from the
other models discussed so far, because it does not require any
hadronic accelerators.

In the early universe, when the symmetry of the Higgs field was broken
at the GUT scale, regions which were causally not connected might have
acquired Higgs fields with different ``orientations''. The boundaries
between such domains with different ground states are called {\em
  topological defects} (TD), because their manifestation is determined
by the topology of these regions. Topological defects may appear as so
called domain walls, cosmic strings, monopoles, or textures
\cite{Gangui:2003uu}.

The energy associated with a topological defect is of the order of
magnitude of the GUT energy, i.e.~10$^{15}$\,GeV to 10$^{16}$\,GeV.
This energy is released during the annihilation or collapse of the
defects in the form of super massive gauge bosons and Higgs bosons,
which are usually referred to as $X$ particles \cite{Yoshida:1996ie}.

These $X$ particles can decay into leptonic and hadronic channels, the
latter one producing ultra high energy neutrinos in hadronic cascades,
which will not be associated with any cosmic accelerators. The
neutrino flux predicted from TD models is shown in
Fig.~\ref{fig:flux_models}.

\section{Summary of expected neutrino fluxes}
\label{sec:flux_summary}

\begin{figure}[ht]
  \centering
  \includegraphics[width=0.9\textwidth]{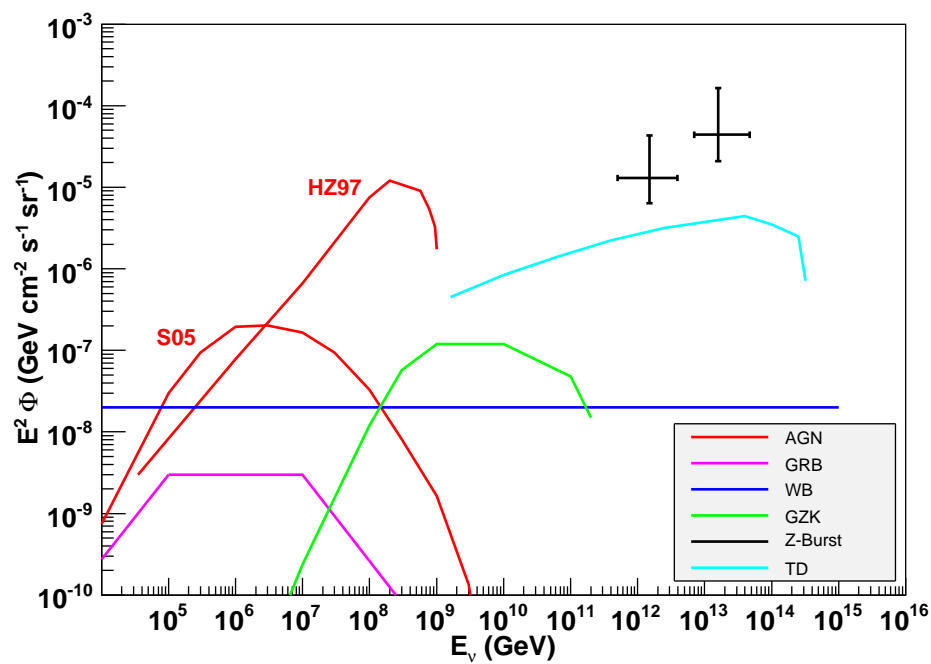}
  \caption[Expected flux of ultra high energy neutrinos]{Total flux
    (all flavours) of ultra high energy neutrinos predicted by the
    various theoretical models. AGNs: S05 from \cite{Stecker:2005hn},
    HZ97 from \cite{Halzen:1997hw}; GRB from \cite{Waxman:1997ti}; WB
    from \cite{Bahcall:1999yr}; GZK from \cite{Engel:2001hd}; Z-Bursts
    from \cite{Fodor:2001qy}; TD from \cite{Yoshida:1996ie} ($m_\nu =$
    1\,eV).}
  \label{fig:flux_models}
\end{figure}

Figure~\ref{fig:flux_models} shows the flux of ultra high energy
neutrinos predicted by the models discussed in this chapter. On the
ordinate the neutrino flux multiplied by $E_\nu^2$ is plotted, so that
models predicting a flux proportional to $E_\nu^{-2}$ as from first
order Fermi acceleration are flat in this representation. The plot
gives the total flux over all (anti-)neutrino flavours, since the
acoustic detection method discussed in this thesis is not sensitive to
lepton flavour. For models where in the literature only the $\nu_\mu +
\bar{\nu}_\mu$ flux at Earth is given, this flux is multiplied by a
factor of three, since due to neutrino oscillations an equipartition
between all neutrino flavours is expected if the neutrinos are
produced in distant and extended sources with a flavour ratio of
$\nu_e : \nu_\mu : \nu_\tau = 1 : 2 : 0$, which is the case for cosmic
neutrinos from pion decay.

\chapter{Detection of ultra high energy neutrinos}
\label{chap:experiments}
\minitoc

\bigskip In this chapter different experimental approaches towards the
detection of ultra high energy neutrinos are discussed and several
experiments already existing or starting in the near future are
presented.

\section{Water-\v{C}erenkov detectors}
\label{sec:water_cerenkov}

There are several experiments currently taking data or being built,
aiming at the detection of cosmic neutrinos in the high-energy range
from about 50\,GeV to 100\,TeV. All these experiments are based on the
Water-\v{C}erenkov (or Ice-\v{C}erenkov) technique, i.e.~on the
detection of \v{C}erenkov photons in the visual frequency band, which
are emitted by charged particles propagating at superluminal
velocities through a transparent medium (water or ice).

Water-\v{C}erenkov detectors consist of arrays of photomultiplier
tubes deeply embedded in water or ice and looking below the horizon
for upward going particles, since at these energies neutrinos are the
only particles able to propagate freely through the Earth. The
overburden of water or ice, typically several kilometres thick, acts
as a shielding against downward going muons from atmospheric air
showers, which are produced in interactions of cosmic rays in the
atmosphere.

\begin{figure}[ht]
  \centering
  \includegraphics[width=0.6\textwidth]{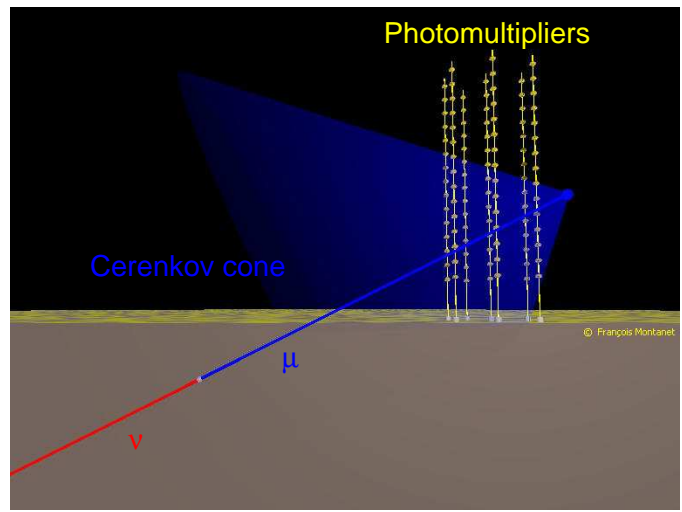}
  \caption[Detection principle of Water-\v{C}erenkov
  detectors]{Detection principle of Water-\v{C}erenkov detectors: A
    neutrino interacts in the surroundings of the detector and the
    resulting muon produces a cone of \v{C}erenkov light, which is
    detected by photomultipliers.}
  \label{fig:water_cerenkov}
\end{figure}

The detectors are designed to detect muon neutrinos interacting via
charged current weak interactions in the detector or in the
surrounding medium or bottom rock. The produced muon propagates freely
through the detector and emits \v{C}erenkov light. The muon path can
be reconstructed by measuring the arrival times and the corresponding
number of causally connected photoelectrons in the different
photomultipliers (cf.~Fig.~\ref{fig:water_cerenkov}). At high
energies, the direction of the muon coincides well enough with the
direction of the initial neutrino, so that the direction of the source
can be determined with an angular resolution better than 1$^\circ$ in
water. Due to the shorter scattering length of light in ice, the
angular resolution of Ice-\v{C}erenkov telescopes is slightly worse;
it is about 2$^\circ$.

The lower threshold energy of Water-\v{C}erenkov detectors, which is
about 50\,GeV, is set by the spacing of the photomultipliers.  For
lower energies the muon tracks become too short to be resolved by the
detector. The upper energy threshold is set by the target mass of the
detector. At higher energies, the neutrino flux becomes so low, that
the number of neutrinos from the different theoretical source models
expected to be measured within the lifetime of the experiment is
insufficient to get reliable information.

Table~\ref{tab:water_cerenkov_detectors} summarises the parameters of
a selection of existing and planned Water-\v{C}erenkov neutrino
telescopes, all of which deploy their photomultipliers fixed to
vertical structures which are either frozen into the ice or set on the
seabed.

The Baikal\footnote{http://www.ifh.de/baikal/baikalhome.html}
experiment is the oldest neutrino telescope which is still taking
data. It is situated in Lake Baikal in Siberia, which freezes in
winter and thus allows to use the ice sheet as a deployment platform.
AMANDA and its successor IceCube\footnote{http://icecube.wisc.edu/}
are located at the south pole. Their strings are deployed by hot water
drilling into the ice.  There are three detectors currently being
constructed in the Mediterranean Sea:
ANTARES\footnote{http://antares.in2p3.fr/},
NEMO\footnote{http://nemoweb.lns.infn.it/}, and
Nestor\footnote{http://www.nestor.org.gr/}, all using different
technical approaches. Further, there are plans for a cubic-kilometre
sized counterpart to IceCube in the northern hemisphere:
KM3NeT\footnote{http://www.km3net.org/}.

\begin{table}[hbt]
  \centering
  \caption[Parameters for a selection of Water-\v{C}erenkov neutrino
  telescopes]{Parameters for a selection of Water-\v{C}erenkov neutrino
    telescopes. ($V$: instrumented volume, $d_{xy}$: horizontal PMT
    spacing, $d_z$: vertical PMT spacing)}
  \label{tab:water_cerenkov_detectors}
  \begin{tabular}{llrrrrr}
    \hline
    & Medium & \# PMTs & \multicolumn{1}{c}{$V$} &
    \multicolumn{1}{c}{$d_{xy}$} & \multicolumn{1}{c}{$d_z$} &
    \multicolumn{1}{c}{Overburden} \\
    & & & \multicolumn{1}{c}{(km$^3$)} & \multicolumn{1}{c}{(m)} &
    \multicolumn{1}{c}{(m)} & \multicolumn{1}{c}{(m)} \\ \hline
    Baikal & fresh water & 192 & 10$^{-4}$ & 21 & 5, 7.5 & 1100 \\
    AMANDA & ice & 677 & 10$^{-2}$ & 50 & 14.0 & 1500 \\
    IceCube & ice & 4800 & 1.0 & 125 & 17.0 & 1400 \\
    ANTARES & sea water & 900 & 10$^{-2}$ & 70 & 14.5 & 2000 \\ \hline
  \end{tabular}
\end{table}

None of the existing experiments could measure an unambiguous signal
of cosmic neutrinos, yet. However, there has been a recent observation
by the AMANDA experiment of two neutrinos in spatial and temporal
coincidence with a TeV gamma flare of the blazar 1ES 1959+650
\cite{Halzen:2005pz}. Unfortunately, those events were only discovered
after the unblinding of the data, and thus their statistical
significance cannot be reliably estimated.

\section{Extensive Air Shower detectors}
\label{sec:eas}

Another promising approach towards the detection of ultra high energy
neutrinos are extensive air shower (EAS) detectors. EAS detectors are
designed for the detection of particle cascades produced by cosmic
rays in the atmosphere, but new telescopes like the Pierre Auger
Observatory\footnote{http://www.auger.org/} have apertures large
enough to detect also very rare neutrino induced cascades.

\begin{figure}[ht]
  \centering
  \includegraphics[width=0.4\textwidth]{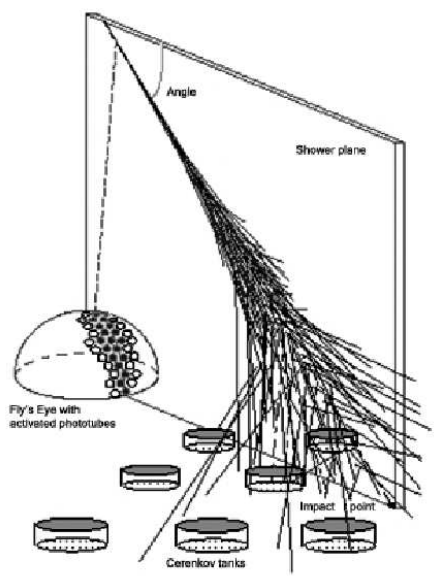}
  \caption[Detection principle of the Pierre Auger hybrid
  detector]{Detection principle of the Pierre Auger hybrid detector:
    The fluorescence light produced by an extensive air shower is
    detected in Fly's Eye imaging telescopes; particles reaching
    ground level are measured in Water-\v{C}erenkov tanks.}
  \label{fig:auger}
\end{figure}

The Pierre Auger Observatory is a hybrid detector, which uses two
independent experimental techniques to measure air showers. The
detection principle is shown in Fig.~\ref{fig:auger}: A surface array
of 1600 Water-\v{C}erenkov tanks is spread over an area of
3000\,km$^2$ in a triangular grid with 1.5\,km spacing. Each tank
contains 12\,m$^3$ of water, which is observed by three
photomultipliers to measure the spatial and temporal distribution of
air shower particles reaching ground level.  Further, the whole area
is observed by four fluorescence detectors, each of which has a field
of view of 180$^\circ$ in azimuth angle. These Fly's Eyes are designed
to detect the fluorescence light emitted by atmospheric nitrogen when
a particle cascade develops in the atmosphere. Their duty cycle is
about 10\% since they can only be operated at night and when the
weather conditions are suitable.

This hybrid setup allows the reconstruction of the primary particle
energy with an error of 9\%, and of its direction better than
1$^\circ$ (Surface detector only: 1.4$^\circ$ direction resolution and
10\% energy resolution).

The Pierre Auger Observatory can also detect ultra high energy
neutrinos if they interact in the atmosphere. Cosmic ray hadrons
interact at the top of the atmosphere\footnote{At an altitude between
  10 and 20\,km for a vertically downward going particle.}, whereas
neutrinos as only weakly interacting particles can interact at any
atmospheric depths. The depth of interaction can be fixed by determining
the depth of the shower maximum, i.e. the depth, where the cascade
particle density is highest. The shower maximum can either be measured
directly by the fluorescence detectors, or can be calculated from the
ratio of the number of muons to electrons arriving at ground level
and seen by the surface array, since the number of high-energy
electrons decreases faster than the number of muons because of their
larger electromagnetic cross section.

Promising neutrino candidates are inclined showers with zenith angles
larger than approximately 60$^\circ$, since those showers must have
been created deeply in the atmosphere when reaching the surface
detector. The neutrino flux predicted to be detectable within one year
of measurement time is plotted in Fig.~\ref{fig:flux_limits}.

\section{Radio-\v{C}erenkov detectors}

Even larger target masses can be observed if the signal to be detected
propagates over wide distances, and thus the number of single sensors
can either be reduced, or they can be distributed using larger
inter-sensor spacing. For example, radio waves can propagate over
large distances in nearly any medium. Such radio waves are emitted
coherently from the excess of negative charge in electromagnetic
cascades.

An electromagnetic cascade developing, especially in dense media,
gathers an excess of electrons, since positrons annihilate in the
medium and additional electrons are gathered via Compton scattering.
This propagating, uncompensated charge emits highly polarised
\v{C}erenkov radiation, bremsstrahlung, and transition radiation,
which are coherent for wavelengths larger than the extensions of the
cascade \cite{Askarian:1962}. Since ultra high energy electromagnetic
cascades in matter typically have a length of several metres, coherent
emission is expected in the radio frequency band.

This is called Askarian effect. It was verified experimentally by
directing picosecond pulses of GeV bremsstrahlung photons into a
silica sand target, where nanosecond radio-frequency pulses in the
0.3\,GHz to 6\,GHz range were measured \cite{Saltzberg:2000bk}. This
allows for simple and cheap experiments, since only conventional radio
antennas are required for detection.

In the next sections different experimental approaches to detect this
coherent Radio-\v{C}erenkov radiation are presented which presently
set the best experimental limits on the flux of ultra high energy
neutrinos. These upper limits on the neutrino flux are summarised in
Sec.~\ref{sec:flux_limits}.

\subsection{Radio detection in ice --- RICE}
\label{sec:rice}

The Radio Ice \v{C}erenkov
Experiment\footnote{http://www.bartol.udel.edu/\~\
  spiczak/rice/rice.html} (RICE) consists of an array of 18 radio
antennas deployed together with the AMANDA detector in the antarctic
ice cap. The antennas are spread over a 200 $\times$ 200 $\times$
200\,m$^3$ cube in a depth of 100 to 300\,m, and are sensitive from
0.2\,GHz to 1\,GHz \cite{Kravchenko:2001id}.

This setup is designed to detect the radio emission and reconstruct
the vertex of electromagnetic cascades induced by UHE neutrinos
($E_\nu \gtrsim$ 10$^7$\,GeV) in an effective volume of 2 $\times$ 2
$\times$ 1\,km$^3$ around the detector. Neutrino signals can be
discriminated from background by the characteristic conical emission
pattern of the cascade: the \v{C}erenkov cone, whereas the background
typically is produced by point sources, which emit spherical waves.
Spherical waves typically produce large-voltage signals in a higher
number of antennas than a signal limited to a cone.

RICE is taking data since 2000, and no neutrino candidates where
measured yet. The resulting limit on the neutrino flux is shown in
Fig.~\ref{fig:flux_limits}.

\subsection{Radio observations of the moon --- GLUE}
\label{sec:glue}

Looking for an even larger monolithic target, which, at the cost of a
higher energy threshold, can be observed with a single antenna, one
inevitably has to consider the moon.

The Goldstone Lunar Ultra-high energy neutrino
Experiment\footnote{http://www.physics.ucla.edu/\~\ moonemp/public/}
(GLUE) uses the 70\,m and the 34\,m Deep Space Network antennas at the
Goldstone Tracking Facility in California to detect 2\,GHz radio
signals from the moon, which are produced by neutrino induced
cascades.

\begin{figure}[ht]
  \centering
  \includegraphics[width=0.6\textwidth]{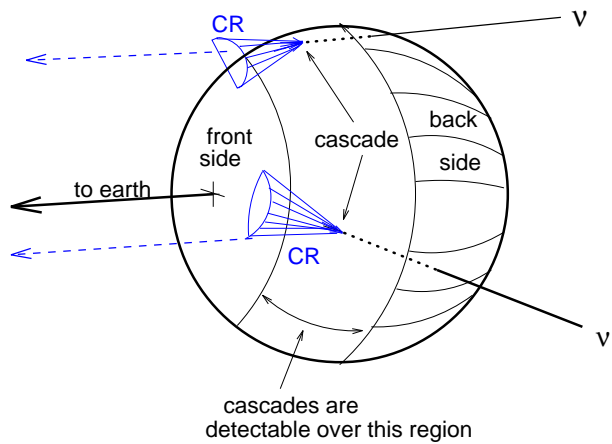}
  \caption[Geometry for lunar neutrino cascade detection]{Schematic of
    the geometry for lunar neutrino cascade detection (from
    \cite{Gorham:2001aj}).}
  \label{fig:glue}
\end{figure}

With this technique cascades up to a depth of 10\,m under the lunar
surface can be observed, which results in an effective volume
exceeding 10$^5$\,km$^3$ at highest energies \cite{Gorham:2001aj}.
Figure~\ref{fig:glue} shows the geometry under which cascades can be
observed, which is limited because the angle for total internal
reflection is to the first order the complement of the \v{C}erenkov
angle of 56$^\circ$, and the emitted \v{C}erenkov cone is collimated
with a spread of 1$^\circ$ FWHM. The upper limit on the diffuse
neutrino flux set by GLUE is shown in Fig.~\ref{fig:flux_limits}.

There are also suggestions to detect the radio emission at lower
frequencies in the 150\,MHz band \cite{Bacelar:2005}, where the
\v{C}erenkov cone is less collimated, and thus the limitations imposed
by total internal reflections are avoided. First estimates show that
the sensitivity to UHE neutrinos might be increased substantially
compared to the experiments working in the 2\,GHz band.

\subsection{Satellite experiments --- FORTE}
\label{sec:forte}

At highest energies the radio emission from particle cascades in ice
can not only be detected in the ice (cf. Sec.~\ref{sec:rice}) itself,
but also with satellites orbiting the Earth.

In 1997 NASA launched the FORTE (Fast On-orbit Recording of Transient
Events)
satellite\footnote{http://nis-www.lanl.gov/nis-projects/forte\_science/}
which carries several instruments for meteorological studies and the
U.S. nuclear detonation detection system. Among these are two
broadband (30 to 300\,MHz) radio-frequency dipole antennas that are
orthogonal to each other, and which can also be used to search for
radio emission from neutrino induced cascades.

FORTE orbits the Earth at an altitude of 800\,km with an inclination
of 70$^\circ$ towards the equator. From this orbit it cannot monitor
Antarctica, but it can detect events in the Greenland ice sheet with a
depth up to 1\,km. This results in an effective volume of
approximately 2 $\cdot$ 10$^6$\,km$^3$.  \v{C}erenkov radiation is
highly polarised which allows to select nanosecond pulses that are
neutrino candidates out of a background of lightning and other natural
phenomena.

The FORTE collaboration has observed one event, which survives all
their selection cuts, and which is still being analysed
\cite{Lehtinen:2003xv}. Assuming that this single event defines the
background rate, one can derive the limit on the neutrino flux shown
in Fig.~\ref{fig:flux_limits}.

\subsection{Balloon experiments --- ANITA}

A new experiment has been proposed which will be able to monitor
large masses of Antarctic ice at much lower cost as a satellite.

\begin{figure}[ht]
  \centering
  \includegraphics[width=0.5\textwidth]{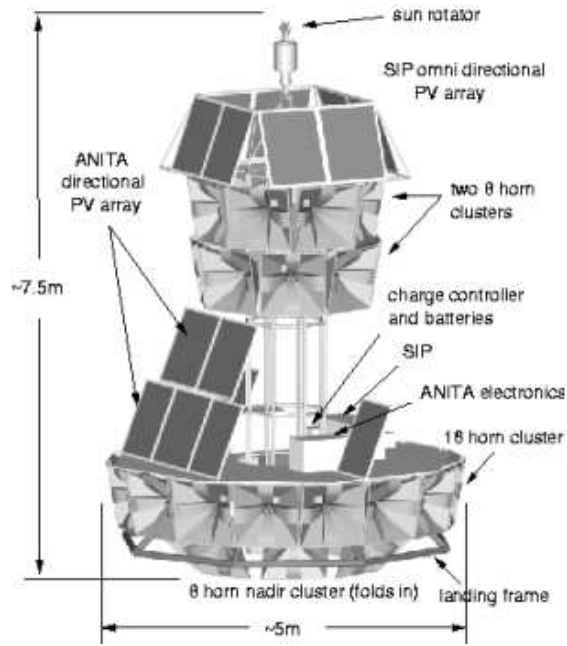}
  \caption[The ANITA instrument]{The ANITA instrument (from
    \cite{Miocinovic:2005jh}).}
  \label{fig:anita}
\end{figure}

The Antarctic Impulse Transient
Antenna\footnote{http://www.ps.uci.edu/\~\ anita/} (ANITA) is an
instrument planned to contain 40 dual polarisation horn antennas (cf.
Fig.~\ref{fig:anita}) operating from 0.2 to 1.2\,GHz
\cite{Miocinovic:2005jh}.

\begin{figure}[ht]
  \centering
  \includegraphics[width=0.5\textwidth]{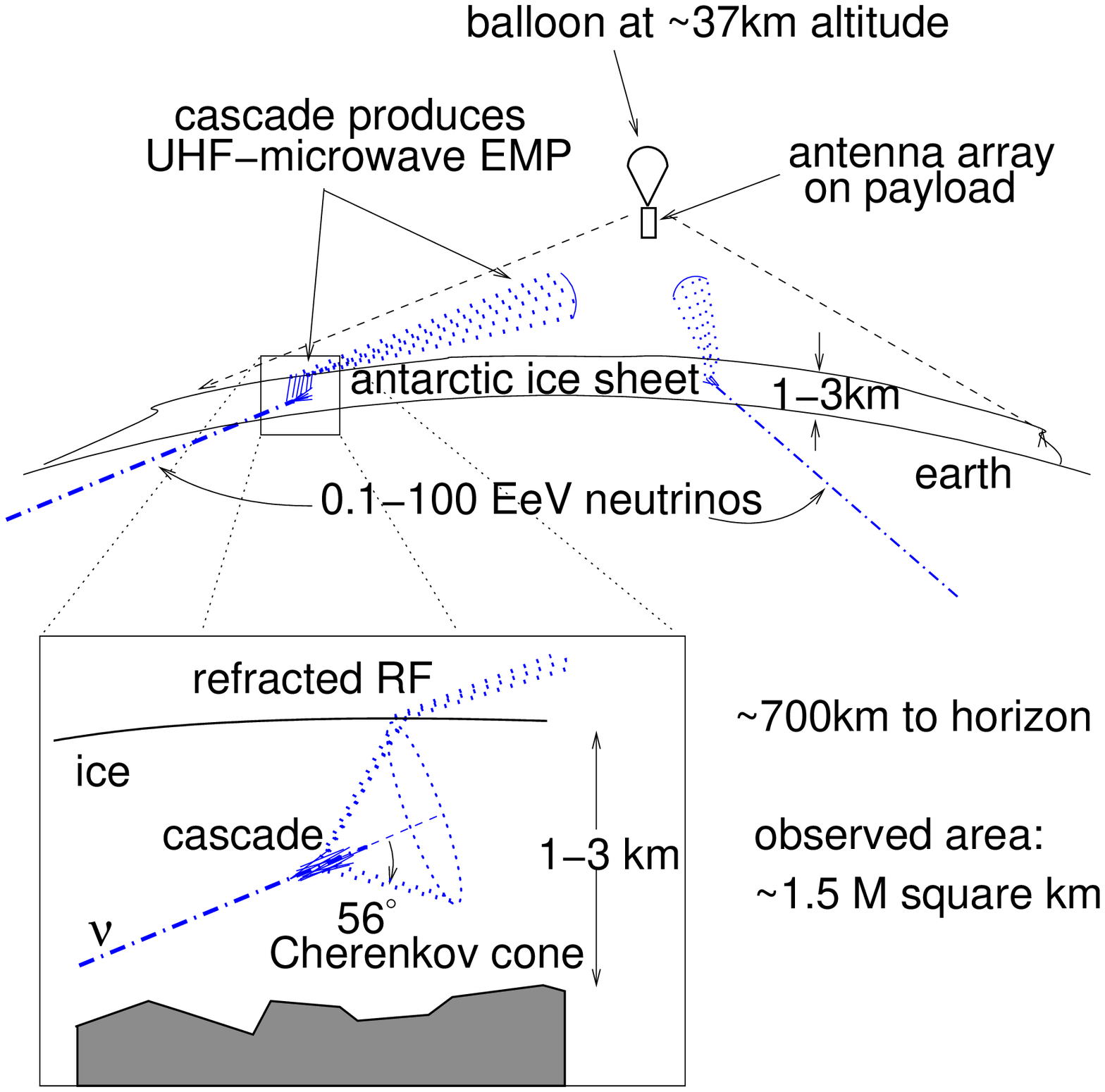}
  \caption[Detection principle of ANITA]{Schematic of the detection
    principle of balloon borne experiments \\
    (from \cite{Miocinovic:2005jh}).}
  \label{fig:anita_principle}
\end{figure}

ANITA will be attached to a balloon and circle the south pole at an
altitude of approximately 40\,km. As sketched in
Fig.~\ref{fig:anita_principle}, it will be able to detect nanosecond
radio pulses emitted from neutrino induced electromagnetic cascades in
the Antarctic ice cap with an effective volume of about
10$^6$\,km$^3$.

Due to the short propagation path of the radio signals compared to
FORTE, ANITA will be able to detect neutrinos with a much lower energy
threshold of 10$^9$\,GeV. The neutrino flux expected to be detectable
with ANITA during three flights of about 15 days each is shown in
Fig.~\ref{fig:flux_limits}. The ANITA collaboration has demonstrated
the feasibility of the concept with the flight of ANITA-lite
\cite{Miocinovic:2005jh}, and expects to be able to perform one to two
flights per Antarctic summer season starting in 2006.

\section{Present limits on the neutrino flux}
\label{sec:flux_limits}

Table~\ref{tab:detectors} gives, for comparison, an overview over some
of the presented existing and planned detectors for ultra high energy
neutrinos, stating the energy range and the effective volumes
$V_\mathrm{eff}$ inside which neutrino events can be detected at the
highest energies. The upper energy threshold $E_\mathrm{max}$ is set
by the fact, that for the theoretically predicted fluxes (cf.
Chap.~\ref{chap:sources}) no events above this energy are expected
during the lifetime of the experiment.

\begin{table}[hbt]
  \centering
  \caption[Summary of some of the presented neutrino
  detectors]{Summary of some of the
    presented neutrino detectors ($V_\mathrm{eff}$: effective volume,
    $E_\mathrm{min}$: lower neutrino-energy threshold,
    $E_\mathrm{max}$: upper energy threshold)}
  \label{tab:detectors}
  \begin{tabular}{llrll}
    \hline
    & Detection principle & \multicolumn{1}{c}{$V_\mathrm{eff}$} &
    \multicolumn{1}{c}{$E_\mathrm{min}$} &
    \multicolumn{1}{c}{$E_\mathrm{max}$} \\
    & & \multicolumn{1}{c}{(km$^3$)} & \multicolumn{1}{c}{(GeV)} &
    \multicolumn{1}{c}{(GeV)} \\ \hline
    Baikal & Water-\v{C}erenkov & & 10$^4$ & 10$^8$ \\
    Auger & Air shower & & 10$^8$ & 10$^{11}$ \\
    RICE & Radio-ice in ice & 4 & 10$^7$ & 10$^{11}$ \\
    GLUE & Radio-moon from Earth & 10$^5$ & 10$^{11}$ & 10$^{14}$ \\
    FORTE & Radio-ice from satellite & 2 $\cdot$ 10$^6$ & 10$^{13}$ &
    10$^{17}$ \\
    ANITA & Radio-ice from balloon & 10$^6$ & 10$^9$ & 10$^{14}$ \\
    \hline
  \end{tabular}
\end{table}

\begin{figure}[ht]
  \centering
  \includegraphics[width=0.9\textwidth]{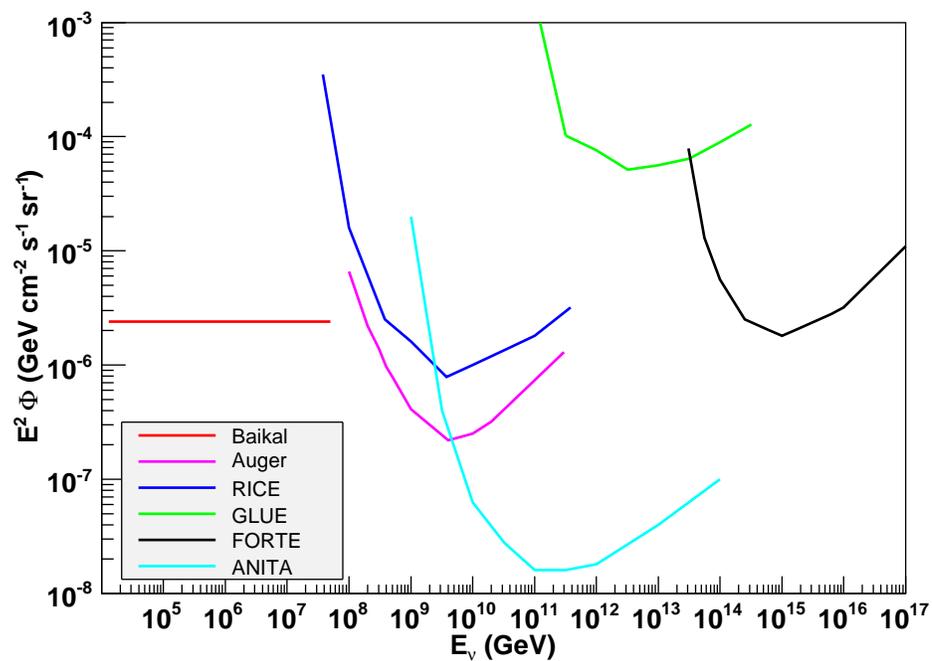}
  \caption[Experimental limits on the UHE neutrino flux]{Experimental
    limits on the flux of ultra high energy neutrinos (Baikal from
    \cite{Wischnewski:2005rr}, Auger from \cite{Bertou:2001vm}, RICE
    from \cite{Miocinovic:2005jh}, GLUE from \cite{Gorham:2003da},
    FORTE from \cite{Lehtinen:2003xv}, ANITA from
    \cite{Miocinovic:2005jh}).}
  \label{fig:flux_limits}
\end{figure}

In Fig.~\ref{fig:flux_limits} the experimental limits on the neutrino
flux, which are set by the different experiments, are summarised. The
calculation of these flux limits is discussed in detail in
Appendix~\ref{chap:limits}. The presented limits are for a 90\%
confidence level (C.L.). The Auger limit is for one year of lifetime;
the ANITA limit is for 45 days, expected to be cumulated during three
flights.

\begin{figure}[ht]
  \centering
  \includegraphics[width=0.9\textwidth]{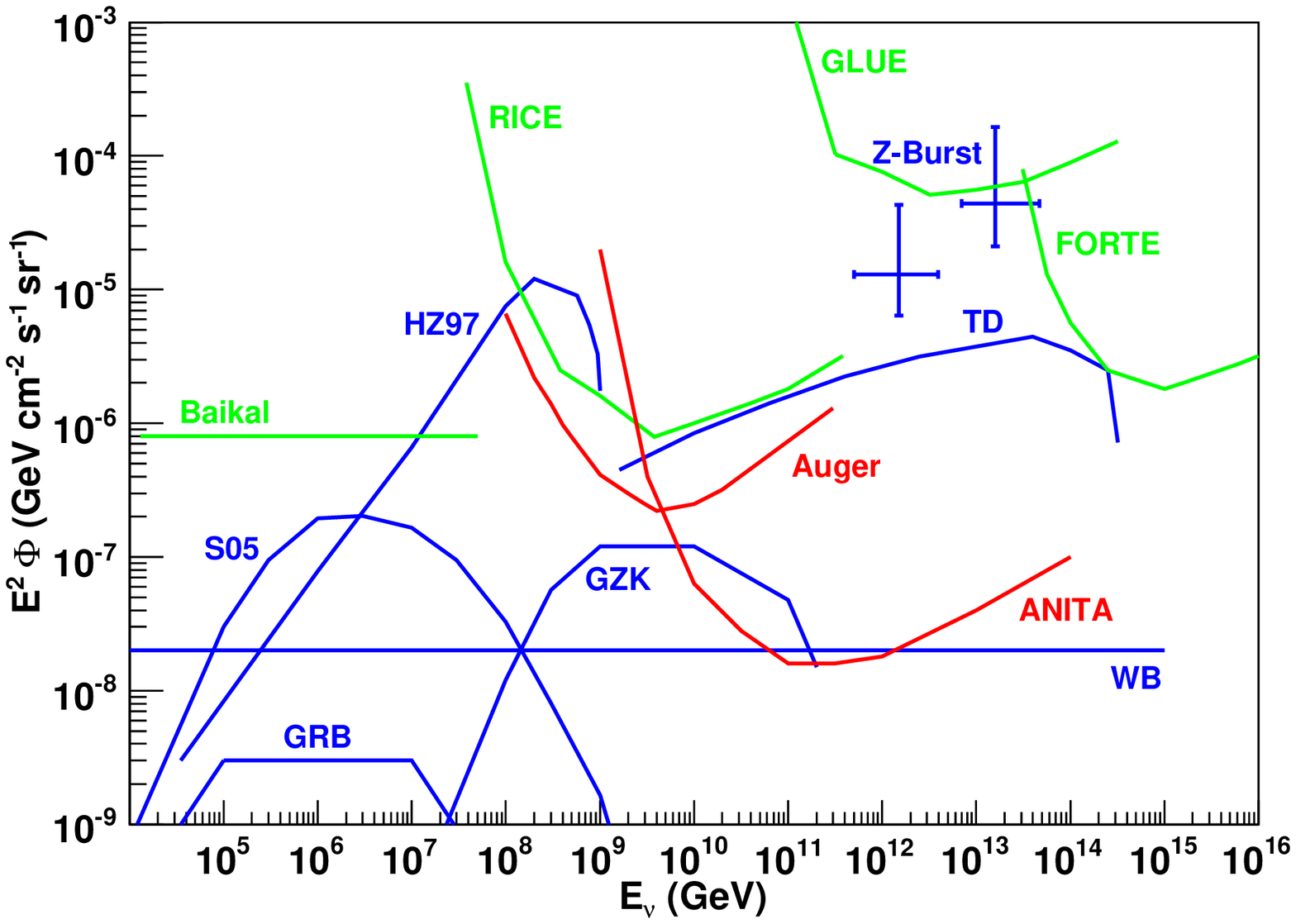}
  \caption[Comparison of theoretical flux models and experimental
  limits]{Comparison of theoretical flux models and experimental
    limits. Blue curves are the theoretically predicted fluxes from
    Fig.~\ref{fig:flux_models}. Green and red curves are the
    experimental flux limits shown in Fig.~\ref{fig:flux_limits}
    (Green curves: limits measured by existing experiments; red
    curves: limits projected for future experiments).}
  \label{fig:models_vs_limits}
\end{figure}

The experimental limits are compared to the theoretically predicted
flux models shown in Fig.~\ref{fig:models_vs_limits}. The AGN model
HZ97 could be excluded by the Baikal and RICE experiments.  GLUE could
already rule out different Z-burst scenarios not presented in this
work. Auger and ANITA will be able to test the Topological Defect
model presented. ANITA will further be sensitive to GZK and WB
neutrinos.

\chapter{Thermoacoustic sound generation}
\label{chap:thermoacoustic_model}
\minitoc

\bigskip In this chapter a new approach towards the detection of
ultra high energy neutrinos is introduced: Acoustic detection.
Thermoacoustic sound generation in general is discussed, and
experimental verifications of the thermoacoustic model are presented.

\section{Theoretical considerations}

Acoustic particle detection is based on the thermoacoustic model -- a
hydrodynamic theory, which was first discussed by G.A.~Askarian in
1957 \cite{Askarian:1957, Askarian:1979zs}.

The basic idea is, that energy deposited locally in a medium, e.g. by
a particle cascade, leads to local heating of the medium, and thus to
a fast expansion, which spreads in form of a sonic wave from the place
of energy deposition. This sound wave is described by a pressure field
$p (\vec{r}, t)$, where $p$ is a small deviation from the static
pressure $p_0$ of the fluid. This pressure field can be calculated
from the energy density $\varepsilon (\vec{r}, t)$ deposited in the
medium. The details are worked out in
Appendix~\ref{chap:hydrodynamics}, and the result describing the
pressure field is the inhomogeneous wave equation
(\ref{eq:thermoacoustic_model_app}):

\begin{equation}
  \label{eq:thermoacoustic_model}
  \frac{1}{c^2} \frac{\partial^2 p}{\partial t^2} - \Delta p =
  \frac{\alpha}{C_p} \frac{\partial^2 \varepsilon}{\partial t^2}
\end{equation}

\noindent Here, $c$ is the speed of sound, $\alpha$ is the bulk
expansion coefficient, and $C_p$ is the specific heat capacity at
constant pressure. Equation~\ref{eq:thermoacoustic_model} is solved by
a {\em Kirchhoff integral}:

\begin{equation}
  \label{eq:kirchhoff_long}
  p (\vec{r}, t) = \frac{\alpha}{4 \pi C_p} \, \int \frac{\mathrm{d}^3
    r^\prime}{\vert \vec{r} - \vec{r^\prime} \vert}
  \frac{\partial^2}{\partial t^2} \, \varepsilon \left(
    \vec{r^\prime}, t - \frac{\vert \vec{r} - \vec{r^\prime} \vert}{c}
  \right)
\end{equation}

This can further be simplified, if we assume an instantaneous energy
deposition at time $t_0$, i.e. an energy deposition on much shorter
timescales, than the propagation of the acoustic wave:

\begin{equation}
  \label{eq:instantaneous}
  \varepsilon (\vec{r}, t) = \tilde{\varepsilon} (\vec{r}) \, \Theta
  (t - t_0) \quad \textrm{or} \quad \frac{\partial}{\partial t} \,
  \varepsilon (\vec{r}, t) = \tilde{\varepsilon} (\vec{r}) \, \delta
  (t - t_0)
\end{equation}

\noindent where $\Theta (t)$ is the Heaviside step function and
$\delta (t)$ is the Dirac delta distribution. This assumption is
certainly true for the sound generation of particle cascades in water
which are considered in this work. Particle cascades, as discussed in
the following sections, typically develop with the speed of light in
vacuum, whereas the thermoacoustic sound generation is governed by the
speed of sound\footnote{The speed of sound in water is $c \approx
  1500\,\mathrm{m}/\mathrm{s}$.}, which is much smaller. Inserting
(\ref{eq:instantaneous}) into (\ref{eq:kirchhoff_long}) leads to the
following equation:

\begin{equation}
  \label{eq:kirchhoff_short}
  p (\vec{r}, t) = \frac{\alpha}{4 \pi C_p} \, c^2 \,
  \frac{\partial}{\partial R} \, \int_{S^{R}_{\vec{r}}} \mathrm{d}^2
  r^\prime \, \frac{\tilde{\varepsilon} (\vec{r^\prime})}{R} \quad
  \quad (t > t_0)
\end{equation}

\noindent where the integration is over spherical surfaces
$S^{R}_{\vec{r}}$ centred at the observation point $\vec{r}$ and with
radius $R = c (t - t_0)$. This means, that for a given position
$\vec{r}$ at any time $t$ the thermoacoustic pressure signal is a
superposition of all the signal contributions produced in a distance
$c (t - t_0)$, i.e. of all the signal contributions that propagate in
the given time interval from the production location considered to the
observation point.

\begin{figure}[ht]
  \centering
  \includegraphics[width=0.9\textwidth]{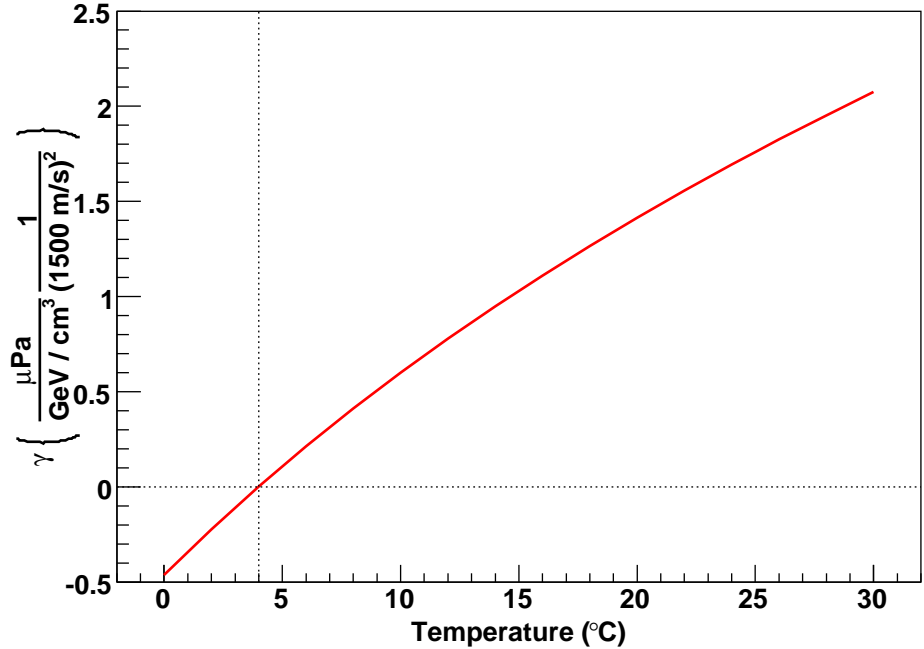}
  \caption[The $\gamma$ parameter of water as a function of
  temperature]{The $\gamma$ parameter of water as a function of
    temperature in units convenient for the purposes of acoustic
    particle detection (Data from \cite{Kretzschmar}).}
  \label{fig:gamma}
\end{figure}

\bigskip Using water as detection medium, one has to take further
considerations into account. The expansion coefficient $\alpha$ and
the heat capacity $C_p$ both strongly depend on the temperature $T$.
Especially, $\alpha$ vanishes at $4^\circ$C due to the anomaly of
water. The temperature dependent factor in (\ref{eq:kirchhoff_short})
is defined as $\gamma$, and the temperature dependence is shown in
Fig.~\ref{fig:gamma}.

\begin{equation}
  \label{eq:gamma}
  \gamma = \frac{\alpha}{4 \pi C_p}
\end{equation}

There are several groups studying the feasibility of using the
acoustic signal produced by particle cascades from neutrino
interactions as a basis for a future large-scale acoustic neutrino
detector. In-situ acoustic measurements using existing arrays (civil
and military) of hydrophones as well as custom made detectors have
been carried out in Lake Baikal \cite{Budnev:2005}, in the Atlantic
Ocean near the Bahamas \cite{Lehtinen:2001km}, and in the
Mediterranean Sea \cite{Riccobene:2004hm, Naumann:2005}.

These experiments mainly allowed, apart from addressing the technical
feasibility of such a detector, to study the important subject of
acoustic background noise in water (cf.~Sec.~\ref{sec:noise}). Also an
experimental limit on the diffuse neutrino flux at energies $E_\nu
\gtrsim 10^{13}$\,GeV could already be derived from the SAUND
experiment at the Bahamas consisting of seven military hydrophones
\cite{Vandenbroucke:2004gv}, but it is not yet on a competitive basis
compared to the FORTE result (Sec.~\ref{sec:forte}), which is at the
moment the best limit in this energy range.

For all those experiments, the vanishing of $\gamma$ at 4$^\circ$C,
and with it of the thermoacoustic signal, poses no problem since they
operate at locations with modest water temperature\footnote{The
  experiment in Lake Baikal has great systematic limitations due to
  the temperature gradient between $0^\circ$C at the frozen surface
  and $4^\circ$C at the bottom.}. For example, at the bottom of the
Mediterranean Sea there is a constant temperature between 12$^\circ$C
and 14$^\circ$C.

\section{Experimental verification of the thermoacoustic model}

In order to study the validity of the thermoacoustic model and to test
custom designed hardware like hydrophones we performed several
experiments.  Different means of energy deposition -- a proton beam
and a laser -- were used to mimic the line-like energy deposition of a
particle cascade under controlled conditions. We investigated the
dependence of the thermoacoustic pulses on the beam parameters, i.e.
intensity and beam width, and on the water temperature. The pulses
were measured with both commercial and custom-made hydrophones and
were compared to Monte Carlo simulations.

\subsection{The proton beam experiment}

\subsubsection{Experimental setup}

To test the applicability of the thermoacoustic model to the detection
of ultra high energy neutrinos by measuring the sound pulse generated
from the neutrino induced particle cascade, an intuitive approach is
to dump a particle beam from an accelerator into a water target. It
turns out that proton beams with particle energies ranging from a
100\,MeV to 1\,GeV are ideal candidates for such an experiment, since
their range in water is a few 10\,cm, a distance most suitable for
laboratory setups.

Several such experiments were performed in the past
\cite{Sulak:1978ay, Hunter:1981, Albul:2001um}, but the results always
had large statistical errors as well as systematic uncertainties.
Thus, the thermoacoustic sound generation model is verified in
principle, but no high precision measurements exist.

We designed and carried out an experiment \cite{Graf:2004, Graf:2005},
that would on the one hand allow for precision measurements of the
thermoacoustic pressure pulses, especially as a function of the
temperature\footnote{The temperature dependence is important to
  distinguish the thermoacoustic signal from pulses produced by other
  possible sound generation mechanisms which are supposed to persist
  at 4$^\circ$C.}, and on the other hand give the possibility to test
acoustic sensors custom-made for in-situ studies of the background
noise in the Mediterranean Sea. In this work I will concentrate on the
Monte Carlo studies used to analyse the measured signals. Details on
the technical realisation of the experiment can be found in
\cite{Graf:2004}.

\begin{figure}[ht]
  \centering
  \subfigure[Experimental setup.]{
    \label{subfig:proton_setup_a}
    \includegraphics[width=0.45\textwidth]
    {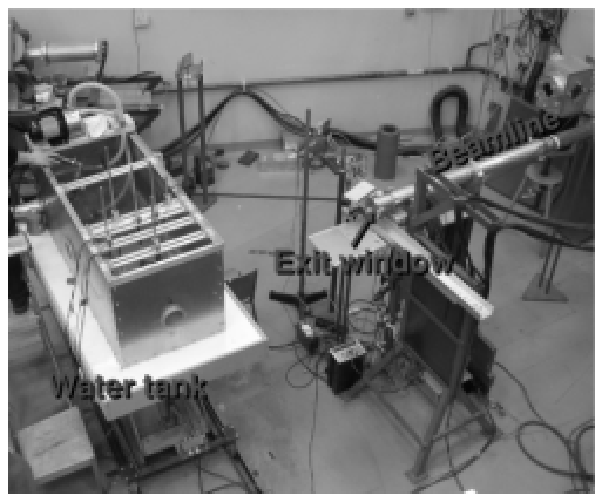}}
  \qquad
  \subfigure[Schematic and coordinate system.]{
    \label{subfig:proton_setup_b}
    \includegraphics[width=0.45\textwidth]
    {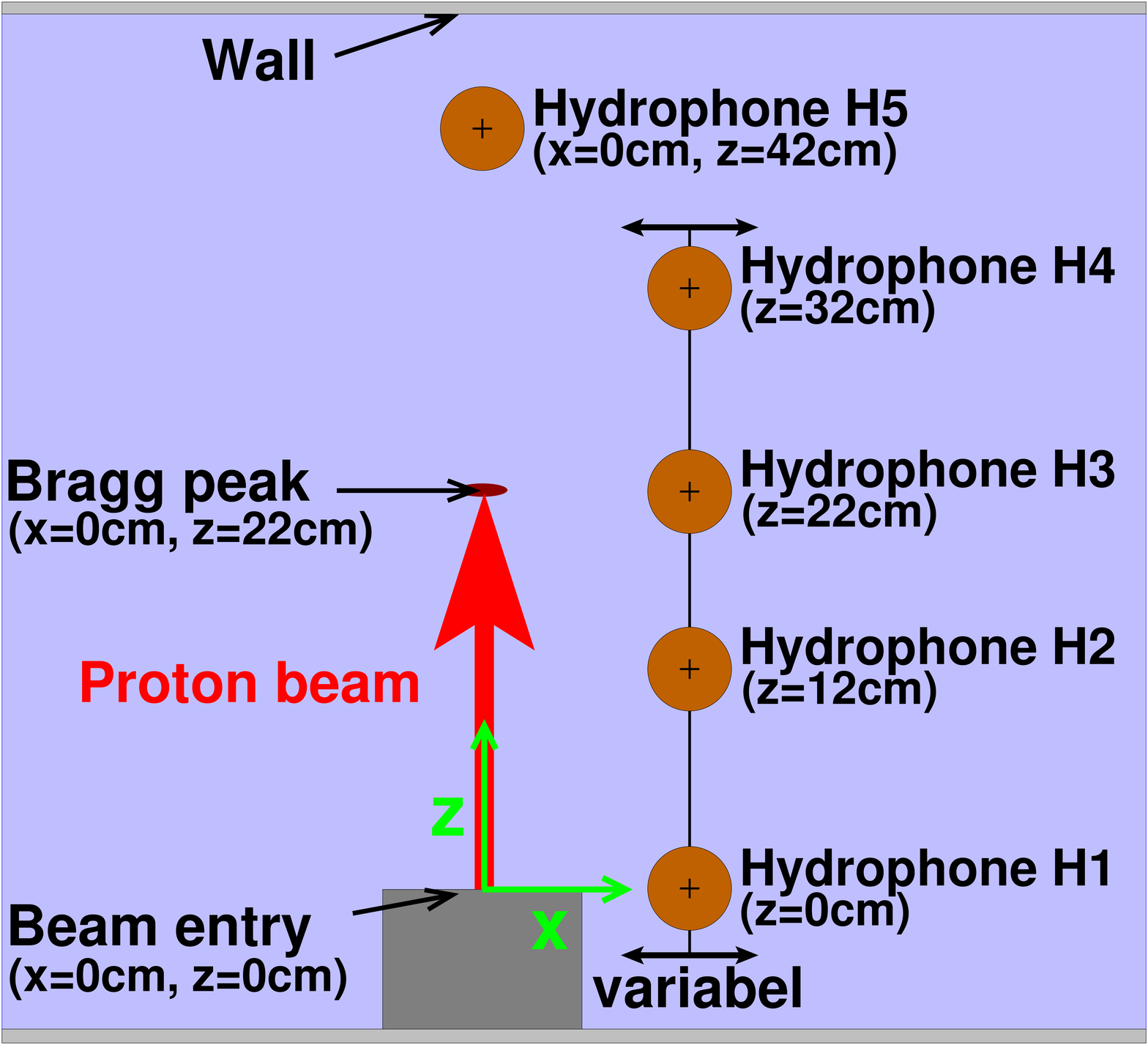}}
  \caption[Experimental setup of the proton beam
  experiment]{Experimental setup of the proton beam experiment:
    \subref{subfig:proton_setup_a} The water tank set up at the end of
    the beam line, and \subref{subfig:proton_setup_b} schematic of the
    hydrophone positions relative to the beam entry and coordinate
    system used.}
  \label{fig:proton_setup}
\end{figure}

\bigskip For the experiment, the 177\,MeV proton beam of the Gustaf
Werner Cyclotron at the The Svedberg
Laboratory\footnote{http://www.tsl.uu.se/} of the University of
Uppsala, Sweden was used. The beam was dumped into a $150 \times 60
\times 60 \, \mathrm{cm}^3$ water tank. The beam entry window is set
about 10\,cm from the tank walls into the water volume to delay the
signal reflections from the walls and separate them from the direct
signal.  This is achieved by a plastic pipe set into the tank wall,
where one end is sealed with plastic foil. The acoustic signals were
recorded by up to five position adjustable hydrophones simultaneously,
and digitised with a rate of $10 \, \mathrm{MS} / \mathrm{s}$ to $100
\, \mathrm{MS} / \mathrm{s}$. This rate corresponds to a time of
100\,ns to 10\,ns between two samples which is sufficiently short for
an expected length of the thermoacoustic pulse in the order of
100\,$\mu$s. The data were stored on disc and analysed off-line. The
experimental setup is shown in Fig.~\ref{fig:proton_setup}.

\subsubsection{Monte Carlo simulations}
\label{sec:proton_mc}

To analyse the recorded data we developed a simulation code. In a
first step the energy deposition of a single proton in water was
studied with GEANT4 \cite{Agostinelli:2002hh}, an Open
Source\footnote{http://geant4.web.cern.ch/geant4/} simulation
package for high energy physics written in C++.

\begin{figure}[ht]
  \centering
  \subfigure[Bragg peak.]{
    \label{subfig:proton_profile_a}
    \includegraphics[width=0.48\textwidth]{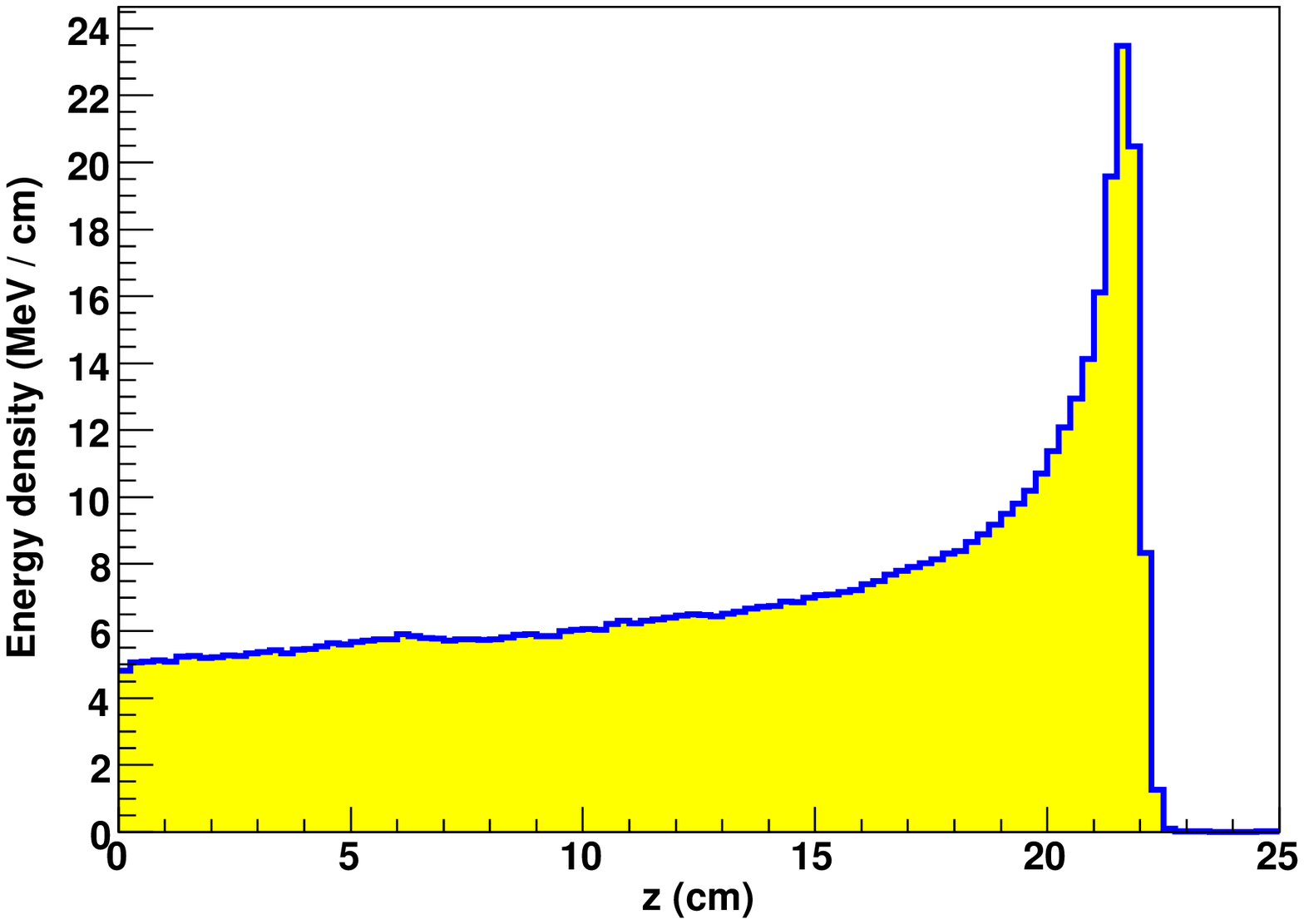}}
  \subfigure[Profile.]{
    \label{subfig:proton_profile_b}
    \includegraphics[width=0.48\textwidth]{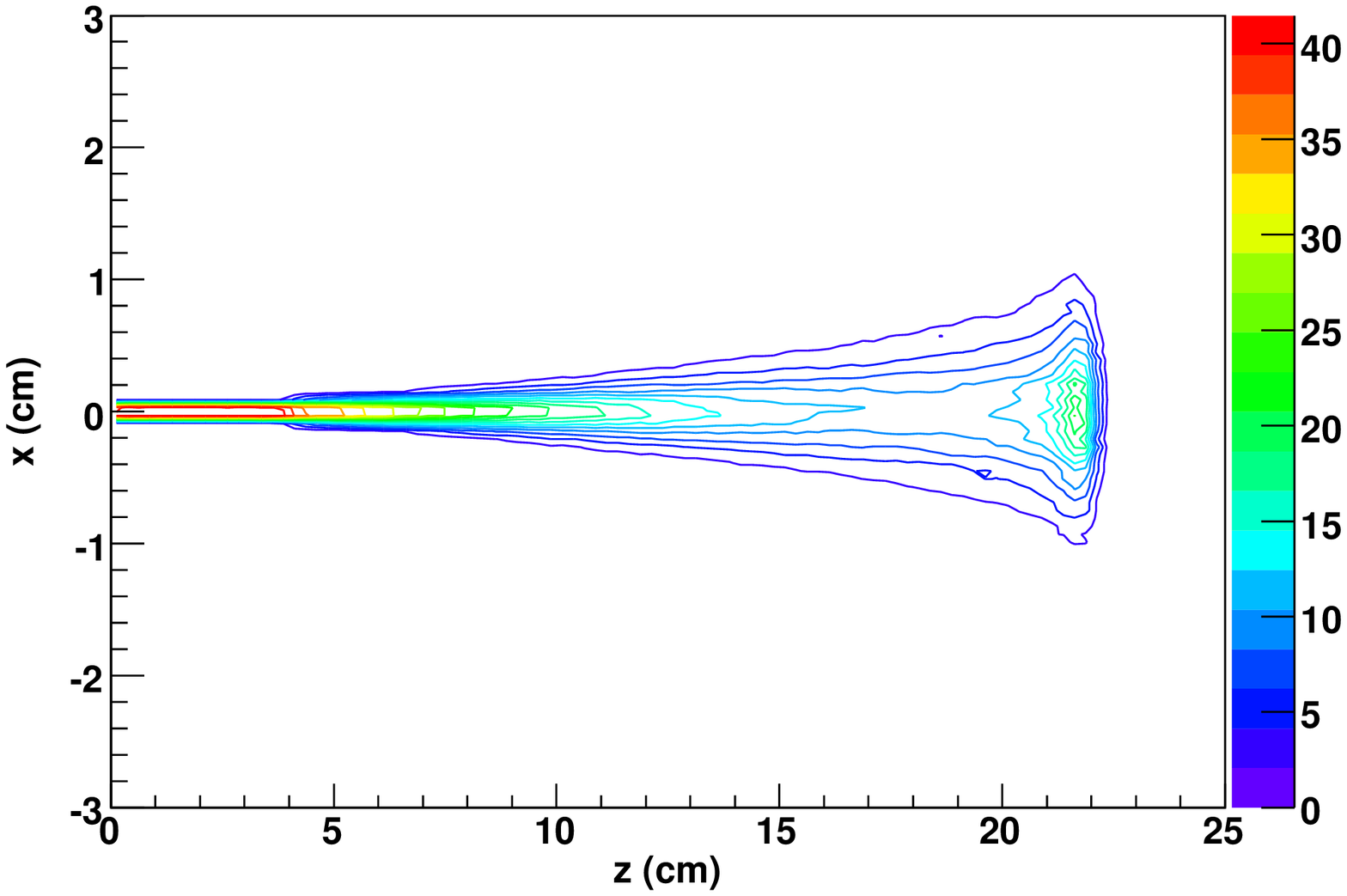}}
  \caption[Energy density deposited by a 180\,MeV proton in
  water]{Energy density deposited by a 180\,MeV proton in water:
    \subref{subfig:proton_profile_a} Projection onto the direction of
    the proton shows dense energy deposition at the end of the track
    (Bragg peak), and \subref{subfig:proton_profile_b} projection onto
    a plane containing the proton path (colour scale in MeV/cm$^2$).}
  \label{fig:proton_profile}
\end{figure}

The main energy loss mechanism for 180\,MeV protons in a medium is
ionisation, which can be simulated with GEANT4 taking into account
also scattering along the path. Figure~\ref{fig:proton_profile} shows
the mean\footnote{Averaged over 10000 single protons.} energy density
deposited by a single 180\,MeV proton in water. The proton enters the
water at the origin of the coordinate system and propagates in
positive $z$-direction.

Figure~\ref{subfig:proton_profile_a} shows that much energy is
deposited towards the end of the track, in the so-called ``Bragg
peak'' at $z = 22$\,cm. But as can be seen in
Fig.~\ref{subfig:proton_profile_b}, which is rotationally symmetric
around the beam axis, this energy is spread over a large volume, since
scattering angles become larger\footnote{The transversal momentum
  gained in each scattering process is fairly constant, but the
  longitudinal momentum of the proton decreases with energy, which
  leads to increasing scattering angles.} in the lower energy part of
the track. Thus, the energy density is highest in the first 5\,cm of
the track. We store the complete 3-D energy information in a histogram
with $50 \times 50 \times 50$ bins.

Also, we used GEANT4 to study the influence of the 1.2\,m of air and
the 2\,mm plastic foil which the protons have to traverse after
leaving the beam line, and before entering the water. We found that
these factors can be completely neglected. Further, we compared the
results to the former gold standard simulation code Geant~3.21 and
found excellent agreement.

\begin{figure}[ht]
  \centering
  \subfigure[Temporal distribution.]{
    \label{subfig:bunch_profile_a}
    \includegraphics[width=0.48\textwidth]
    {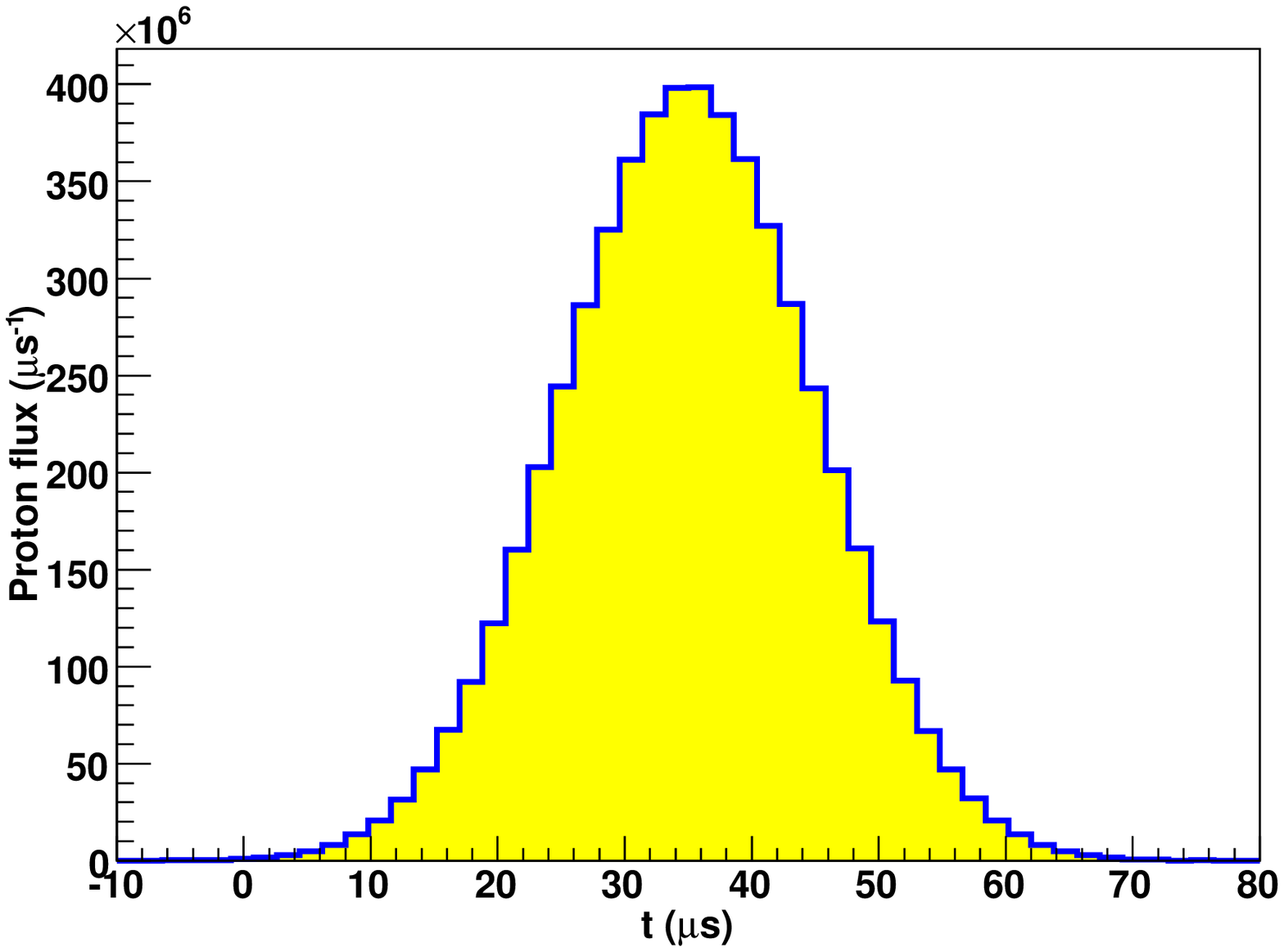}}
  \subfigure[Profile.]{
    \label{subfig:bunch_profile_b}
    \includegraphics[width=0.48\textwidth]
    {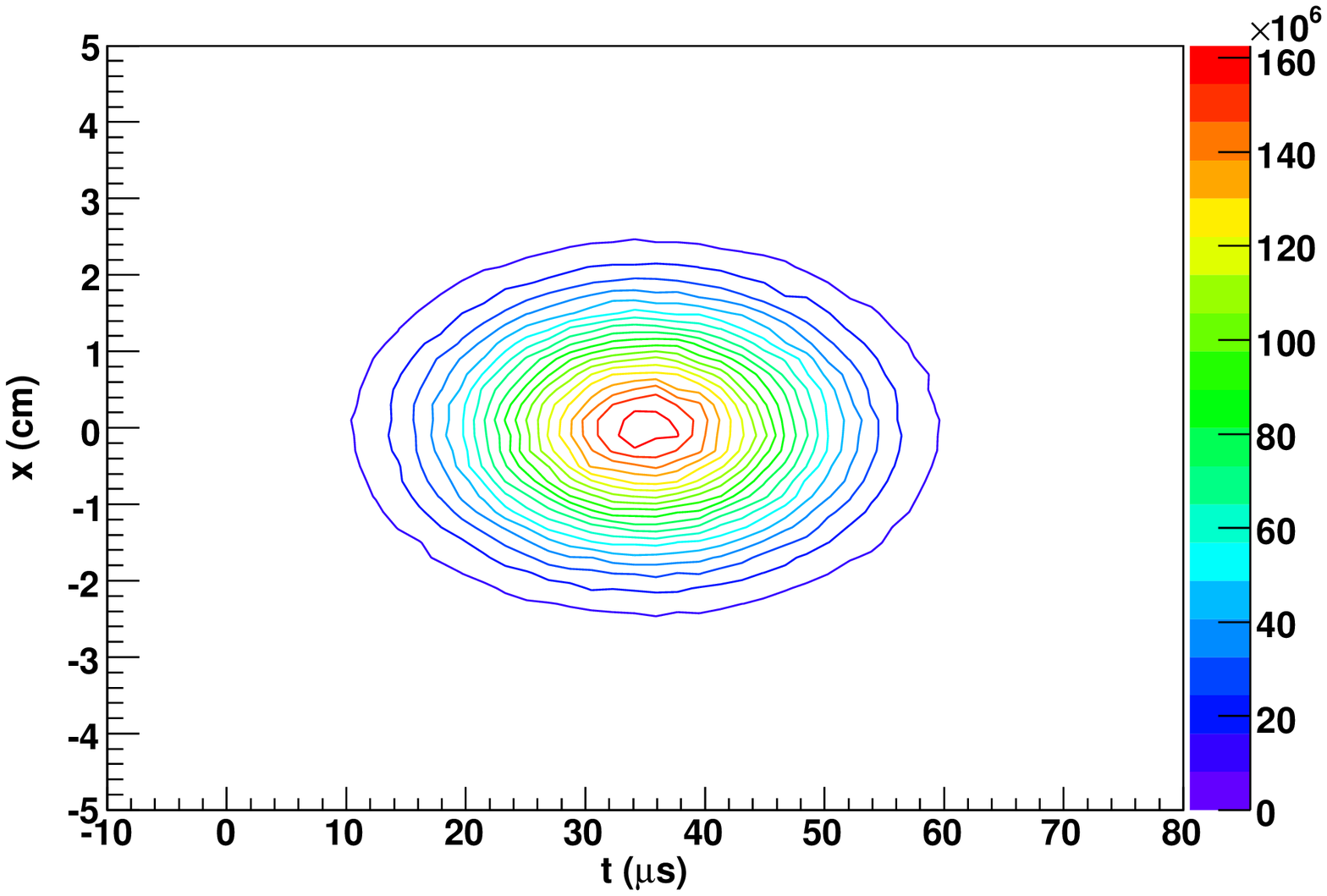}}
  \caption[Profile of the proton bunch]{Profile of the bunch of
    $\approx 10^{10}$ protons (bunch energy 1.8\,EeV) delivered by the
    cyclotron: \subref{subfig:bunch_profile_a} Temporal distribution
    of the protons and \subref{subfig:bunch_profile_b} profile of the
    proton bunch.}
  \label{fig:bunch_profile}
\end{figure}

The simulated energy deposition from a single proton is used to
calculate the energy deposited by a proton bunch delivered from the
accelerator. We assume, that there is no interaction between the
protons within a bunch, and thus get the total energy deposition as a
convolution between the single proton deposition and the proton flux
in the bunch. A typical bunch used for this experiment is shown in
Fig.~\ref{fig:bunch_profile}.

The bunch width is variable, and typically a width of 2\,cm FWHM was
used. The temporal distribution of the protons is fixed by the
accelerator and can be described as a Gaussian with 24\,$\mu$s FWHM.
The convolution leads to a set of 50 histograms $h (t_i)$ of $50
\times 50 \times 50$ bins, one for each time slice $t_i$. Each
histogram contains the energy deposited in time slice $i$. Since all
energy dissipation mechanisms are much slower than the energy
deposition, the total energy deposited until time $t_i$ can be
calculated as the sum over all histograms $\varepsilon (t_i) = \sum_{j
  \le i} h (t_j)$.

\begin{figure}[ht]
  \centering
  \subfigure[Pressure vs. time.]{
    \label{subfig:p_pulses_sim_a}
    \includegraphics[width=0.48\textwidth]
    {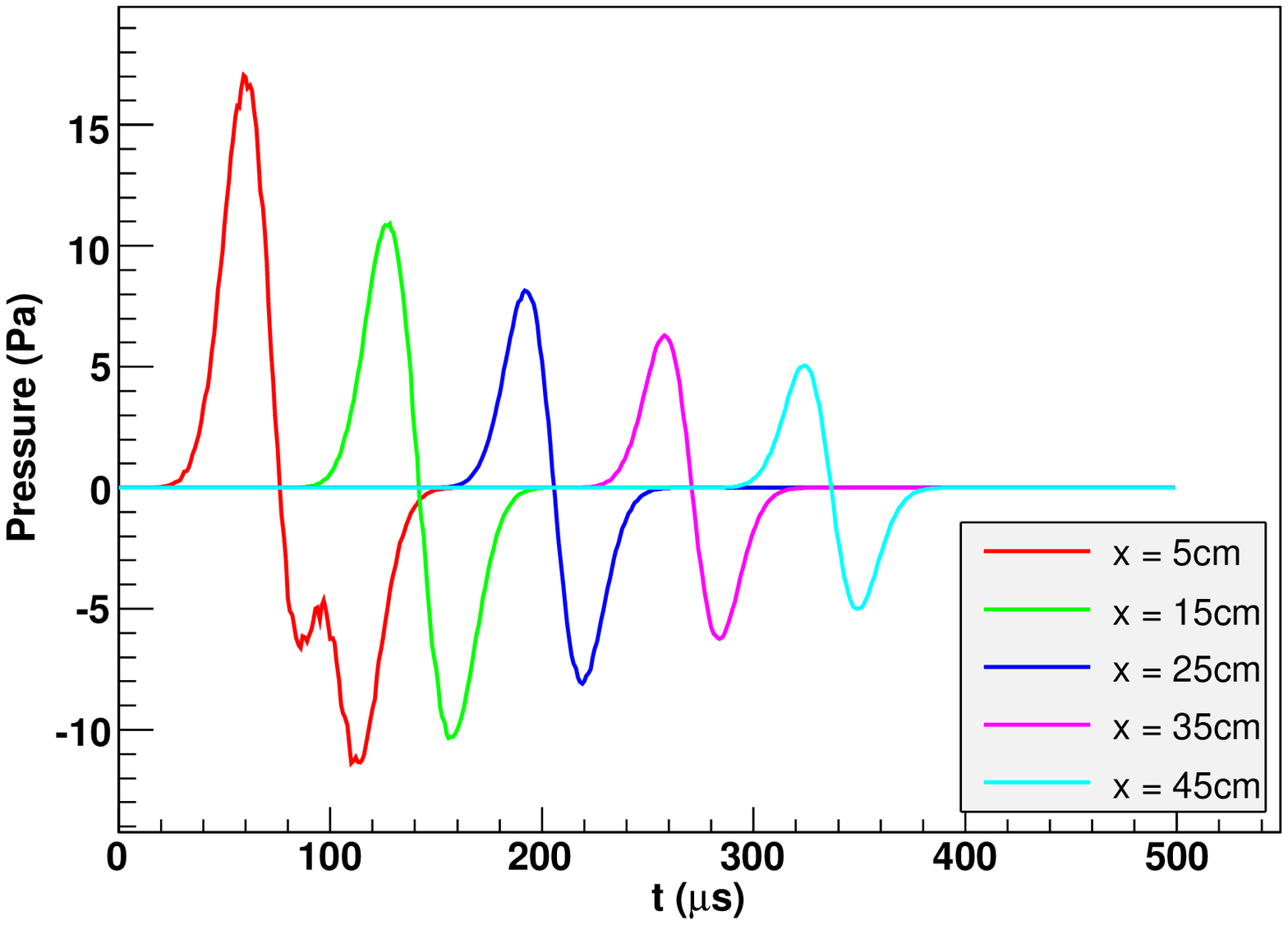}}
  \subfigure[Pressure vs. position.]{
    \label{subfig:p_pulses_sim_b}
    \includegraphics[width=0.48\textwidth]
    {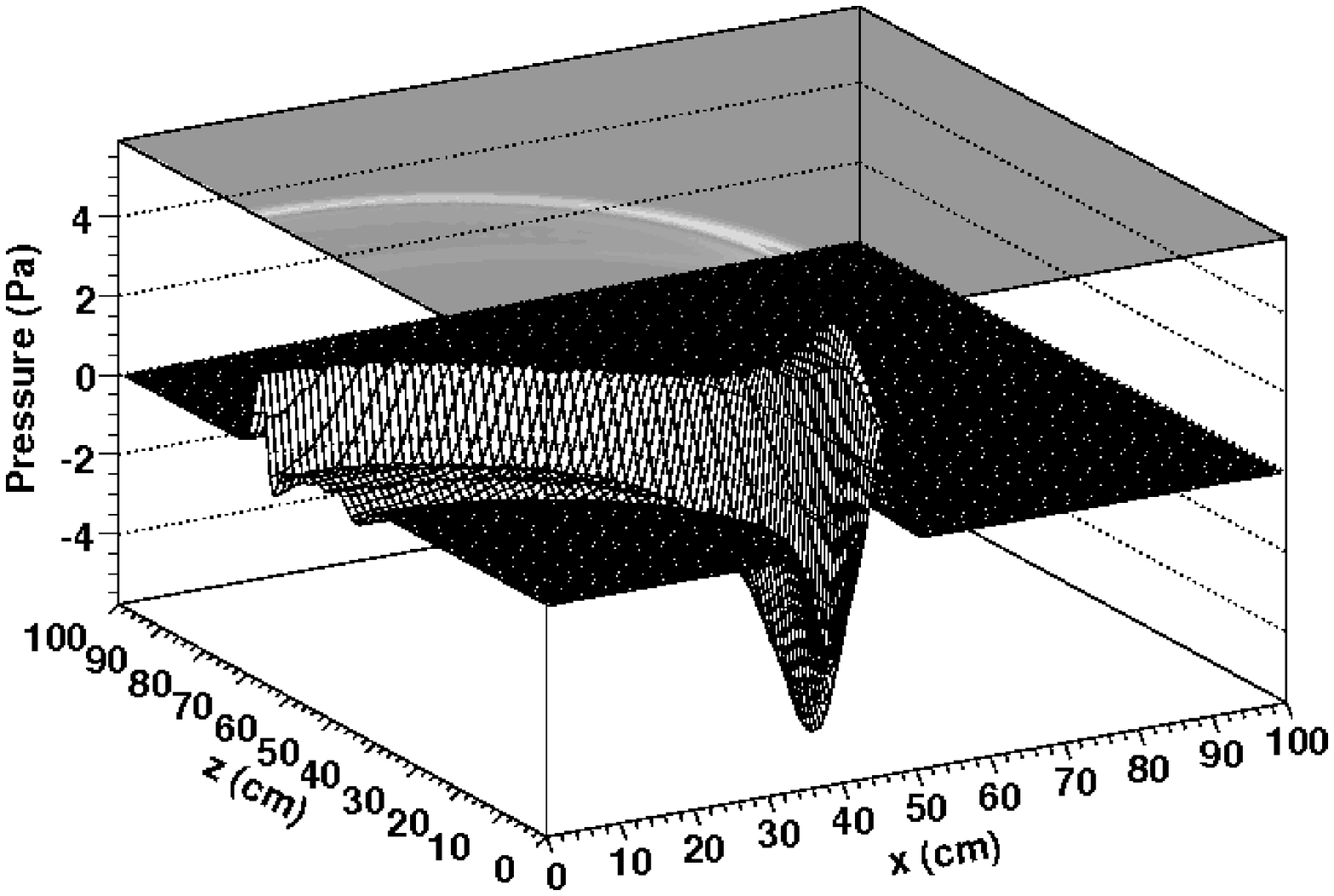}}
  \caption[Simulated pressure pulses produced by the proton
  beam]{Simulated pressure pulses produced by the proton beam:
    \subref{subfig:p_pulses_sim_a} Pressure as a function of time for
    different distances $x$ perpendicular to the proton beam ($z$
    fixed at 12\,cm), and \subref{subfig:p_pulses_sim_b} pressure
    versus position at $t = 333\,\mu$s (the proton bunch starts to
    enter into the water at $t_0 = 0$ in positive $z$ direction).}
  \label{fig:p_pulses_sim}
\end{figure}

Compared to the expected length of the pressure pulse of a few
10\,$\mu$s the energy deposition of the proton bunch over 24\,$\mu$s
cannot be assumed instantaneous, and the full equation
(\ref{eq:kirchhoff_long}) has to be used to calculate the
thermoacoustic signal. It is solved using a discrete version of
(\ref{eq:kirchhoff_long}) where the integration is replaced by a sum
over histogram bins, and the time derivative is calculated as
difference between histograms of the set at different times divided by
the time step.  Figure~\ref{fig:p_pulses_sim} shows the resulting
pressure field.

The bipolar acoustic pulse at different distances perpendicular to the
proton beam can be seen in Fig.~\ref{subfig:p_pulses_sim_a}. The
pressure field at a fixed time is shown in
Fig.~\ref{subfig:p_pulses_sim_b}, where it is clearly visible, that
the signal is highest perpendicular to the beam, leading to the
expected disc shaped signature. The ringing of the signal which can be
seen in Fig.~\ref{subfig:p_pulses_sim_a} in the curve for $x = 5$\,cm
and in Fig.~\ref{subfig:p_pulses_sim_b} is a near field effect caused
by the different characteristic of different parts of the energy
distribution. In contrast to a ultra high energy particle shower, which
is a true line source as we will see, the proton beam signal is a
superposition of a line source with the highest energy density in the
first few centimetres of the track and a point source constituted by
the Bragg peak.

\begin{figure}[p]
  \centering
  \includegraphics[angle=90,width=0.92\textwidth]
  {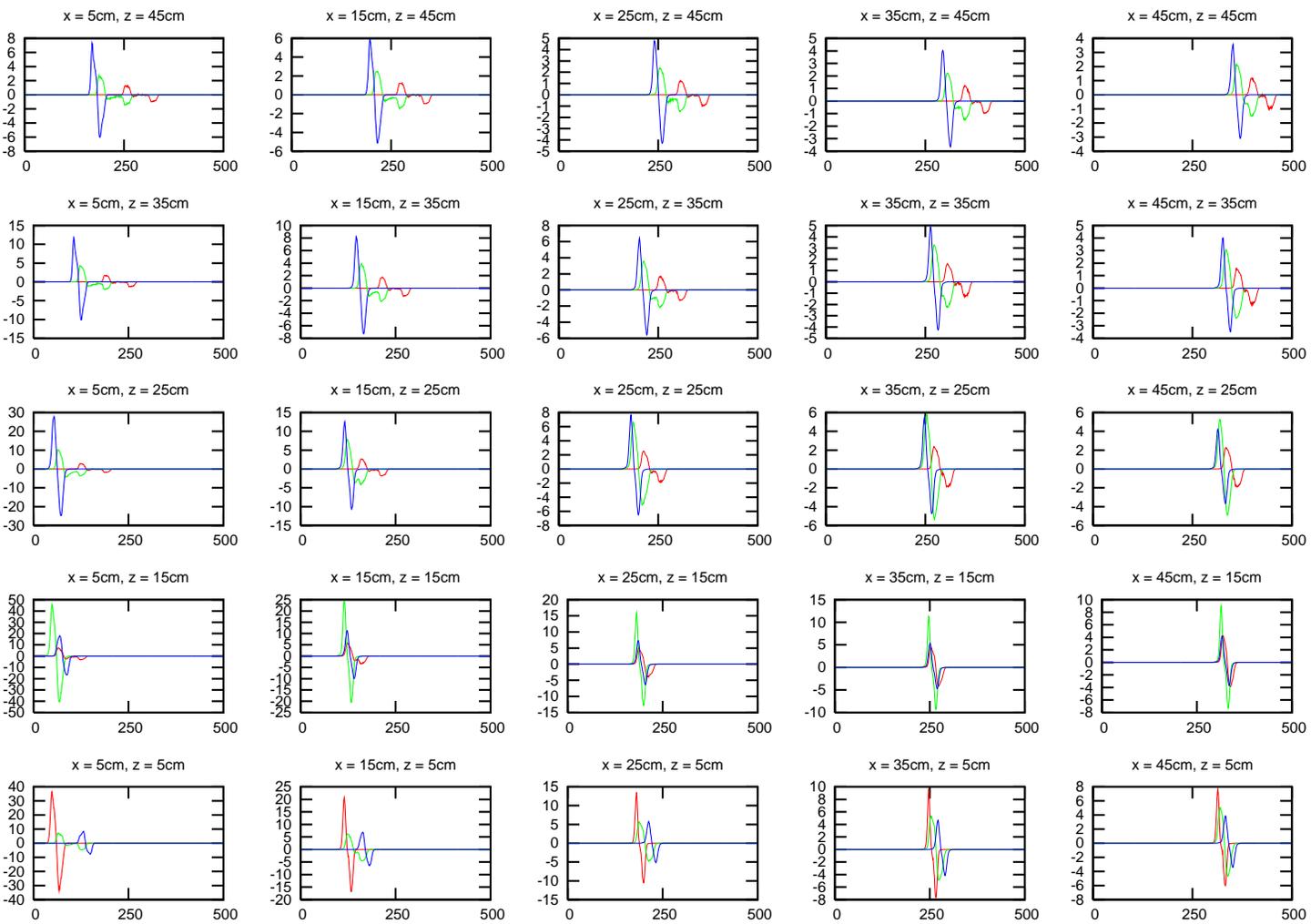}
  \caption[Contribution of different parts of the energy distribution
  to the acoustic signal]{Contribution of different parts of the
    energy distribution to the acoustic signal. Red curves: $z <
    10\,$cm; green curves $10 \, \mathrm{cm} \le z < 20 \,
    \mathrm{cm}$; blue curves $z \ge 20\,$cm (for each plot the
    abscissa is time in $\mu$s; the ordinate is pressure in Pa).}
  \label{fig:signal_contributions}
\end{figure}

We studied the contribution of the different parts by calculating the
signal emitted from different regions of the energy deposition
separately. The regions where chosen such that approximately the same
total amount of energy is deposited in each of them. The investigated
regions are $z < 10$\,cm, 10\,cm $\le z < 20$\,cm, and $z \ge
20$\,cm. The contribution of these parts of the energy deposition to
the signal can be seen in Fig.~\ref{fig:signal_contributions} showing
the resulting acoustic pulses at different positions.

It can be seen that in the near field for small $z$ the ringing comes
from the later arriving signal of the Bragg peak, and for larger $z$
it is determined by the late arriving signal from the line like
part. In the far field all the signal contributions start to add
coherently. This behaviour was also observed in the measured signals.

\subsubsection{Results}

\begin{figure}[ht]
  \centering
  \includegraphics[width=0.9\textwidth]
  {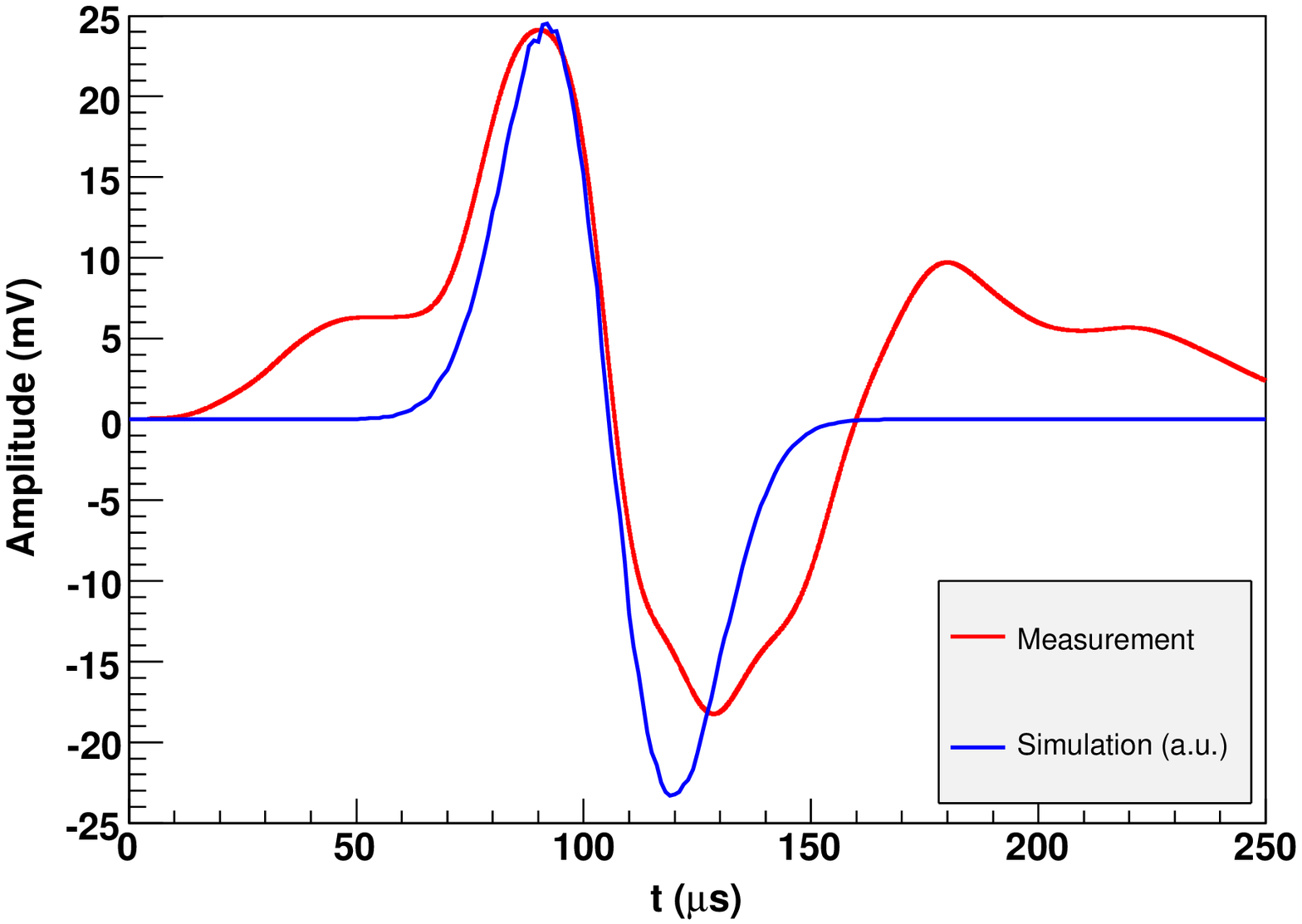}
  \caption[Comparison of measured and simulated signal]{Comparison of
    measured and simulated signal at $x = 10$\,cm and $z =
    12$\,cm. The bunch parameters are as in
    Fig.~\ref{fig:bunch_profile}; the simulated signal is rescaled to
    fit the amplitude of the measured signal.}
  \label{fig:p_puls_measured}
\end{figure}

For the analysis, signals averaged over 1000 proton bunches were used
to suppress noise. Figure~\ref{fig:p_puls_measured} shows a typical
measured acoustic signal compared to the simulation (The simulated
signal is rescaled to fit the amplitude of the measured signal).

The measured signal still includes the response of the hydrophone,
nonetheless the bipolar structure is clearly visible, and the signal
length matches the expectation. Thus, measured and simulated signals
match reasonably well.

The plateau at the beginning of the measured signal, which is not of
acoustic origin -- the travel time of an acoustic signal for the
10\,cm distance to the hydrophone is 67\,$\mu$s -- can be explained by
a flow of charge from the proton bunch into the measurement system
\cite{Graf:2004}, since it has a fixed shape but its height is
proportional to the proton bunch energy. The broadening of the first
minimum as well as the ringing of the signal can be explained by a
very early reflection of the signal at the proton beam entrance window
into the tank, and by the response function of the hydrophone.

Further we could show \cite{Graf:2004}, that in complete agreement
with our expectations, the time of the first maximum of the signal is
proportional to the distance of the hydrophone perpendicular to the
proton tracks, and that the signal amplitude is proportional to the
total bunch energy, whereas the signal length is energy independent.

We determined the signal speed to be $(1458 \pm 4) \, \mathrm{m} /
\mathrm{s}$ which is in excellent agreement with the textbook value
for the speed of sound in fresh water ($(1455 \pm 1) \, \mathrm{m} /
\mathrm{s}$ at the temperature considered).

\begin{figure}[ht]
  \centering
  \subfigure[Simulation.]{
    \label{subfig:amplitude_dist_a}
    \includegraphics[width=0.48\textwidth]
    {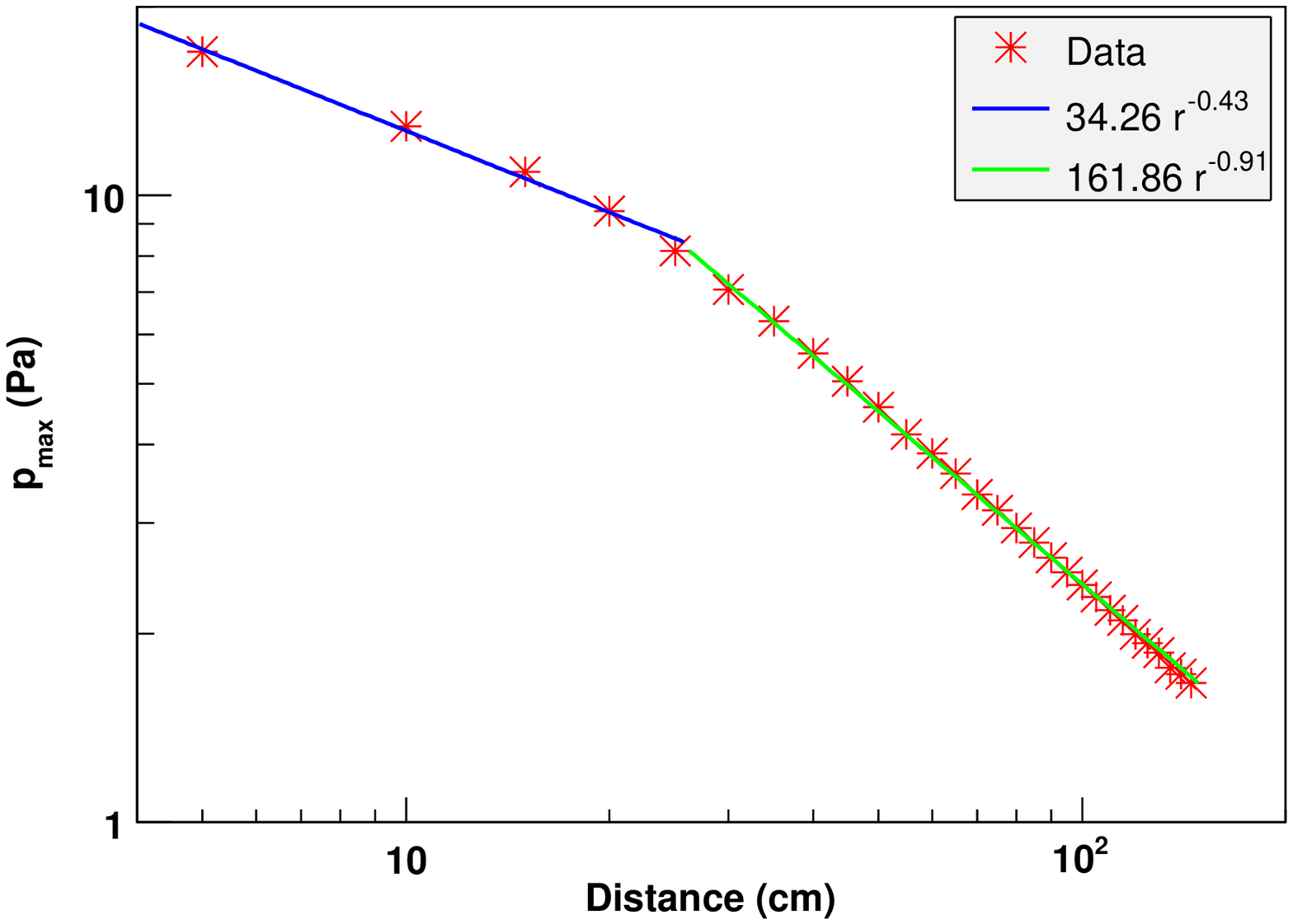}}
  \subfigure[Measurement.]{
    \label{subfig:amplitude_dist_b}
    \includegraphics[width=0.48\textwidth]
    {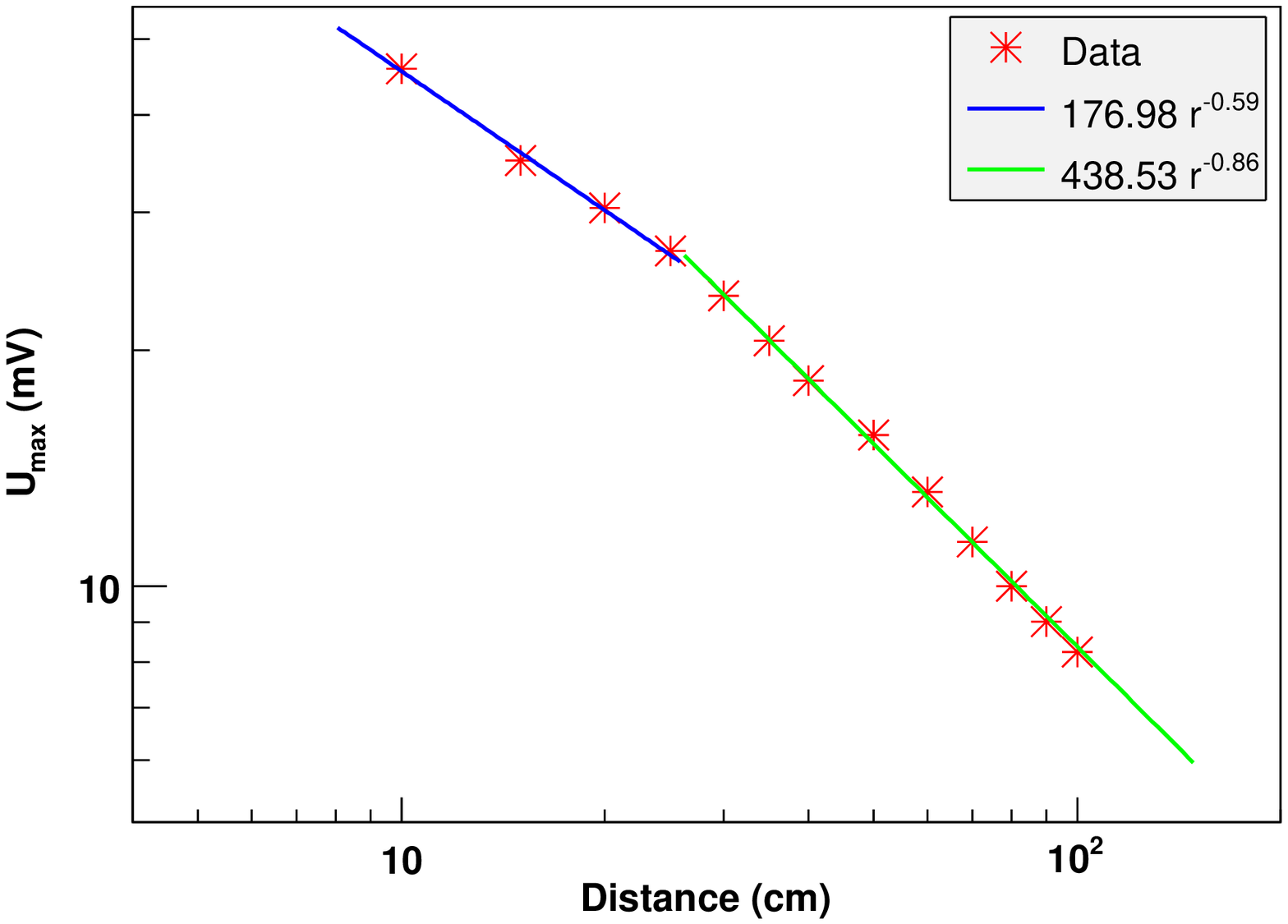}}
  \caption[Dependency of the signal amplitude on the
  distance]{Dependency of the signal amplitude on the distance
    perpendicular to the proton beam ($z = 12$\,cm). The solid lines
    are fit curves to the data. The transition between near and far
    field at $r \approx 26$\,cm is clearly visible.}
  \label{fig:amplitude_dist}
\end{figure}

An interesting feature to test the thermoacoustic model is the
distance behaviour of the signal amplitude $p_\mathrm{max}$. The model
predicts a disc-shaped signal where the symmetry axis is given by the
line like energy deposition. Thus, in the near field region --
compared to the size of the energy deposition region -- we expect that
the amplitude decreases as $1 / \sqrt{r}$, where $r$ is the distance
perpendicular to the axis. In the far field region, where the energy
deposition region can be assumed as point-like, the amplitude should
decrease like $1 / r$. For the proton beam experiment the transition
region between near and far field lies within the water tank.
Figure~\ref{fig:amplitude_dist} shows the signal amplitude
$p_\mathrm{max}$ as a function of $r$ for the simulation and the
measurement.

The transition between the near field region and the far field region
at $r \approx 26$\,cm is clearly visible. The observed slopes agree
well with the simple model prediction of a line like source in the
near field, and a point source in the far field.

Further, we studied the dependence of the signal on the water
temperature. A water temperature which is homogeneous over the whole
tank was achieved by cooling the whole water volume and subsequent
controlled reheating. Temperatures between 1$^\circ$C and 15$^\circ$C
were set up with a precision of 0.1$^\circ$C.

It was observed that, as expected, the pressure pulse is inverted at
4$^\circ$C where the bulk expansion coefficient $\alpha$ changes its
sign. Below 4$^\circ$C heating of the water leads to a contraction,
which propagates as an inverted bipolar pressure pulse from the energy
deposition region. However, we found that the signal never vanishes
completely as one would expect from the thermoacoustic model. At
$4^\circ$C, there is a residual signal of the order of 1\,mV (cf.
scale in Fig.~\ref{fig:p_puls_measured}), which is obviously not of
thermoacoustic origin. The exact nature of this residual signal is
unclear, but the laser beam experiment (cf.~Sec.~\ref{sec:laser})
suggests, that it is a feature specific to the proton beam.

\begin{figure}[ht]
  \centering
  \includegraphics[width=0.9\textwidth]
  {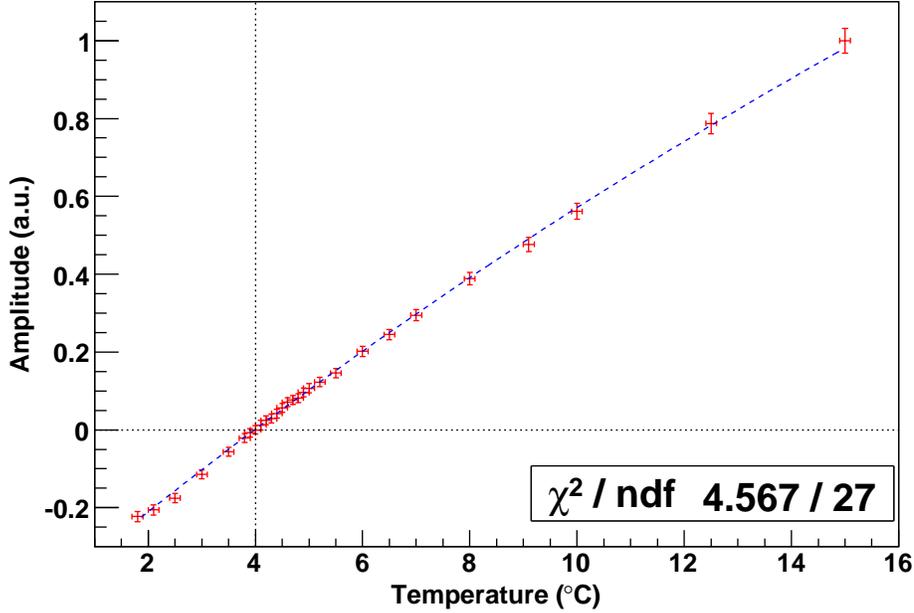}
  \caption[Amplitude of the proton beam signal as a function of
  temperature]{Amplitude of the proton beam signal at $x = 10$\,cm and
    $z = 12$\,cm as a function of the water temperature after
    subtraction of the residual signal at $4^\circ$C from all signals.
    The amplitude is normalised to unity at 15$^\circ$C. The blue
    curve shows the $\gamma$ factor from Fig.~\ref{fig:gamma} fitted
    with an overall scaling factor.}
  \label{fig:p_amplitude_temp}
\end{figure}

For the analysis of the data we subtracted the residual signal from
all signals for the different temperatures. The resulting amplitude
temperature relation is shown in Fig.~\ref{fig:p_amplitude_temp}.

Inverted signals below 4$^\circ$C, where an increase of the
temperature leads to a contraction of the water, are defined to have
negative amplitudes. The measured values are in good agreement with
the predictions from material properties represented by $\gamma$ as
defined in (\ref{eq:gamma}) and shown in Fig.~\ref{fig:gamma}.

\medskip Summarising, we could verify the thermoacoustic
model with high accuracy. We showed that a rapid, line like energy
deposition in a medium leads to a bipolar pressure pulse which mainly
propagates perpendicular to the energy deposition region. The acoustic
nature of the measured pulses could be proven by the determination of
their velocity which is compatible with the speed of sound. The pulse
variation with temperature is as expected from the model.

\subsection{The laser experiment}
\label{sec:laser}

In order to address some questions left unanswered by the proton beam
experiment presented in the previous section, especially the question
for the nature of the residual signal observed at 4$^\circ$C, we
performed acoustic measurements with a laser beam dumped in water
\cite{Schwemmer:2005}.

We used the same experimental setup as for the proton beam experiment
which is shown in Fig.~\ref{fig:proton_setup}, only the beam entrance
window was replaced by silica glass. An infrared, pulsed Nd:YAG laser
with a wavelength of 1064\,nm, a maximum pulse energy of 15\,EeV, and
a beam diameter of 1\,cm is induced into the water, and the resulting
bipolar pressure pulses are recorded by the position adjustable
hydrophones. Since the laser has an absorption length of 7\,cm in
water, the tank is large enough that the laser beam is completely
absorbed.

The main differences to the proton beam experiment are as follows:

\begin{itemize}
\item {\bf Pulse length.} The laser we used can produce pulses with a
  length between 6\,ns and 8\,ns compared to the 24\,$\mu$s bunch
  length of the proton pulse. Thus, for the laser beam an
  instantaneous energy deposition can be assumed which is comparable
  to the particle cascades induced by ultra high energy neutrinos, and
  the time structure of the resulting acoustic signal is determined by
  the spatial extension of the energy deposition, and not by the bunch
  length as it was the case for the proton beam\footnote{With an
    instantaneous energy deposition mechanism, the resulting pressure
    pulse is a superposition of spheric waves emitted {\em
      simultaneously} from the whole deposition region; for the case
    of an energy deposition over a time window the acoustic signal is
    a superposition of spheric waves emitted in an interval of space
    and time.}.
\item {\bf Energy deposition mechanism.} The infrared laser beam is
  absorbed in water by exciting rotational and vibrational modes of
  the water molecules, whereas the protons transfer their energy to
  the water by ionisation and proton nucleus scattering. These
  completely different energy transfer mechanisms allow us to
  distinguish between thermoacoustic signals and contributions
  characteristic for the particular process of energy transfer, like
  the non acoustic plateau observed in the proton beam signals, which
  cannot be reproduced at the laser beam, and might thus well be an
  effect related to the transfer of electric charge.
\item {\bf Availability.} The availability of the laser setup in our
  laboratory in Erlangen allows us to repeat measurements at any time
  and to scrutinise effects which are not completely understood, being
  independent of the availability of an accelerator.
\end{itemize}

The simulation of the acoustic signals is done with the same methods
as discussed for the proton beam experiment. The energy density
deposited by the laser beam can be parameterised as a Gauss function
with a width of 1\,cm perpendicular to the axis, and has an
exponential decay in the direction of the beam characterised by the
absorption length of 7.1\,cm. In contrast to the proton beam
experiment the energy deposition can be assumed as instantaneous, and
thus (\ref{eq:kirchhoff_short}) can be used for the calculation of the
signals.

\begin{figure}[ht]
  \centering
  \includegraphics[width=0.9\textwidth]
  {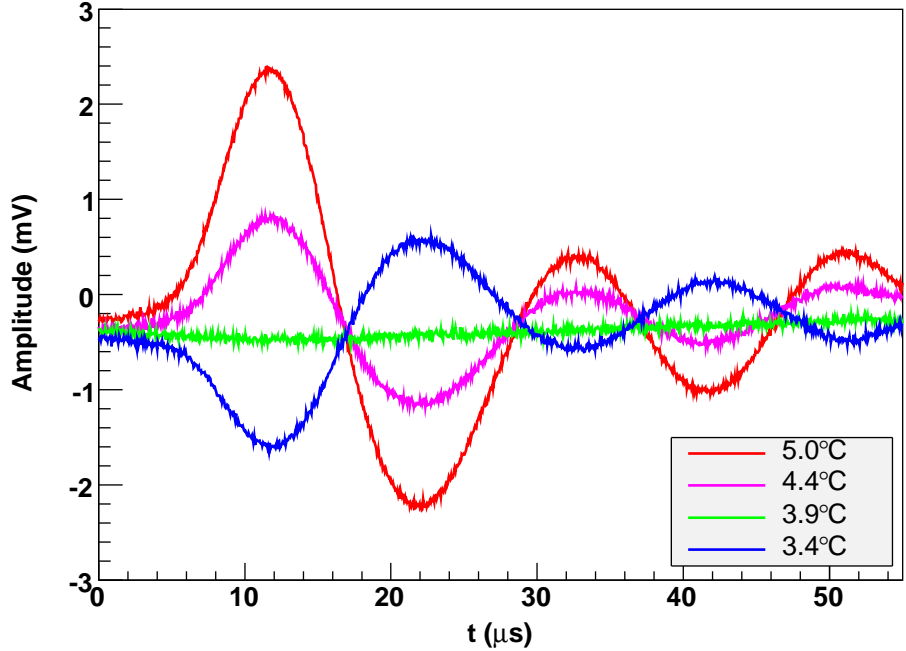}
  \caption[Laser induced pressure pulses]{Laser induced pressure
    pulses measured at $x = 10$\,cm and $z = 12$\,cm at different
    water temperatures. The signal vanishes completely at
    3.9$^\circ$C, and changes its polarity below this temperature.}
  \label{fig:laser_signals}
\end{figure}

We explored the dependence of the thermoacoustic signal on the sensor
position and on the pulse energy, which are all in good agreement
with the expectations from the simulations based on the thermoacoustic
model \cite{Schwemmer:2005}. Figure~\ref{fig:laser_signals} shows
signals measured at different temperatures around 4$^\circ$C.

\begin{figure}[ht]
  \centering
  \includegraphics[width=0.9\textwidth]
  {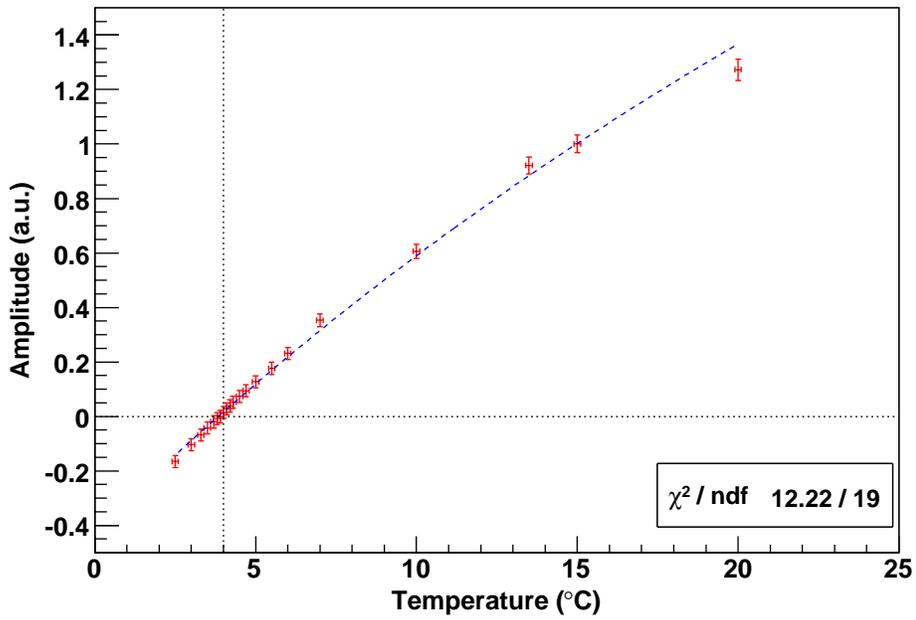}
  \caption[Amplitude of the laser induced signal as a function of
  temperature]{Amplitude of the laser induced signal at $x = 10$\,cm
    and $z = 12$\,cm as a function of the water temperature.
    The amplitude is normalised to unity at 15$^\circ$C. The blue
    curve shows the $\gamma$ factor from Fig.~\ref{fig:gamma} fitted
    with an overall scaling factor and a temperature offset. The fit
    yields a zero crossing of the signal at 3.85$^\circ$C.}
  \label{fig:laser_amplitude_temp}
\end{figure}

The first thing to notice is that the initial bipolar signal, which is
followed by ringing in the hydrophone, has a length of approximately
30\,$\mu$s, and is much shorter than the proton beam induced signal
which has a length of about 100\,$\mu$s. The reason is, as was
previously discussed, that with an instantaneous energy deposition the
signal length is defined by the extension of the energy deposition
region, which is about 10\,cm in length only. Further the signal
vanishes completely at about 3.9$^\circ$C. This supports the
assumption, that the residual signal observed in the proton beam
experiment is specific to this energy deposition mechanism, and that
it is valid to subtract this residual signal for the evaluation of the
thermoacoustic model. The dependence of the pulse amplitude on the
temperature is shown in Fig.~\ref{fig:laser_amplitude_temp}.

The fit of the expected amplitude behaviour proportional to the
$\gamma$ factor (\ref{eq:gamma}), where an overall scaling factor and
a temperature offset are left as free parameters, yields a zero
crossing of the signal at $(3.85 \pm 0.06_\mathrm{stat.} \pm
0.1_\mathrm{syst.})^\circ$C, which is in excellent agreement with
predictions based on the thermoacoustic model.

\medskip Since we have shown the validity of the thermoacoustic model,
and that it is possible to simulate pressure pulses for a given energy
deposition, we will continue in the next chapter with the
investigation of acoustic pulses produced by neutrino induced particle
cascades.


\chapter{Acoustic signals from ultra high energy neutrinos}
\label{chap:tam_uhe_neutrinos}
\minitoc

\bigskip After we have shown the validity of the thermoacoustic model
in the previous chapter and have established methods to calculate
pressure signals produced by charged particles propagating through
water, we will now discuss the production of acoustic signals by ultra
high energy neutrinos. First we will review the propagation of the
neutrinos to, and their interaction in the detector. Then we will
investigate acoustic signals produced by ultra high energy hadronic
and leptonic cascades induced by the neutrino.

\section{Propagation and interaction of ultra high energy neutrinos}
\label{sec:nu_propagation}

In the introduction to this work we stated that neutrinos are the
ideal messengers from cosmological sources since, as only weakly
interacting particles, they arrive at the Earth unperturbed. Further
we pointed out that Water-\v{C}erenkov neutrino telescopes are built
to look downwards to use the Earth as a shielding for everything
except neutrinos. These statements remain only partially valid when
looking at neutrinos at highest energies. It is still true that these
neutrinos can travel unperturbed through the void of the universe,
but, since their cross section for interactions with nucleons
increases with energy, already the Earth constitutes enough matter to
completely shield neutrinos.

In this work we use the neutrino nucleon total cross section from
\cite{Ralston:1996bb}:

\begin{equation}
  \label{eq:cross_section}
  \sigma_\mathrm{tot} = 1.2 \cdot 10^{-32} \, \mathrm{cm}^2 \, \left(
    \frac{E_\nu}{10^9 \, \mathrm{GeV}} \right)^{0.4}
\end{equation}

Equation \ref{eq:cross_section} is valid for $E_\nu \gtrsim
10^4$\,GeV, where $E_\nu$ is the energy of the incident neutrino in
the rest frame of the nucleon. It is derived by extrapolating measured
cross sections to higher energies under the assumption that the quark
density in the nucleon follows a negative power law when going to
smaller values of Bjorken $x$\footnote{The Bjorken $x$ variable gives
  the momentum fraction of the quark in the nucleon. At higher
  energies the neutrino ``sees'' quarks with lower momentum, i.e. if
  the quark density is a negative power law function of $x$, it sees
  more quarks it can interact with, and the cross section rises
  accordingly.}. Further, (\ref{eq:cross_section}) neglects the
Glashow resonance at $6.7 \cdot 10^6$\,GeV, which describes the
resonant production of $W^-$ gauge bosons by electron anti-neutrinos
interacting with electrons in the Earth. This resonance is narrow
(width $\Gamma \approx 2.5 \cdot 10^5$\,GeV) and, as will be discussed
later, at an energy well below the energies relevant for this work.
Thus its neglect is well acceptable.

\begin{figure}[ht]
  \centering
  \includegraphics[width=0.9\textwidth]{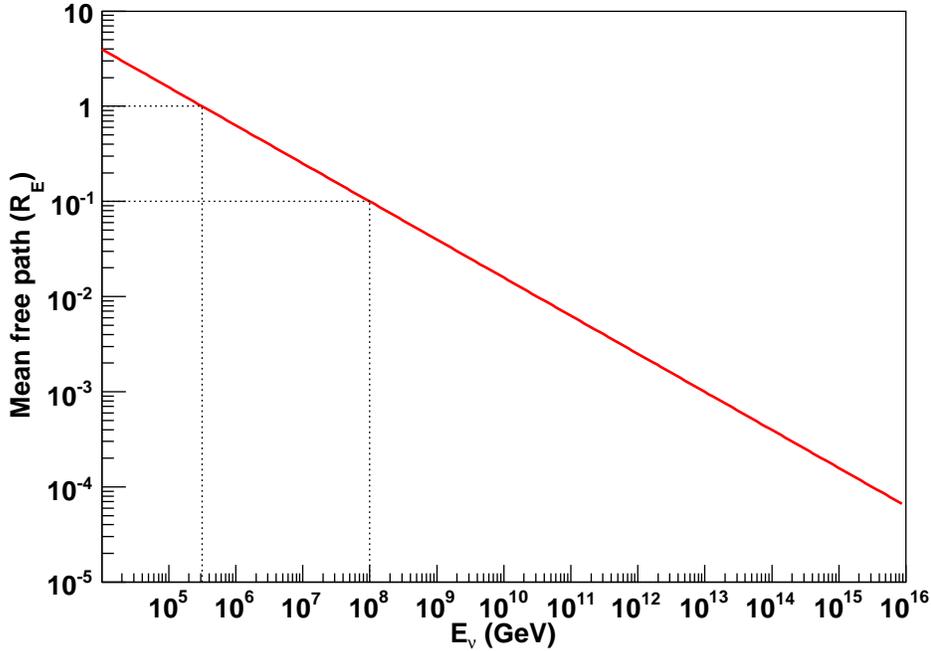}
  \caption[Mean free path length for neutrinos traversing the
  Earth]{Mean free path length for neutrinos traversing the Earth in
    Earth radii. Indicated are the energies where the mean free path
    is one and one tenth of the Earth's radius.}
  \label{fig:mean_free_path}
\end{figure}

Using this cross section we can now calculate the mean free path $L$
of neutrinos traversing the Earth,

\begin{equation}
  L = \frac{m_n}{\rho \, \sigma_\mathrm{tot}}
\end{equation}

\noindent where $m_n$ is the mass of the nucleon and $\rho = 5.52 \,
\mathrm{g} / \mathrm{cm}^3$ is the mean density of the Earth.
Figure~\ref{fig:mean_free_path} shows $L$ in Earth radii $R_E$ as a
function of the neutrino energy. The energies where $L$ is one Earth
radius and where $L$ is one tenth of the Earth radius, and thus the
energy range where the Earth becomes opaque to neutrinos, are
indicated. Since we want to design a detector for neutrinos with
energies $E_\nu \gtrsim 10^9$\,GeV we will consider only neutrinos
coming from above the horizon, i.e.~with zenith angles $\theta <
90^\circ$.

On the other hand, if we consider an underwater neutrino telescope we
have to ensure that the neutrinos can propagate freely through the
water overburden down to the detector. Table~\ref{tab:mean_free_path}
gives the mean free path of the neutrino in water ($\rho = 1.0 \,
\mathrm{g} / \mathrm{cm}^3$) and the zenith angle
$\theta_\mathrm{max}$ for which the neutrino path length is equal to
the mean free path under the assumption that the ocean is flat and
that the detector is at a depth of 2\,km, a typical depth for existing
neutrino telescopes.

\begin{table}[hbt]
  \centering
  \caption[Mean free path length of neutrinos in water]{Mean free path
    length $L$ of neutrinos in water and zenith angle
    $\theta_\mathrm{max}$ for which the neutrino path length is equal
    to $L$ assuming a flat ocean and a detector depth of 2\,km.}
  \label{tab:mean_free_path}
  \begin{tabular}{rrr}
    \hline
    \multicolumn{1}{c}{$E_\nu$} & \multicolumn{1}{c}{$L$} &
    \multicolumn{1}{c}{$\theta_\mathrm{max}$} \\
    \multicolumn{1}{c}{(GeV)} & \multicolumn{1}{c}{(km)} \\ \hline
    $10^9$ & 1390 & $89.9^\circ$ \\ 
    $10^{10}$ & 553 & $89.8^\circ$ \\
    $10^{11}$ & 220 & $89.5^\circ$ \\
    $10^{12}$ & 88 & $88.7^\circ$ \\
    $10^{13}$ & 35 & $86.7^\circ$ \\
    $10^{14}$ & 14 & $81.7^\circ$ \\ \hline
  \end{tabular}
\end{table}

Based on the presented data we will assume throughout this work for a
first simulation of an acoustic neutrino telescope, that all neutrinos
with $\theta < 90^\circ$ can propagate freely down to the detector,
and that all neutrinos from below the horizon are absorbed inside the
Earth\footnote{For an exact treatment of the neutrino propagation
  through the Earth one will especially have to take care of the tau
  neutrino since, in contrast to the muon which is stopped in the
  Earth, the tau lepton decays rapidly producing another high energy
  tau neutrino.}.

\medskip After the neutrino has reached the detector we need to study
the neutrino interaction with a nucleon in the water. In this weak
interaction an outgoing charged lepton (charged current interaction,
CC) or an outgoing neutrino (neutral current interaction, NC) and a
hadronic jet are produced. Here, as will become clear when we discuss
the acoustic signals from hadronic and electromagnetic cascades in the
following sections, it is especially interesting which fraction of the
energy is transferred into these two channels.

We investigated this question by using the ANIS (All Neutrino
Interaction Simulation) \cite{Gazizov:2004va} simulation program. ANIS
is an Open Source\footnote{http://www.ifh.de/nuastro/anis/anis.html}
code written in C++ to simulate the propagation of ultra high energy
neutrinos through the Earth to the detector and calculate the first
interaction vertex in the detector. It can handle neutrino energies
ranging from 10\,GeV to 10$^{12}$\,GeV.

\begin{figure}[ht]
  \centering
  \subfigure[Distribution of $y$.]{
    \label{subfig:y_dist_a}
    \includegraphics[width=0.48\textwidth]
    {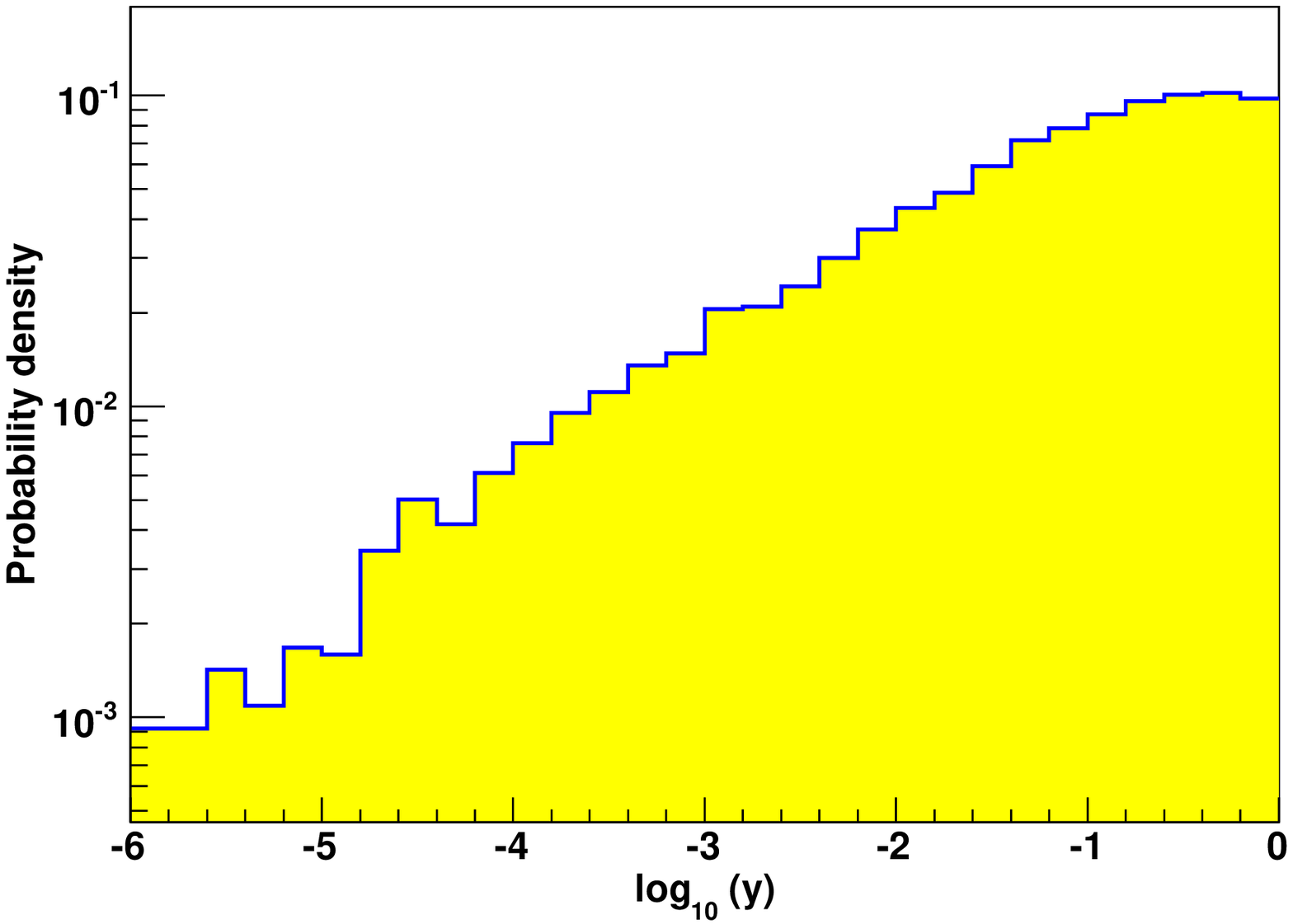}}
  \subfigure[Energy dependence.]{
    \label{subfig:y_dist_b}
    \includegraphics[width=0.48\textwidth]
    {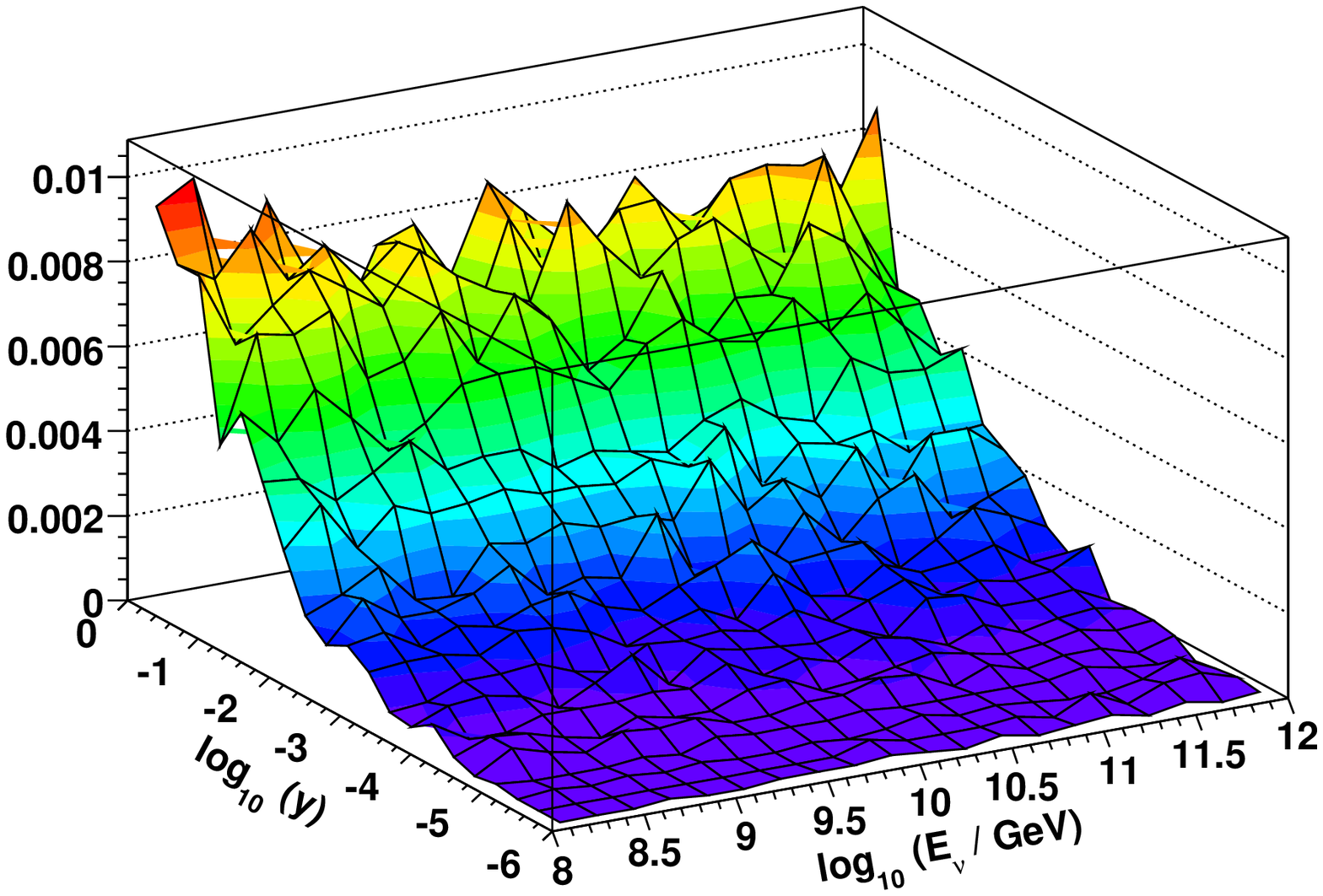}}
  \caption[Distribution of the kinematic variable $y$]{Distribution of
    the kinematic variable $y$ which is the fraction of energy
    transferred in a neutrino nucleon interaction from the neutrino to
    the hadronic system in the rest frame of the nucleon.
    \subref{subfig:y_dist_a} shows the probability density function
    for $\log y$; \subref{subfig:y_dist_b} shows that this function is
    independent of the energy of the incident neutrino.}
  \label{fig:y_dist}
\end{figure}

For our study we only investigated the kinematic parameter $y$ which
describes the energy transfer from the neutrino to the hadronic
system:

\begin{equation}
  \label{eq:y}
  y = 1 - \frac{E_\ell}{E_\nu}
\end{equation}

\noindent where $E_\ell$ is the energy of the outgoing lepton in the
rest frame of the nucleon. We studied the distribution of $y$ as a
function of the energy $E_\nu$ of the incident neutrino, and found
that the distribution is energy independent. It is shown in
Fig.~\ref{fig:y_dist}.

The energy independence can be seen in Fig.~\ref{subfig:y_dist_b}, and
the resulting probability density function for $\log y$ is shown in
Fig.~\ref{subfig:y_dist_a}. The median of this distribution is -1.1,
i.e. in half of the neutrino interactions less then $10^{-1.1} = 8\%$
of the energy of the incident neutrino is transferred to the hadronic
system.

\section{Hadronic cascades}
\label{sec:hadronic_cascades}

An ultra high energy hadronic jet produced in a neutrino interaction
deposits its energy in a medium through subsequent strong interactions
with the nuclei in the medium. Such hadronic cascades can be simulated
with Monte Carlo packages. For example, GEANT4 can calculate hadronic
cascades in water for primary energies up to $10^5$\,GeV.

The main problem in extrapolating shower simulations to higher
energies is posed by the fact, that the development of particle
cascades is governed by interactions with small transversal momentum
$p_T$. But differential cross sections at small scattering angles, and
thus small $p_T$, are not very well known since they are difficult to
measure. Common particle detectors at accelerator experiments are
designed cylindrically around the beam line to measure at high $p_T$.
Their acceptance for small angles is naturally limited by the aperture
of the beam line in the forward and backward directions. There are
efforts\footnote{see e.g.~http://totem.web.cern.ch/Totem/} to measure
these low $p_T$ cross sections at the LHC with ``Roman Pots'':
additional detectors far behind the interaction point and very near to
the beam line.

\begin{figure}[p]
  \centering
  \subfigure[10$^5$\,GeV $\pi^+$ energy distribution.]{
    \label{subfig:pi_plus_100TeV_zx}
    \includegraphics[width=0.48\textwidth]
    {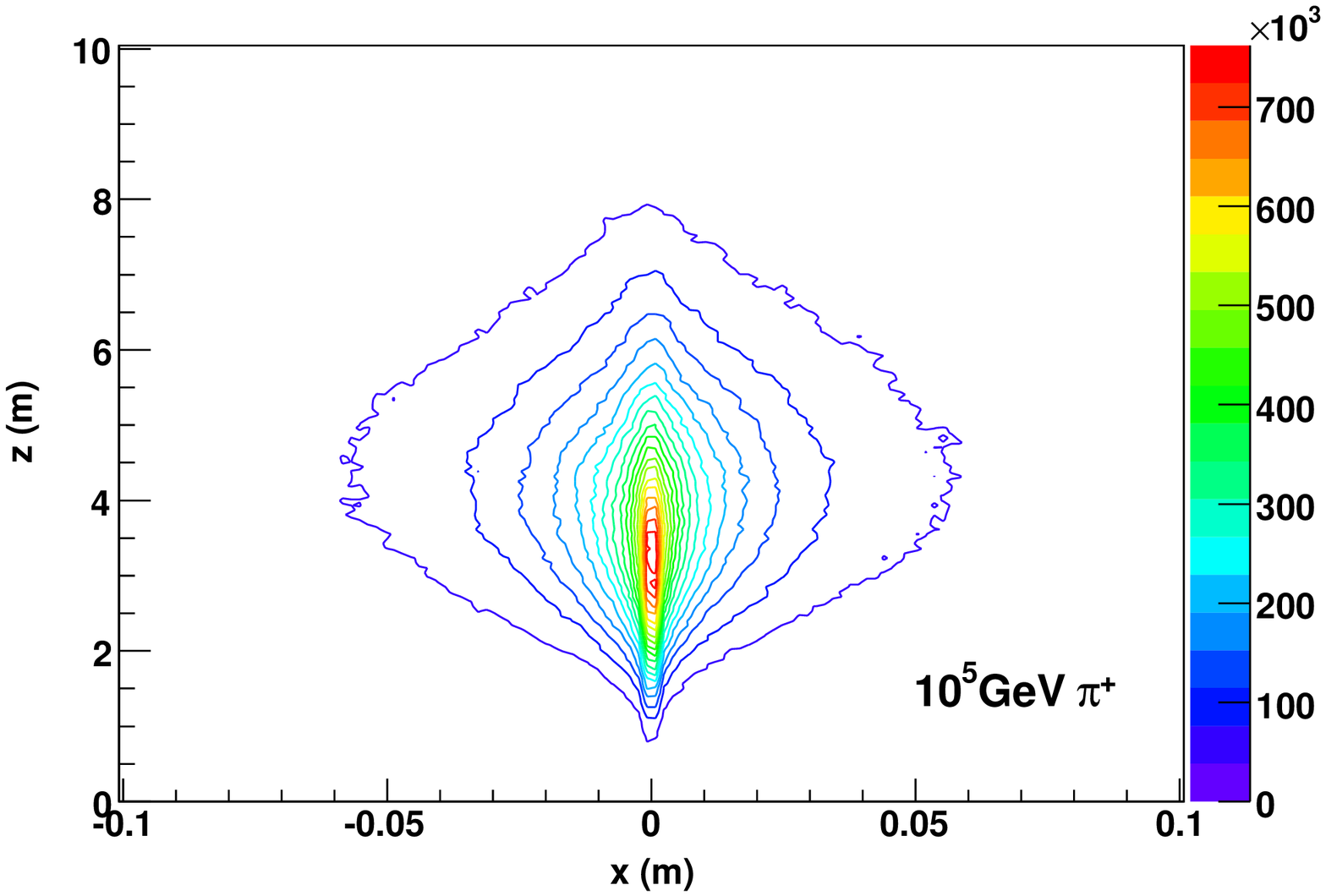}}

  \subfigure[10$^5$\,GeV $\pi^+$ longitudinal and transversal profile.]{
    \label{subfig:pi_plus_100TeV}
    \includegraphics[width=0.48\textwidth]{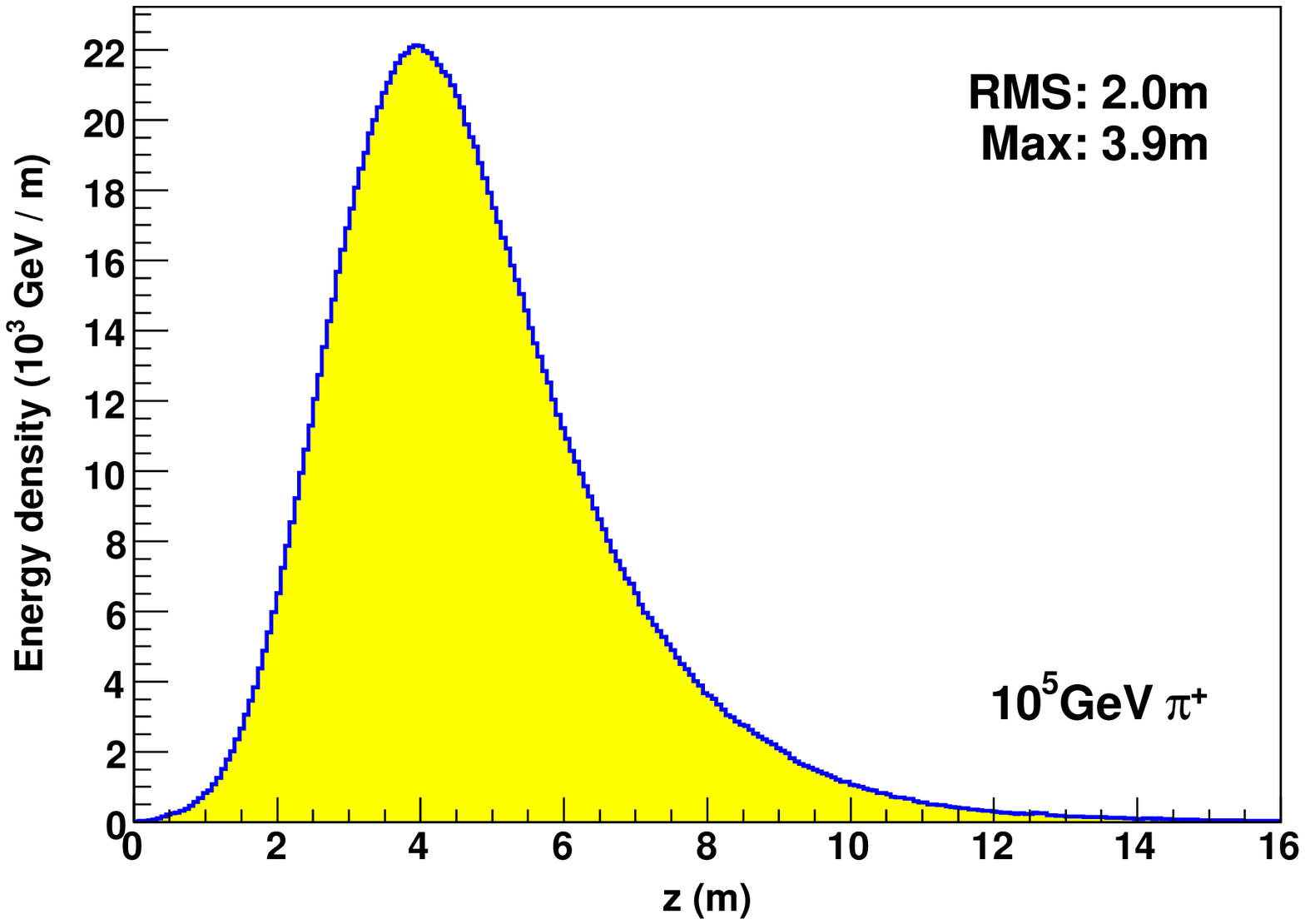}
    \includegraphics[width=0.48\textwidth]{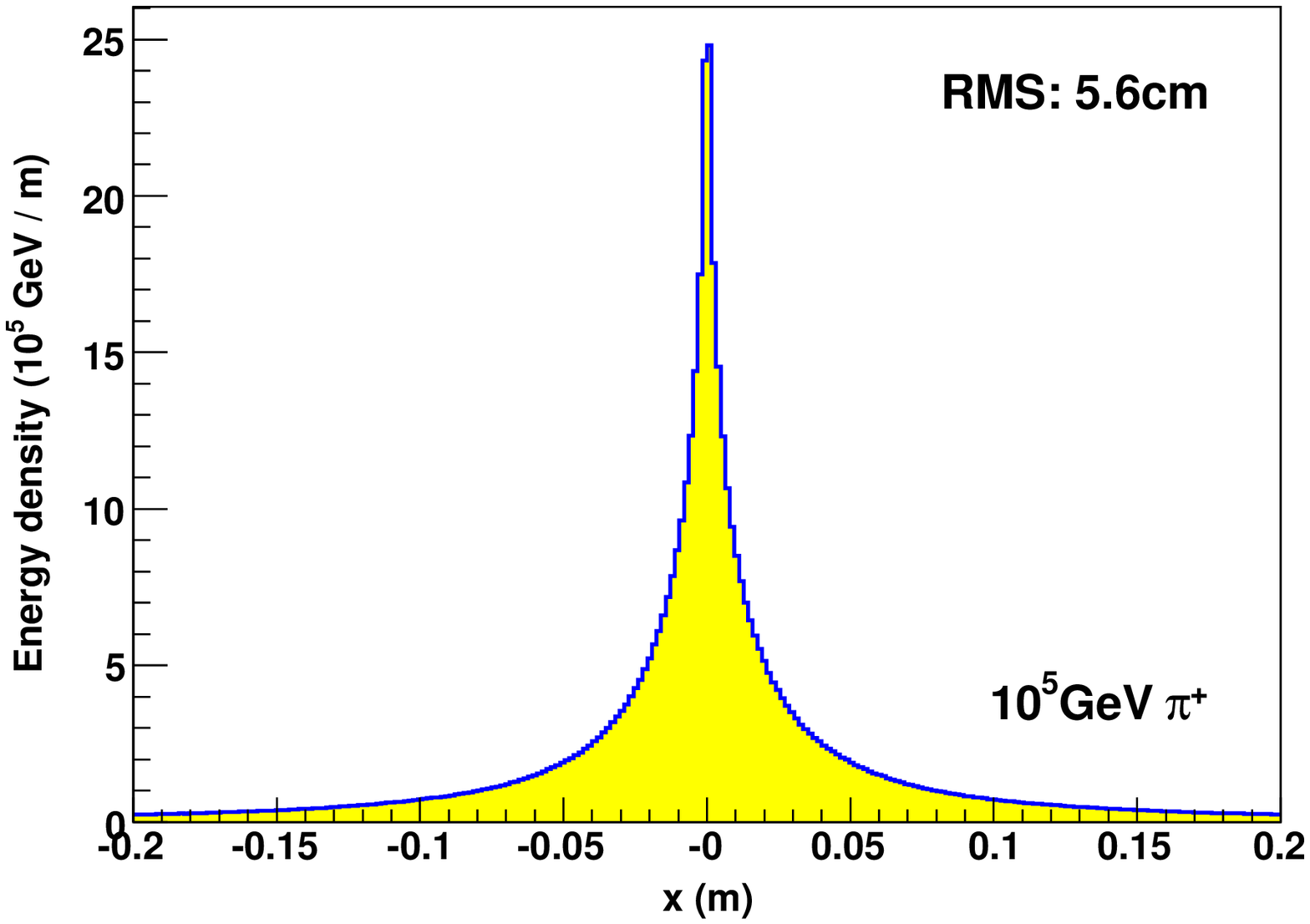}}
  \subfigure[10$^3$\,GeV $\pi^+$ longitudinal and transversal profile.]{
    \label{subfig:pi_plus_1TeV}
    \includegraphics[width=0.48\textwidth]{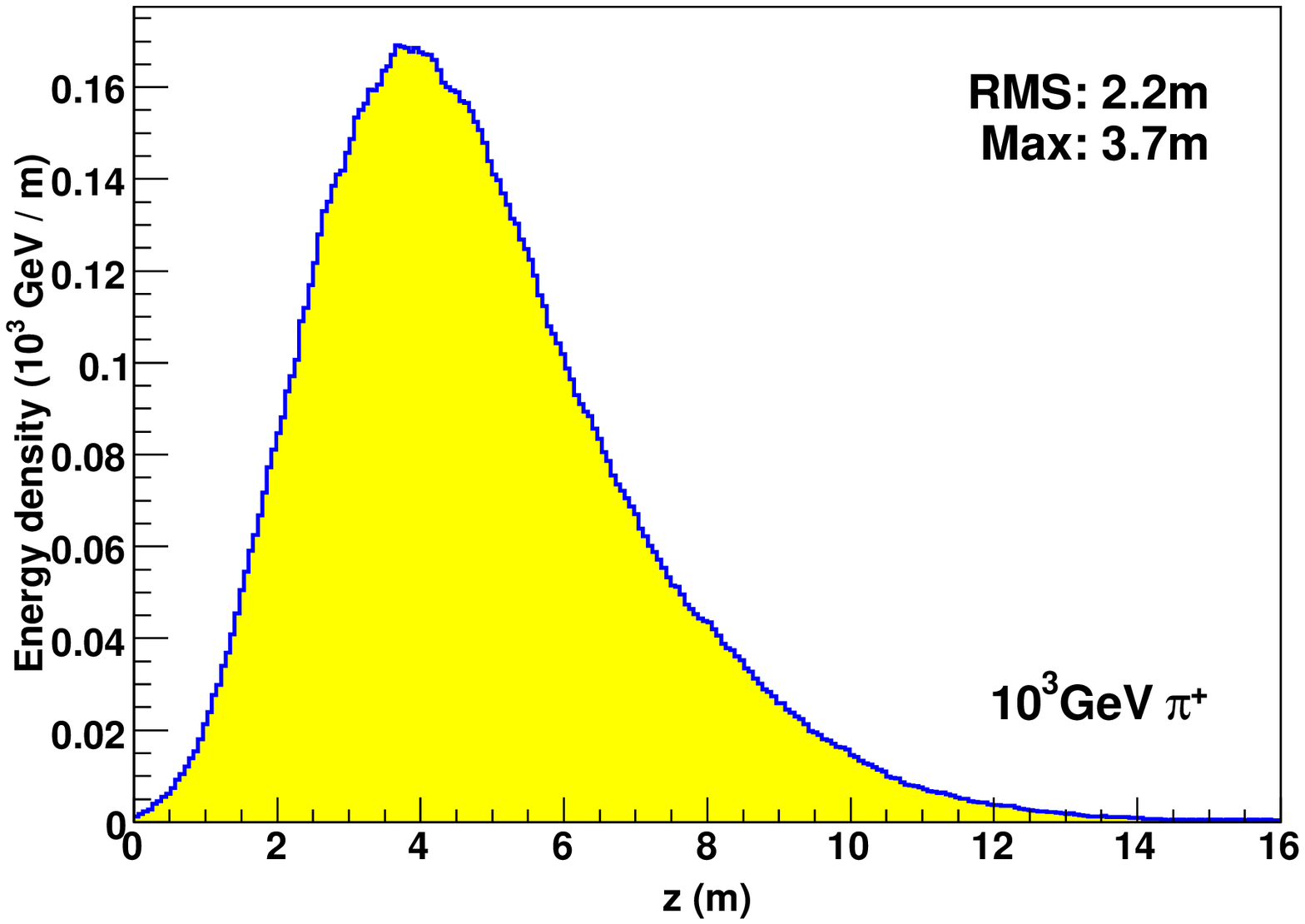}
    \includegraphics[width=0.48\textwidth]{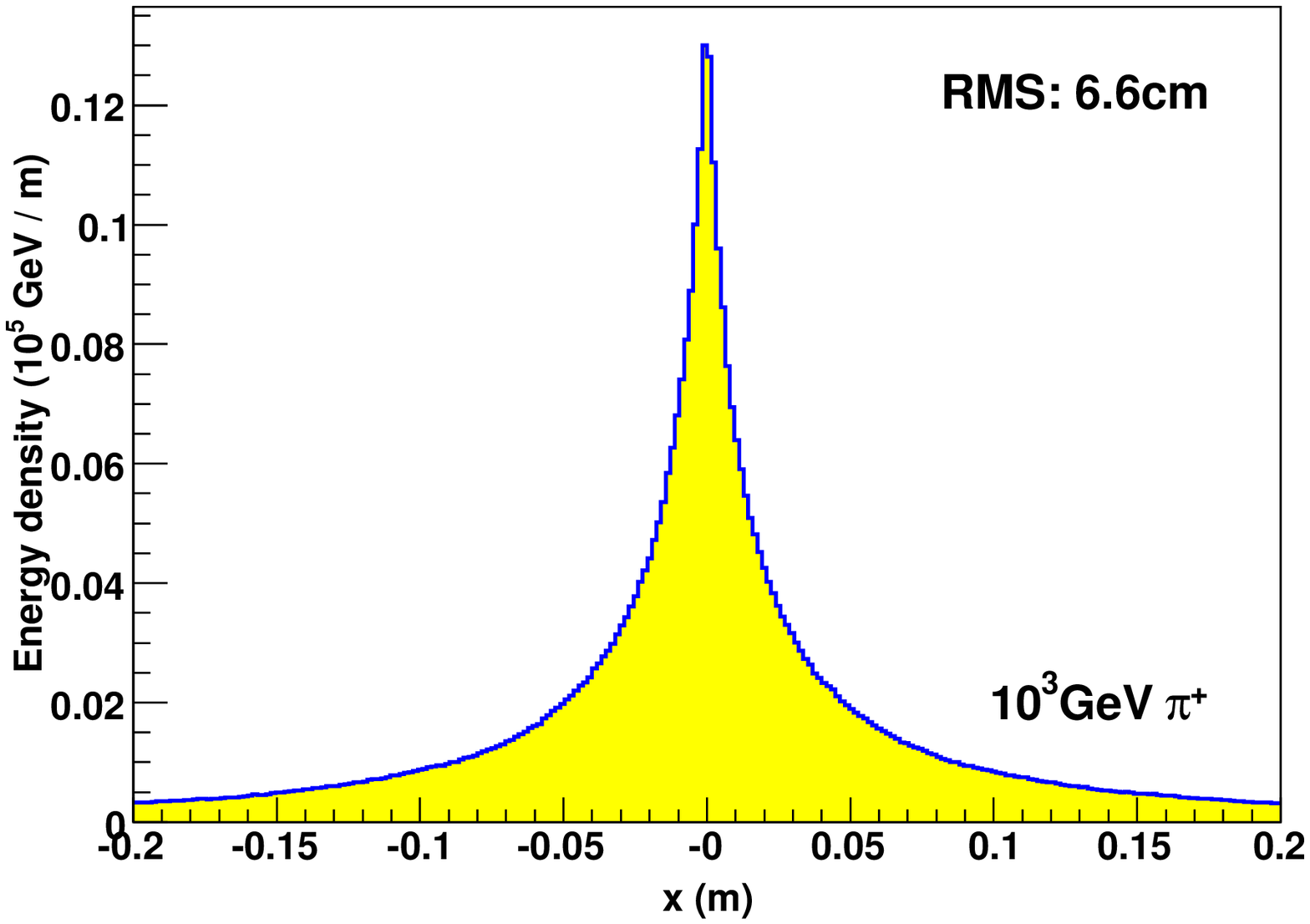}}
  \caption[Energy density deposited by $\pi^+$ mesons in water]{Energy
    density deposited by $\pi^+$ mesons in water.
    \subref{subfig:pi_plus_100TeV_zx} projection onto a plane
    containing the direction of the incident pion; the colour scale
    gives the energy density in GeV / m$^2$.
    \subref{subfig:pi_plus_100TeV} and
    \subref{subfig:pi_plus_1TeV} comparison of the energy density
    deposited by a 10$^5$\,GeV and a 10$^3$\,GeV $\pi^+$ respectively.}
  \label{fig:pi_plus_profiles}
\end{figure}

Due to this, we restrict ourselves to the simulation of showers with
energies up to 10$^5$\,GeV, and extrapolate the deposited energy
density to higher primary energies. The profile of a typical hadronic
cascade in water induced by a 10$^5$\,GeV $\pi^+$ meson is shown in
Figs.~\ref{subfig:pi_plus_100TeV_zx} and
\subref{subfig:pi_plus_100TeV}. The energy of the incident pion is
deposited in a cylindrical volume with a length of about 5\,m FWHM,
and a diameter of approximately 2\,cm FWHM. Systematic studies over
two decades of energy from 10$^3$\,GeV to 10$^5$\,GeV, where GEANT4
delivers valid results, show that the {\em functional form} of the
energy density is independent of the primary energy within about
fifteen percent (cf.~Figs.~\ref{subfig:pi_plus_100TeV} and
\subref{subfig:pi_plus_1TeV}). With increasing energy, the maximum of
the longitudinal shower profile is at higher $z$ values (deeper inside
the medium). The shower length\footnote{The shower length is defined
  as the interval of the longitudinal projection of the energy density
  where 90\% of the energy are deposited. It is calculated as the
  difference of the longitudinal coordinates below which 5\% and 95\%
  of the total energy are deposited respectively.}  and the shower
width slightly decrease with energy.

In this work we will assume, that for {\em all} energies the shape,
i.e.~the functional form, of the energy deposition density is given by
Figs.~\ref{subfig:pi_plus_100TeV_zx} and
\subref{subfig:pi_plus_100TeV}, and that the magnitude of the
histograms presented scales linearly with the energy of the hadronic
system.

\begin{figure}[ht]
  \centering
  \includegraphics[width=0.9\textwidth]
  {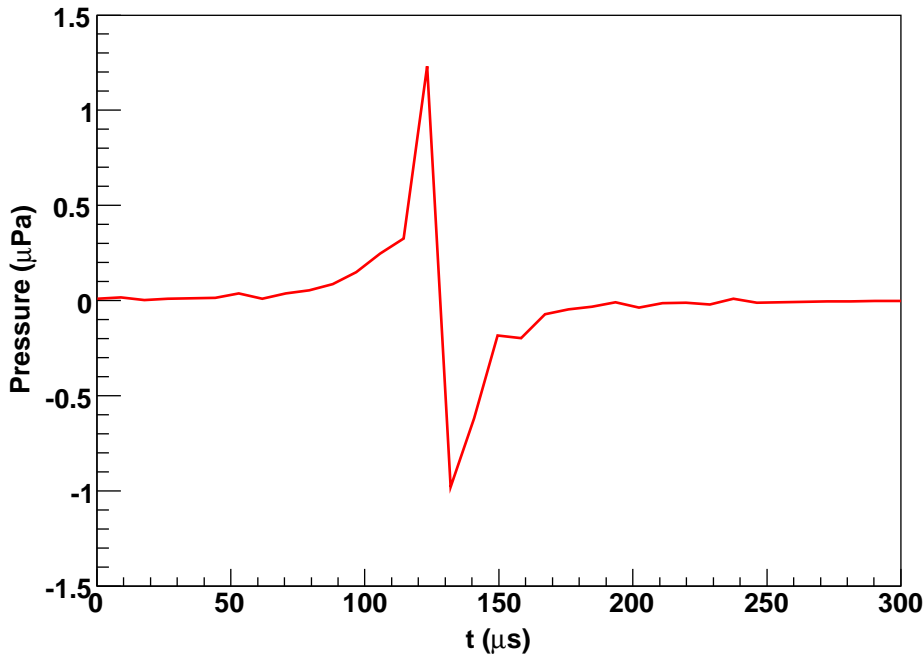}
  \caption[Bipolar acoustic signal produced by a $10^5$\,GeV $\pi^+$
  meson in water]{Bipolar acoustic signal produced by a $10^5$\,GeV
    $\pi^+$ meson in water calculated in a distance of 400\,m from the
    shower centre perpendicular to the shower axis. The cascade was
    produced at time $t = -266.5 \, \mathrm{ms}$, which corresponds to
    a sound travel path of 400\,m.}
  \label{fig:signal_100TeV_pi_plus}
\end{figure}

\medskip Using this energy deposition density we are now in a position
to calculate acoustic pulses emitted from hadronic cascades. Since we
look at cascades with ultra high energies, these cascades develop with
the speed of light in vacuum and we can assume the energy deposition
to be instantaneous and use (\ref{eq:kirchhoff_short}) to determine
the acoustic signal. A typical acoustic signal is shown in
Fig.~\ref{fig:signal_100TeV_pi_plus}.

It is calculated at a position 400\,m away from the cascade centre
perpendicular to its axis, the direction where the amplitude is
expected to be highest. The rather coarse time resolution of the
signal is caused by the bin width with which the energy deposition
density of the cascade was recorded -- a histogram with $250 \times
250 \times 250$ bins, which is already at the upper limit of what can
be stored on common computers. However, we have confirmed that the
maximum amplitude of the signal, which is the parameter important for
the simulations of an acoustic neutrino telescope as presented in
Chap.~\ref{chap:detectors}, is stable with respect to variations of
the number of bins.

In the simulations resulting in Fig.~\ref{fig:signal_100TeV_pi_plus}
no absorption of the signal during the propagation through the water
is included, yet. The properties of the signal, including the
attenuation of sound in water will be discussed in detail in
Chap.~\ref{chap:acoustics}.

\section{Electromagnetic cascades and the LPM effect}

Electromagnetic cascades originate from high energy photons or
electrons and develop in matter, above approximately 100\,MeV, mainly
through bremsstrahlung and pair production. They solely consist of
photons, electrons, and positrons, and thus show less variations than
hadronic cascades. Further, their transversal extension is smaller
than for hadronic showers, so that at a comparable shower length the
energy density in an electromagnetic cascade is higher than in a
hadronic cascade, and the acoustic signal produced has a higher
amplitude.

Unfortunately, this is not true for electromagnetic showers at highest
energies. It was first noted by Landau, Pomeranchuk
\cite{Landau:1953um, Landau:1953gr}, and Migdal \cite{Migdal:1956tc},
that above some threshold energy the cross sections for bremsstrahlung
and pair production would decrease rapidly with energy ({\em LPM
  effect}), which leads to strong elongation of the cascade, and a
decrease of the energy density, and thus of the amplitude of the
acoustic pulse.

Usually the cross sections for bremsstrahlung and pair production are
calculated by assuming momentum transfer to a single scattering centre
(i.e.~nucleus in the medium). At ultra high energies, there arise
destructive quantum mechanical interferences between amplitudes from
different scattering centres \cite{Eidelman:2004}, which reduce the
cross sections. For the case of bremsstrahlung one can picture the
photon as being emitted over a certain interaction length, which is
set by the temporal length of the wave train and the path length the
electron propagates during this time. If this interaction length
becomes comparable to the inter-atom distance in the medium, screening
effects and multiple scattering during the interaction start to play a
significant role.

\begin{figure}[ht]
  \centering
  \includegraphics[width=0.8\textwidth]{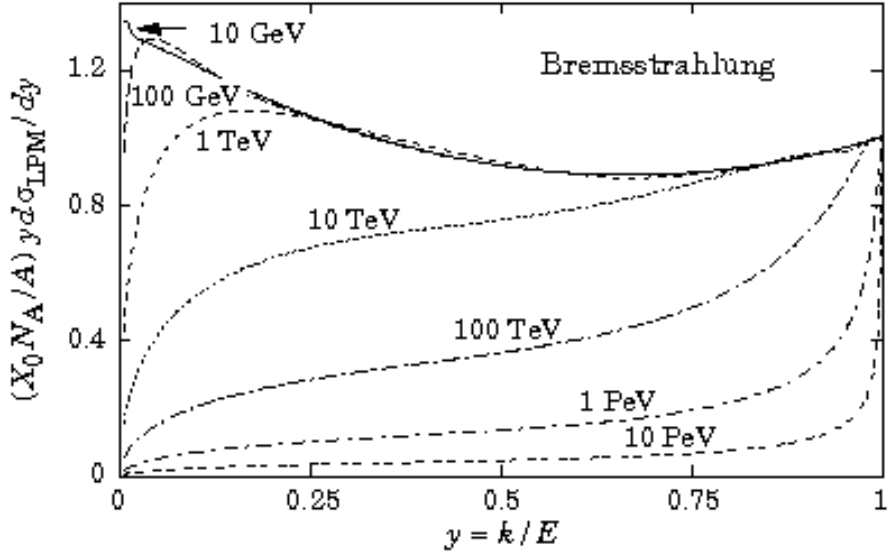}
  \caption[Normalised cross section for bremsstrahlung]{Normalised
    cross section for bremsstrahlung for different electron energies
    $E$. The energy of the emitted photon is denoted by $k$. The
    decrease of the total cross section, which is given by the area
    under the corresponding curves, with increasing energy is clearly
    visible (from \cite{Eidelman:2004}).}
  \label{fig:cross_section_brems}
\end{figure}

\begin{figure}[ht]
  \centering
  \includegraphics[width=0.8\textwidth]{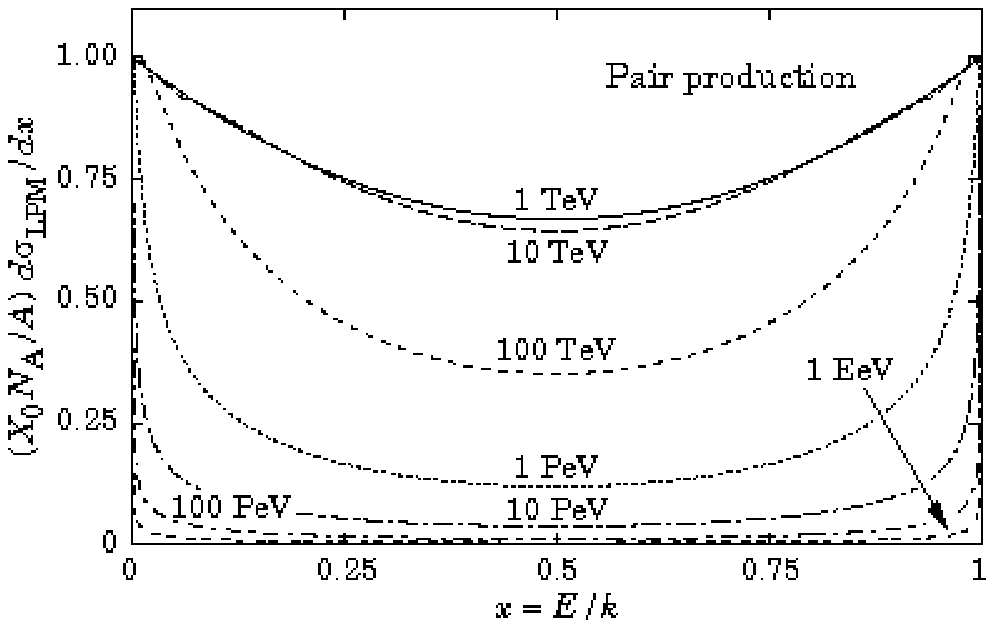}
  \caption[Normalised cross section for pair production]{Normalised
    cross section for pair production for different photon energies
    $k$. The energy of the emitted electron (or positron) is denoted
    by $E$. The decrease of the total cross section with energy is
    again clearly visible (from \cite{Eidelman:2004}).}
  \label{fig:cross_section_pair}
\end{figure}

Detailed calculations in the framework of QED
(e.g.~\cite{Migdal:1956tc}) result in the cross sections presented in
Figs.~\ref{fig:cross_section_brems} and \ref{fig:cross_section_pair}.
Differential cross sections for various electron and photon energies
are shown, normalised to eliminate the material properties. The
parameters and their values for water are as follows: Radiation length
$X_0 = 36 \, \mathrm{g} / \mathrm{cm}^2$, atomic mass $A = 18 \,
\mathrm{g} / \mathrm{mol}$, and Avogadro's number $N_A$, resulting in
the following scaling factor for the cross section:

\begin{displaymath}
  \frac{A}{X_0 \, N_A} = 8.3 \cdot 10^{-25} \, \mathrm{cm}^2
\end{displaymath}

Obviously, the total cross sections decrease with an increasing energy
of the incident particle or photon.

Simulations of electromagnetic cascades using the LPM cross sections
yield the following results \cite{Alvarez-Muniz:1997sh}:

\begin{itemize}
\item The shower length increases dramatically (between one and two
  orders of magnitude). Thus, assuming a constant shower width, the
  energy deposition region is increased, and the energy deposition
  density in the medium is reduced by the same factor. The resulting
  thermoacoustic signals, which scale linearly with the energy density,
  are strongly suppressed compared to signals from hadronic cascades
  with the same total energy. Simulations of acoustic signals from a
  $10^{11}$\,GeV hadronic cascade and a $10^{11}$\,GeV electromagnetic
  LPM cascade have shown, that the amplitude of the acoustic signal
  from the hadronic cascade has five times the amplitude of the pulse
  produced by the electromagnetic cascade \cite{Lehtinen:2001km}.
\item LPM cascades show large variations, since they develop
  sub-showers at various distances along the cascades when particles
  from the main shower drop below the LPM threshold energy. Thus a
  parameterisation of LPM cascades is impossible, and a simulation on
  an event-by-event basis is necessary.
\end{itemize}

\medskip In summary, we get the following conditions for the
simulation of acoustic signals in an acoustic neutrino telescope: Even
if generally electromagnetic cascades are easier to treat numerically
since they are dominated by two types of interactions only,
bremsstrahlung and pair production, this is not longer true at highest
energies due to the LPM effect. There are simulation codes to assess
the longitudinal profile of LPM showers\footnote{These simulation
  codes are mostly developed by EAS experiments like Pierre Auger
  looking at the fluorescence light from air showers. The output of
  fluorescence light can completely be determined from the
  longitudinal profile, since, due to the much shorter wavelengths
  involved, there is no interference expected from contributions in
  the transversal shower extension.} also in water, but no fully 3-D
simulation is known to the author. But this full 3-D energy profile is
required to scrutinise the acoustic signal produced by such a cascade
since the signal is very sensitive to the transversal energy profile.
Although we have shown in Sec.~\ref{sec:nu_propagation} that on
average about 80\% of the neutrino energy is transferred into the
electromagnetic cascade, estimations of the acoustic signals
\cite{Lehtinen:2001km} predict that, for equal energies, the pressure
amplitude from the electromagnetic cascade is only about 20\% of the
amplitude of the hadronic cascade. Thus, we expect for ultra high
energy neutrino interactions on average pressure amplitudes of the
same order of magnitude from the electromagnetic and the hadronic
cascade.

Since in the case of ultra high energies, hadronic cascades are
expected to show less variations than electromagnetic LPM showers, we
decided to completely neglect electromagnetic cascades in our study,
and only utilise the acoustic signals emitted from the hadronic
shower. So all limits derived for the diffuse neutrino flux are solid
upper limits and should be improved by the inclusion of the
electromagnetic cascades into the study.


\chapter{Sound propagation and detection in water}
\label{chap:acoustics}
\minitoc

\bigskip In this chapter we discuss the propagation of the acoustic
signal from the cascade to the sensors and develop a parameterisation
of the signal amplitude for any sensor position. Possibilities to
extract the signal out of the background noise are discussed and
sensitivity estimates for single sensors are presented.

\section{Attenuation in fresh water and sea water}
\label{sec:attenuation}

The main cause for the degradation of the acoustic signal between the
place of production at the hadronic cascade and a sensor is
attenuation. There are two types of attenuation which have to be taken
into account.

The first type is due to dispersion of the signal during propagation
and has already been discussed for the proton beam experiment
(cf.~Fig.~\ref{fig:amplitude_dist}) where we have shown that in the
near field -- compared to the extension of the cascade producing the
signal -- the pressure field has a disc-like shape around the shower
axis, and the pressure amplitude thus decreases like $1 / \sqrt{r}$
when $r$ is the distance from the cascade. In the far field the
cascade can be assumed as a point source, and the signal amplitude
drops like $1 / r$.

\begin{figure}[ht]
  \centering
  \includegraphics[width=0.9\textwidth]{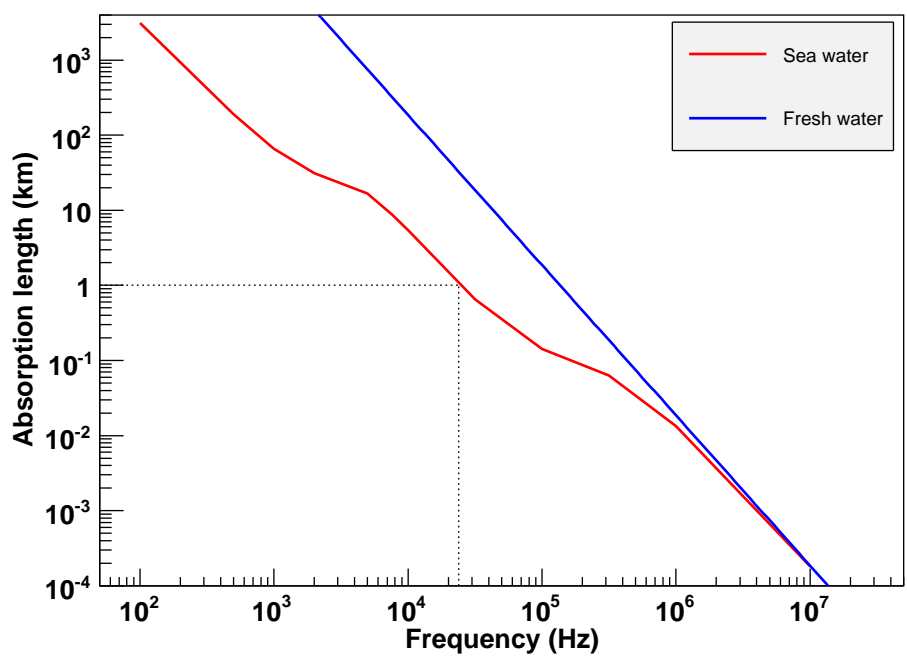}
  \caption[Sonic attenuation length in water]{Sonic absorption length
    in sea water and fresh water as a function of frequency (data from
    \cite{Urick:1983}).}
  \label{fig:attenuation_length}
\end{figure}

The second type of attenuation, which has to be accounted for, is the
absorption in the water, which is strongly frequency dependent and
very different for fresh water and sea water.
Figure~\ref{fig:attenuation_length} shows the absorption length for
both media as a function of frequency. The absorption length is
defined as the distance after which the {\em energy} in the acoustic
wave has decreased by a factor $1 / e$, i.e.~the pressure amplitude
has dropped by a factor $1 / \sqrt{e}$.

In fresh water the absorption is due to the viscosity of the water,
which means that energy is transferred from the sonic wave into the
water when the water molecules are displaced against each other under
the inhomogeneously changing pressure. Let $\mu = 0.01$\,P be the
viscosity of the water\footnote{The SI unit for viscosity is the
  Poise, named after J.L.~Poiseuille (1799 - 1869): 1\,P = $1 \,
  \mathrm{Ns} / \mathrm{m}^2$.}. Then the absorption length $L$ is
given by \cite{Urick:1983}:

\begin{equation}
  \label{eq:absorption_length}
  L = \frac{10}{\ln 10} \, \frac{3 \rho c^3}{16 \pi^2 \mu} \,
  \frac{1}{f^2}
\end{equation}

\noindent where $\rho$ is the density of the water, $c$ is the speed
of sound, and $f$ is the frequency of the acoustic signal.

For frequencies above 1\,MHz the absorption in fresh water and sea
water are nearly the same. Below 1\,MHz the absorption in sea water
increases rapidly due to ionic relaxation processes. The decrease of
the absorption length between about 5\,kHz and 1\,MHz, the range
important for acoustic particle detection, is caused by MgSO$_4$
dissolved in the sea water (see \cite{Verma:1959} for a review of
sound absorption due to ionic relaxation).

In the unperturbed sea there is an equilibrium between the molecular
and the ionic system:

\begin{displaymath}
  \mathrm{MgSO}_4 \rightleftarrows \mathrm{Mg}^{2+} +
  \mathrm{SO}_4^{2-}
\end{displaymath}

\noindent The equilibrium concentration of the components is pressure
dependent, and a perturbation of the equilibrium state through the
pressure of the acoustic wave leads to a dissociation of MgSO$_4$
molecules. The energy for this decomposition is deprived from the
sonic wave. This energy is released only after some relaxation time
$\tau$ mainly as thermal energy, thus degrading the acoustic signal.
Although MgSO$_4$ amounts to only 4.7\% of the weight of all salts
dissolved in the sea \cite{Urick:1983}, this effect causes the
absorption length to decrease by a factor of 30 compared to fresh
water in the frequency range considered.

Below 5\,kHz, a frequency range of minor importance for acoustic
neutrino detection, the sonic absorption length drops by about another
factor of 10. This is caused by an ionic relaxation process involving
boric acid

\begin{displaymath}
  \mathrm{B}(\mathrm{OH})_3 + \mathrm{H}_2\mathrm{O} \rightleftarrows
  \mathrm{B}(\mathrm{OH})_4^- + \mathrm{H}^+
\end{displaymath}

\noindent which is very complicated since the equilibrium state
strongly depends on the pH and on other ions dissolved in the water
\cite{Urick:1983}.

\medskip At a frequency of about 20\,kHz, which is the central
frequency of the bipolar pressure pulse expected from a neutrino
induced particle shower, the sonic absorption length in sea water is
1\,km. This is about a factor of 15 larger than the absorption length
of \v{C}erenkov light which is around 60\,m in both sea water
\cite{Aguilar:2004nw} and ice \cite{Askebjer:1994yn}. This distance
governs the spacing between the photomultiplier tubes in todays water
\v{C}erenkov neutrino telescopes (cf. Sec.~\ref{sec:water_cerenkov}).
Thus, the inter-sensor spacing in an acoustic neutrino telescope could
be larger by a factor of 10 than in an optical neutrino telescope,
which allows to instrument much larger target masses with the same
number of sensors.

We will use the presented attenuation of sound in water in
Sec.~\ref{sec:signal_parameterisation} to develop a parameterisation
of neutrino induced acoustic pulses for any sensor position, which is
a prerequisite for the implementation of a simulation of an acoustic
neutrino telescope.

\section{Refraction and Reflection}
\label{sec:refraction}

Two other mechanisms leading to a degradation of an acoustic pulse
propagating through the sea are refraction and reflection of the
signal on the sea surface and bottom.

\begin{figure}[ht]
  \centering
  \includegraphics[width=0.5\textwidth]{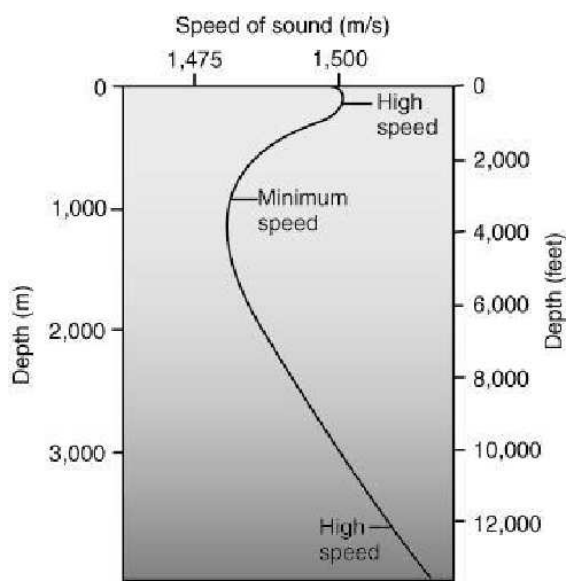}
  \caption[Typical deep sea sound velocity profile]{Typical deep sea
    sound velocity profile (http://www.punaridge.org/).}
  \label{fig:sound_velocity}
\end{figure}

The refraction of the signal is caused by the fact, that the speed of
sound in the sea is not constant, but varies slightly with depth. The
sound velocity is a function of temperature and pressure. A typical
deep sea sound velocity profile is shown in
Fig.~\ref{fig:sound_velocity}.

In the first few ten metres beneath the surface the sound velocity
undergoes strong seasonal variations due to the varying water
temperature. Below this, down to about 1000\,m, the speed of sound
decreases with the water temperature. Further below the temperature
remains fairly constant and the sound velocity starts to increase
again, following the water pressure linearly.

\begin{figure}[ht]
  \centering
  \subfigure[Definition of angles.]{
    \label{subfig:refraction_a}
    \includegraphics[width=0.48\textwidth]{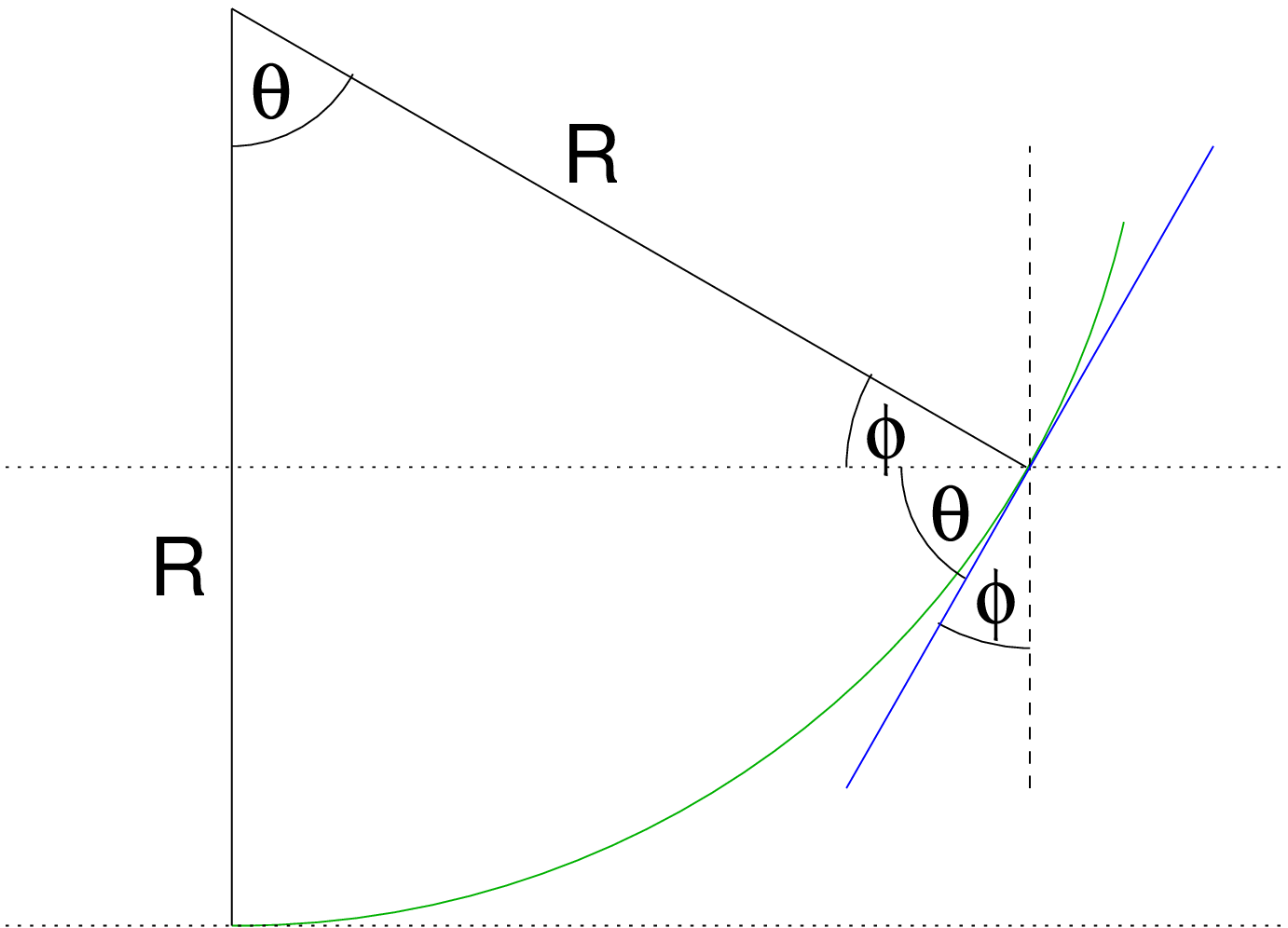}}
  \subfigure[Geometry.]{
    \label{subfig:refraction_b}
    \includegraphics[width=0.48\textwidth]{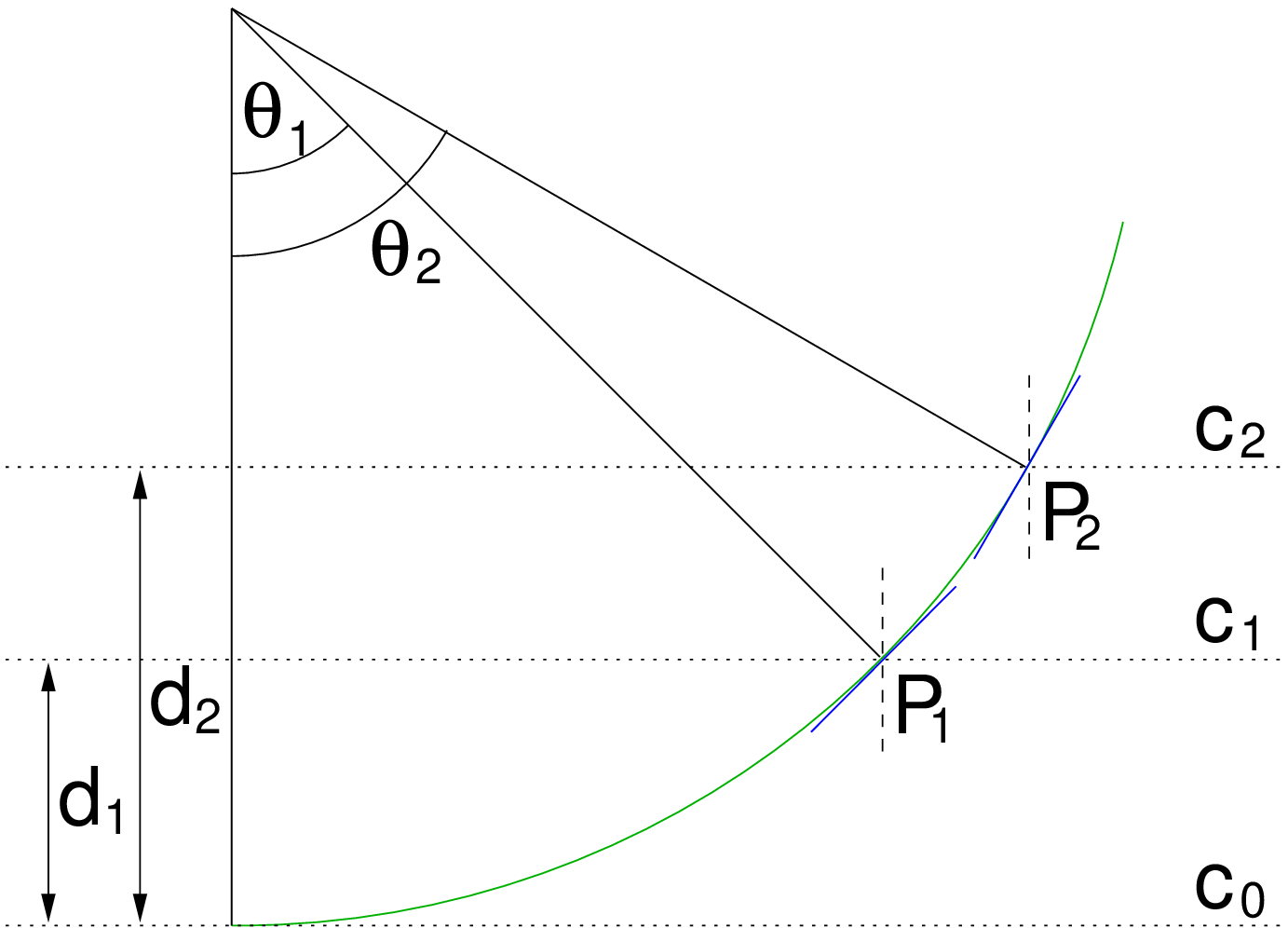}}
  \caption[Refraction in a medium with a linear velocity
  gradient]{Refraction in a medium with a linear velocity gradient.}
  \label{fig:refraction}
\end{figure}

We will now show that a wave in a medium, in which its velocity
changes linearly in one direction, propagates on a circular path
\cite{Urick:1983}. We will use the geometry and the symbols defined in
Fig.~\ref{fig:refraction}.

The sound velocity is supposed to increase linearly from top to bottom
with some rate $k$ so that $c_i = c_0 - k d_i$ ($k > 0$), where $c_0$
is the speed of sound at some reference depth where the ``sonic ray''
would be perpendicular to the velocity gradient (cf.
Fig.~\ref{fig:refraction}). Let us consider a sonic ray at an
arbitrary point $P1$ with an angle of incidence $\phi_1$ which
propagates to $P2$ where its incident angle is $\phi_2$. Then we know
from the law of refraction that $n_i \sin \phi_i = n_i \cos \theta_i =
\mathrm{const.}$, where the index of refraction $n_i$ is proportional
to the inverse of the wave velocity $c_i$:

\begin{displaymath}
  \frac{1}{c_0} = \frac{\cos \, \theta_1}{c_1} = \frac{\cos \,
    \theta_2}{c_2}
\end{displaymath}

\noindent Here, we have used our prerequisite that in the depth where
$\theta = 0$ the speed of sound is $c_0$. Thus, from the law of
refraction follows:

\begin{equation}
  \label{eq:circle_refraction}
  c_1 - c_2 = c_0 \cos \, \theta_1 - c_0 \cos \, \theta_2
\end{equation}

\noindent From the definition of the gradient in the sound velocity we
get:

\begin{equation}
  \label{eq:circle_gradient}
  c_1 - c_2 = -k (d_1 - d_2)
\end{equation}

\noindent Combining (\ref{eq:circle_refraction}) and
(\ref{eq:circle_gradient}) we get the following equation, where we
have made no assumptions in the shape of the sonic ray, yet:

\begin{equation}
  \label{eq:circle_physics}
  d_1 - d_2 = - \frac{c_0}{k} \, \cos \, \theta_1 + \frac{c_0}{k} \,
  \cos \, \theta_2
\end{equation}

If we now only consider the geometry as it is sketched in
Fig.~\ref{subfig:refraction_b}, which assumes a circular trajectory,
we get:

\begin{displaymath}
  R - d_i = R \, \cos \, \theta_i
\end{displaymath}

\noindent or

\begin{equation}
  \label{eq:eq:circle_geometry}
  d_1 - d_2 = -R \, \cos \, \theta_1 + R \, \cos \, \theta_2
\end{equation}

Since we have not made any assumptions about the points $P1$ and $P2$
we can deduce from comparing (\ref{eq:circle_physics}) and
(\ref{eq:eq:circle_geometry}), where the first equation was derived
only from basic physics without making any assumptions on the
trajectory, and the second equation was deduced from the geometry
only, that all points a ray passes in a linear velocity gradient lie
on a circle with radius $R = c_0 / k$. For example for the
Mediterranean Sea the velocity gradient $k = 16.5 \, \mathrm{ms}^{-1}
/ \mathrm{km}$ \cite{Niess:2005}, and thus the radius of the sonic
rays is $R = 91$\,km.

This has a great impact on the design of an acoustic neutrino
telescope. Let us assume a water depth of 2500\,m, which is a typical
value for the Mediterranean Sea, and a neutrino telescope built on the
sea bed with an instrumented volume reaching 1\,km above the sea
floor. We define $\Delta x$ as the horizontal distance a sonic ray,
which is emitted horizontally near the bottom of the sea, has to
travel to gain the vertical distance $\Delta y$:

\begin{equation}
  \Delta x = \sqrt{R^2 - \left( R - \Delta y \right)^2}
\end{equation}

\begin{figure}[ht]
  \centering
  \includegraphics[width=0.9\textwidth]{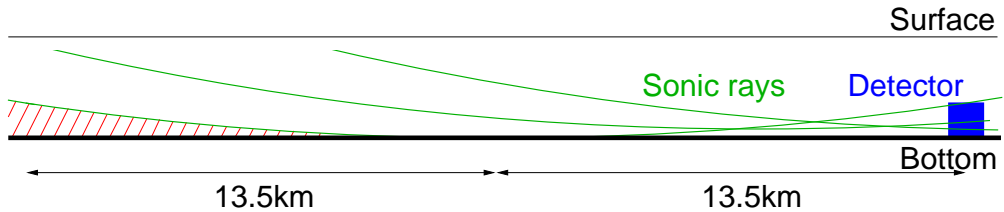}
  \caption[Geometrically accessible detection volume]{Geometrically
    accessible detection volume. The inaccessible volume is indicated
    in red.}
  \label{fig:accessible_volume}
\end{figure}

Figure~\ref{fig:accessible_volume} shows that after a distance of
13.5\,km a ray emitted horizontally at the sea floor will no longer
pass a detector of 1\,km height. Further it can be seen, that for
distances greater than $2 \cdot 13.5$\,km there practically do not
exist any rays which will pass the detector. All rays emitted inside
the volume indicated in red will be bent above the detector. This
introduces a natural cutoff to the water volume observable with a
single sensor\footnote{Not taking this effect into account would allow
  to detect pressure pulses produced by highest energy neutrinos
  ($E_\nu \gtrsim 10^{14}$\,GeV) from distances even greater than
  500\,km, i.e.~from practically everywhere in the Mediterranean
  Sea.}.

For the simulation of an acoustic neutrino telescope, which we will
present in Chap.~\ref{chap:detectors}, we will assume that the sound
propagates on straight lines, since developing a dedicated ray-tracing
algorithm\footnote{Detailed treatment of the sound propagation
  requires ray tracing, since the complicated sound velocity profile
  in the upper water layers leads to deviations from the circular
  path. Especially the minimum of the sound velocity below the surface
  leads to channelling of acoustic waves which is completely neglected
  in this work.} is not part of this work. We introduce the effects of
refraction into the simulation by assuming, that no acoustic signal
from neutrino interactions taking place more than 27\,km ($2 \cdot
13.5$\,km) away from the instrumented volume can reach the detector.
Further, we neglect any signals that are reflected at the sea surface
or bottom. There is no information whether a neutrino induced bipolar
acoustic pulse, which will be degraded when reflected at those uneven
surfaces, can still be extracted from the background noise. By
assuming straight line propagation, signals crossing the sea surface
or bottom simply do not further contribute to the simulation. The
influence of reflection on short acoustic pulses on rough surfaces
will have to be investigated in a separate study.

By using straight line propagation, the zenith distribution of the
signals reaching the detector is slightly changed towards vertically
downward going neutrinos. In the straight line case mostly acoustic
signals produced by vertical cascades will reach the detector, whereas
in the case of refracted trajectories those signals will be bent above
the detector. This is balanced by a contribution of inclined cascades,
whose signals are absorbed at the bottom when assuming straight line
propagation, and which will otherwise be refracted over the seabed and
back into the detector. In this work we {\em assume} that for an
isotropic neutrino flux from the upper hemisphere the total number of
signals reaching the detector is equal for both cases.

\section{Background noise and signal extraction}
\label{sec:noise}

We will now discuss the acoustic background noise which is always
present in the sea, and out of which a neutrino induced bipolar
pressure pulse has to be extracted in an acoustic neutrino telescope.

\subsection{Properties of the background noise}

Acoustic noise in the sea has been studied extensively in the 20th
century (cf. \cite{Urick:1986} for a review), mostly during World War
II for the development and tracking down of submarines. The noise can
be divided into three frequency bands:

\begin{itemize}
\item {\bf below 200\,Hz.} This frequency range, which is unimportant
  with respect to acoustic particle detection, is dominated by
  shipping and similar anthropogenic noise, and thus shows a high
  variability.
\item {\bf 200\,Hz to 50\,kHz.} In this frequency band neutrino
  induced acoustic signals are expected. The noise in this band
  strongly depends on the wind speed at the sea surface, even for deep
  sea measurements. The main sources of noise are wind turbulence on
  the rough sea surface, the motion of the sea surface, interactions
  of surface waves travelling in different directions, and spray and
  cavitation, i.e. air bubbles trapped in the water near the surface
  \cite{Urick:1986}.
\item {\bf above 50\,kHz.} Here, the noise spectrum is dominated by
  thermal noise. It is caused by the thermal motion of the water
  molecules which collide with the sensor.
\end{itemize}

\begin{figure}[ht]
  \centering
  \includegraphics[width=0.9\textwidth]{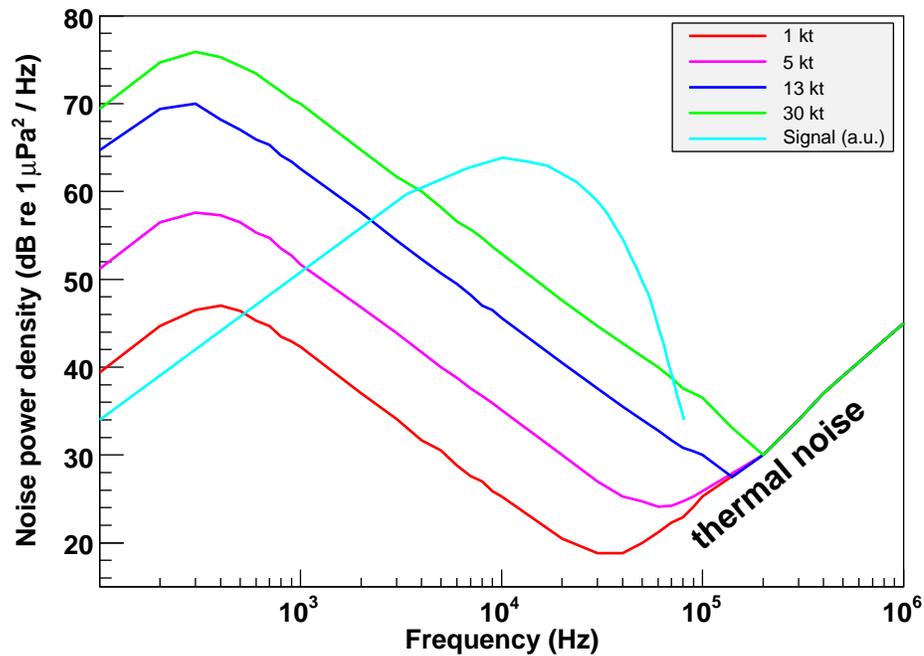}
  \caption[Acoustic noise power density in the sea]{Acoustic noise
    power density in the sea for different wind speeds
    \cite{Lehtinen:2001km} (1\,kt = 1.852\,km/h). Imposed is the
    spectrum of the acoustic signal produced by a hadronic cascade at
    a distance of 1000\,m perpendicular to the shower axis. The
    absolute amplitude of the signal spectrum depends on the cascade
    energy.}
  \label{fig:noise}
\end{figure}

Typical deep sea acoustic noise spectra measured with the AUTEC
hydrophone array \cite{Lehtinen:2001km} near the Bahamas at different
wind speeds are shown in Fig.~\ref{fig:noise}.

The wind noise has a maximum at around 300\,Hz, and decreases for
higher frequencies until the thermal noise, which increases linearly
with frequency, starts to dominate the spectrum. This leads to a
minimum in the noise spectrum between 30\,kHz and 200\,kHz, depending
on the wind speed.  Also shown is the shape of the frequency spectrum
of a neutrino induced bipolar pressure pulse. The absolute amplitude
of this spectrum depends, of course, on the energy of the particle
cascade and its distance to the sensor. It is noteworthy, that the
maximum of the signal spectrum at about 20\,kHz falls into the
decreasing part of the noise spectrum, and even close to its minimum
for low wind speeds.  This leads to an improved signal-to-noise ratio
in an appropriate frequency band around 20\,kHz, and thus increases
the detectability of neutrino induced pressure pulses.

Table~\ref{tab:noise_rms} gives the RMS values of the noise calculated
from the presented spectra in different frequency bands, and also, as
a rough guide, the cumulative probability $w$ to find the indicated
weather conditions.

\begin{table}[hbt]
  \centering
  \caption[RMS of acoustic noise in the sea]{RMS of acoustic noise in
    the sea in mPa for different wind speeds and frequency bands
    (1\,kt = 1.852\,km/h). The probability $w$ to have the indicated
    or better conditions is taken from \cite{Lehtinen:2001km}, and is
    valid for the AUTEC site near the Bahamas.}
  \label{tab:noise_rms}
  \begin{tabular}{rrrrr}
    \hline
    & 0 -- 500\,kHz & 0 -- 100\,kHz & 3 -- 100\,kHz & $w$ \\ \hline
    1\,kt & 36.0 & 7.9 & 4.5 & 0.01 \\
    5\,kt & 42.0 & 23.1 & 9.9 & 0.2 \\
    13\,kt & 90.3 & 83.2 & 41.4 & 0.6 \\
    30\,kt & 188.0 & 184.0 & 82.0 & 1 \\ \hline
  \end{tabular}
\end{table}

These values now allow us to develop a single sensor trigger for
bipolar pulses, and to estimate the sensitivity of a single sensor
system to neutrino induced pressure pulses.

\subsection{Filtering in the time domain}

The most simple trigger algorithm is to use a threshold trigger. The
pressure signal in the sensor is monitored, and when the pressure
rises above some threshold pressure $p_\mathrm{th}$ the time and the
maximum amplitude of the signal are recorded.

The distribution of wind speeds, and thus the temporal variation of
the noise is known (e.g.~\cite{Lehtinen:2001km}). It can be
calculated, that, if one uses single hydrophones and an appropriate
bandpass filter to eliminate the noise in the frequency bands where no
signal contribution is expected, a threshold $p_\mathrm{th} = 35$\,mPa
has to be used if one allows for one event triggered only from
background in 10 years with a five fold coincidence
\cite{Danaher:2005}.

Since especially the thermal noise is expected to have a very short
correlation length\footnote{The Erlangen ANTARES group will equip two
  sectors of the ANTARES detector with acoustic sensors to study the
  properties of the noise at the ANTARES site in the Mediterranean
  Sea, especially the correlation lengths at different frequencies,
  in-situ. The arrangement of the sensors will allow to measure
  correlations over two orders of magnitude in length, from 1\,m to
  100\,m.} it will be possible to further increase the signal-to-noise
ratio, and correspondingly decrease the threshold pressure
$p_\mathrm{th}$ by using sensor clusters where the signals from the
single sensors are summed up coherently. For completely uncorrelated
noise the signal-to-noise ratio in a sensor cluster of $N$ sensors is
expected to increase by $\sqrt{N}$.

For the simulation study presented in this work we will use such a
threshold trigger, and assume that, with appropriate techniques, the
detection threshold for neutrino induced bipolar acoustic pulses can
be lowered down to $p_\mathrm{th} = 5$\,mPa. The dependence of the
detector sensitivity on $p_\mathrm{th}$ will be discussed in
Sec.~\ref{sec:sensitivity}.

More sophisticated filtering techniques like the usage of correlation
functions between the measured data stream and the expected bipolar
signal shape are under study in Erlangen.

\subsection{Filtering in the frequency domain}

A common method is to analyse time slices with a length, which is a
few times the length of the expected signal. Calculating the total
energy in a frequency band between for example 3\,kHz and 100\,kHz
where the signal is expected to dominate over the noise will allow to
distinguish between time slices that contain a signal and those that
contain no signal. The advantage of this method is, that neither the
exact position of the signal in the time slice has to be known to
identify it, nor the exact shape of the signal, since the energy can
be distributed over the considered frequency band. If a time slice is
identified to contain a signal, more sophisticated algorithms with
higher CPU time requirements can be used to determine the exact time
and amplitude of the signal.

\bigskip In the following we will restrict ourselves to detectors
where the problem of signal extraction out of the noise is already
solved. We define the {\em acoustic module} (AM) to be the basic
element of our acoustic neutrino telescope:

\begin{verse}
  An acoustic module (AM) is a device that can unambiguously trigger \\
  on bipolar acoustic pulses, as they are for example produced by
  neutrino \\
  induced particle cascades, which have a maximum pressure amplitude
  \\
  greater then some threshold pressure $p_\mathrm{th}$, and can
  measure and transmit \\
  to shore their arrival time and amplitude.
\end{verse}

An acoustic module will most probably be some local array of acoustic
sensors like hydrophones or naked piezo ceramics mounted into a
pressure tight vessel \cite{Karg:2005} combined with a dedicated
filtering algorithm for bipolar pulses.

\section{Parameterisation of the acoustic signal}
\label{sec:signal_parameterisation}

\begin{figure}[ht]
  \centering
  \includegraphics[width=0.9\textwidth]{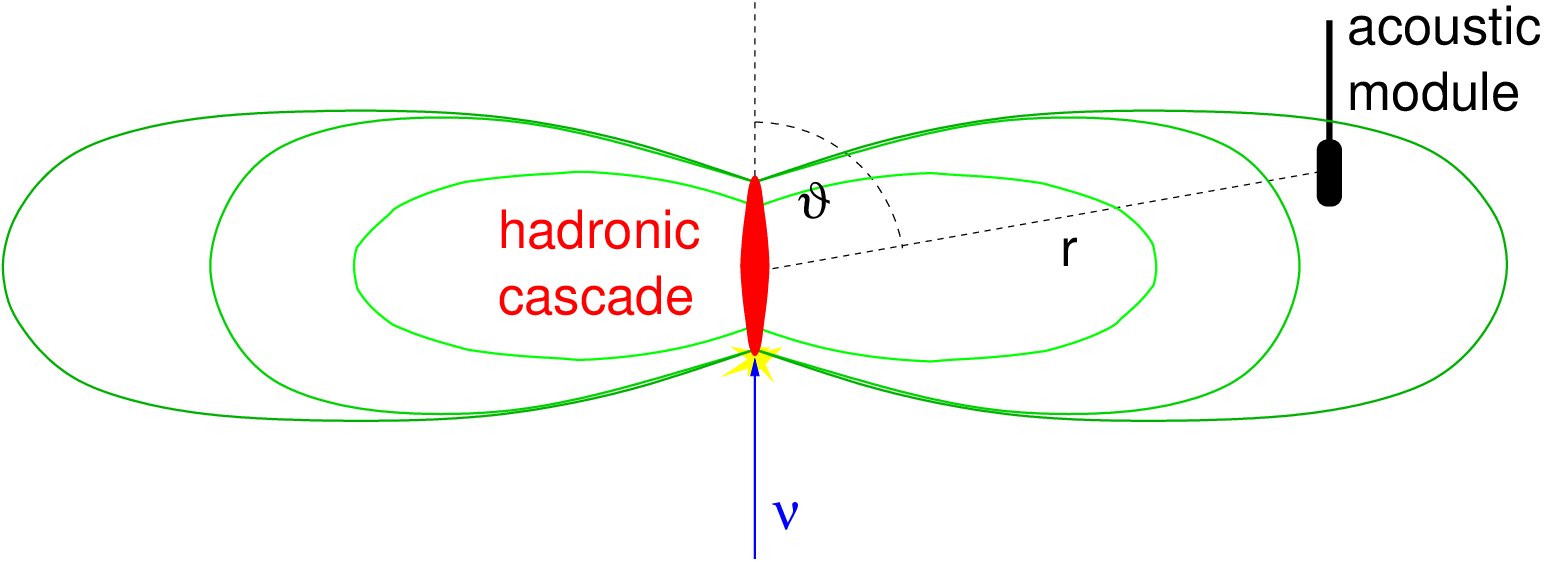}
  \caption[Coordinate system used for the parameterisation of the
  pressure field]{Coordinate system used for the parameterisation of
    the pressure field. The relevant coordinates are the distance from
    the shower centre and the angle to its axis. The system is
    rotationally symmetric around the cascade axis.}
  \label{fig:disc_coordinates}
\end{figure}

In this section we will develop a parameterisation of the maximum
amplitude of neutrino induced bipolar pressure pulses as a function of
the sensor position, which is a prerequisite for the simulation of an
acoustic neutrino telescope presented in Chap.~\ref{chap:detectors}.
We utilise the fact, that the pressure field is rotationally
symmetric around the axis of the cascade. Thus, the pressure field can
be described by the two coordinates distance $r$ from the centre of
the shower to the acoustic module and the angle $\vartheta$ between
the shower axis and the vector pointing to the AM. The coordinate
system used is shown in Fig.~\ref{fig:disc_coordinates}.

We have simulated acoustic pressure pulses produced by a 10$^5$\,GeV
hadronic cascade for various sensor positions using the methods
described in Secs.~\ref{sec:proton_mc} and
\ref{sec:hadronic_cascades}.

Attenuation of the signal due to the geometric spread is included in a
natural way in the wave equation (\ref{eq:thermoacoustic_model}) from
which the signals are calculated. Absorption of the signal is
introduced into the simulation by first calculating the signal at the
sensor position without absorption, and then using a Discrete Fourier
Transform to determine its frequency components. Each frequency $f$ is
then attenuated by the frequency dependent absorption length presented
in Fig.~\ref{fig:attenuation_length}. The signal is then transformed
back into the time domain.

For every signal the peak pressure amplitude $p_\mathrm{max}$ and the
corresponding arrival time $t_\mathrm{max}$ were determined. To
minimise the influence of binning effects of the energy deposition
density by the hadronic cascade and of numerical errors of the
integration of this density, all signals were calculated at 20
positions with constant $r$ and $\vartheta$, distributed at equal
distances around the cascade axis. For the parameterisation the
average values from these 20 signals were used.

\begin{figure}[ht]
  \centering
  \includegraphics[width=0.9\textwidth]{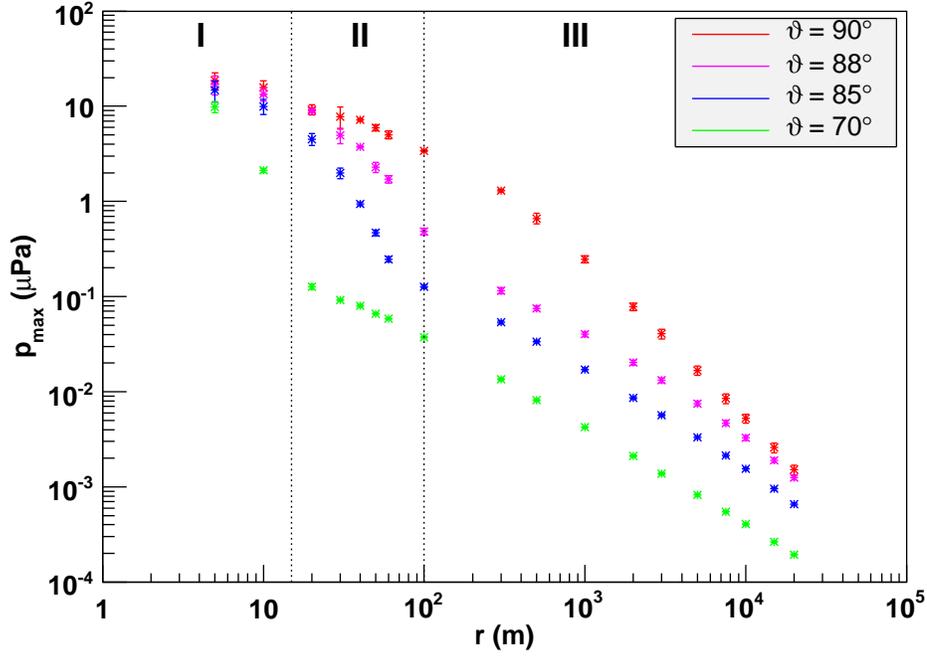}
  \caption[Dependence of the signal amplitude on the
  distance]{Dependence of the signal amplitude on the distance from
    the shower centre. The values are calculated for the $10^5$\,GeV
    hadronic cascade shown in Fig.~\ref{fig:pi_plus_profiles}. The
    different sets of data points are for different angles relative to
    the shower axis.}
  \label{fig:pressure}
\end{figure}

We discovered that the most simple\footnote{Here, {\em simple} is
  meant in the sense that a simple functional form can be found and
  fitted to the derived values.} parameterisation can be achieved
by calculating $p_\mathrm{max}$ and $t_\mathrm{max}$ as a function of
$r$ at a fixed angle $\vartheta$, and repeat this procedure for
different $\vartheta$. Figure~\ref{fig:pressure} shows a compilation
of the results obtained. The error bars indicate the RMS of the 20
values calculated at positions around the shower axis.

As is expected for a disc-shaped signal propagation, the signal is
highest perpendicular to the shower axis ($\vartheta = 90^\circ$) and
decreases rapidly in other directions. Further it can be seen, that
the distance dependence can roughly be divided into three sections: A
near field region $r < 15$\,m (I) where the signal behaviour strongly
depends on the source distribution, and thus on the direction where
the measurement is performed, an intermediate transition region $15 \,
\mathrm{m} < r < 100$\,m (II), and a far field region $r > 100$\,m
(III) where the amplitude decreases like $1 / r$ as expected from a
point source. Each of these sections shows a nearly linear behaviour
in the double logarithmic plot. We use this to introduce the following
parameterisation of the signal amplitude:

\begin{equation}
  \label{eq:parameterisation_p}
  p_\mathrm{max} (r, \vartheta) = p_{0,i} (\vartheta) \left(
    \frac{r}{1 \, \mathrm{m}} \right)^{a_i (\vartheta)} \qquad i \in
  \{\mathrm{I}, \mathrm{II}, \mathrm{III} \}
\end{equation}

\begin{figure}[ht]
  \centering
  \subfigure[Fit parameter $a$.]{
    \label{subfig:pressure_fits_a}
    \includegraphics[width=0.48\textwidth]{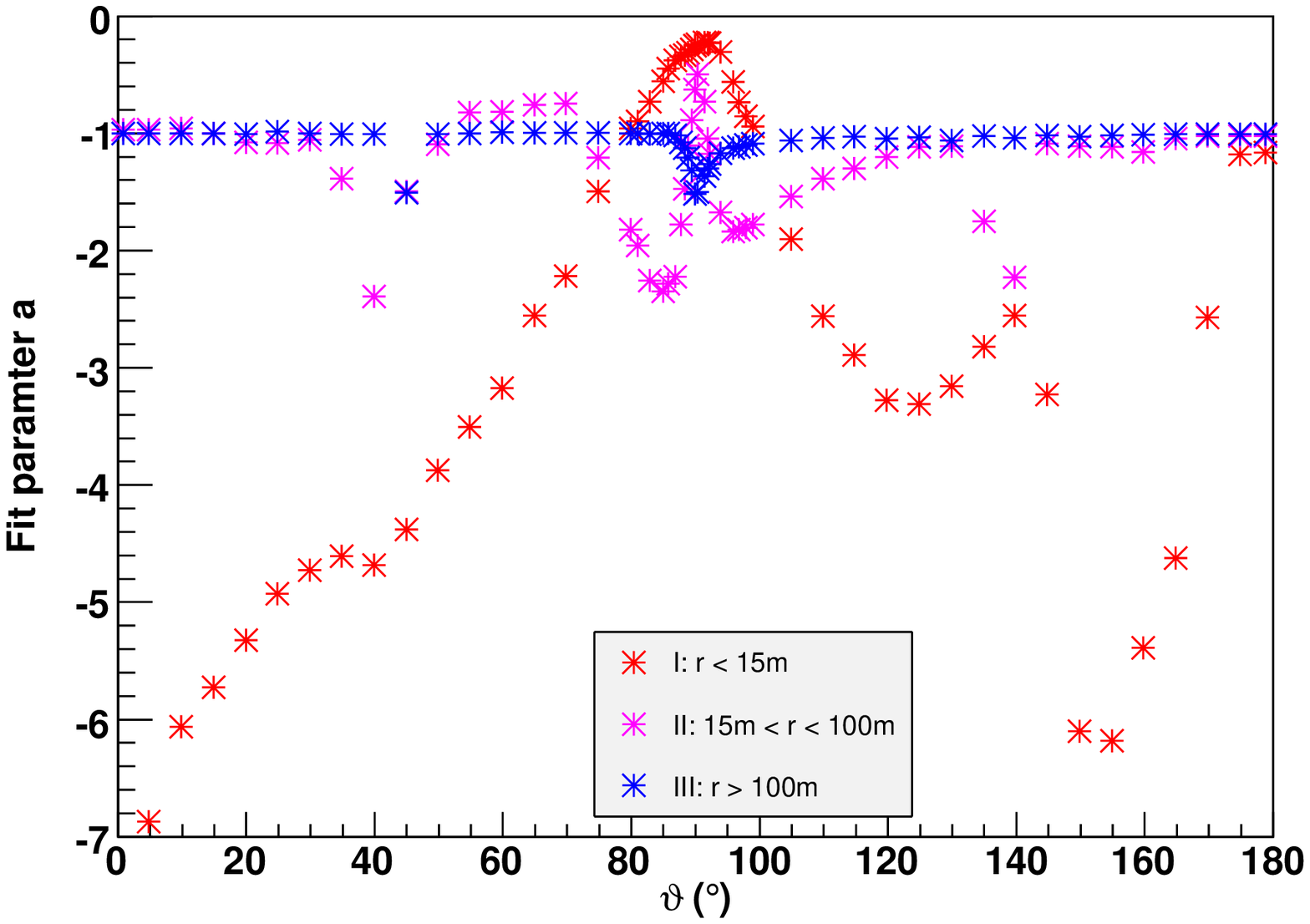}}
  \subfigure[Fit parameter $p_0$.]{
    \label{subfig:pressure_fits_b}
    \includegraphics[width=0.48\textwidth]{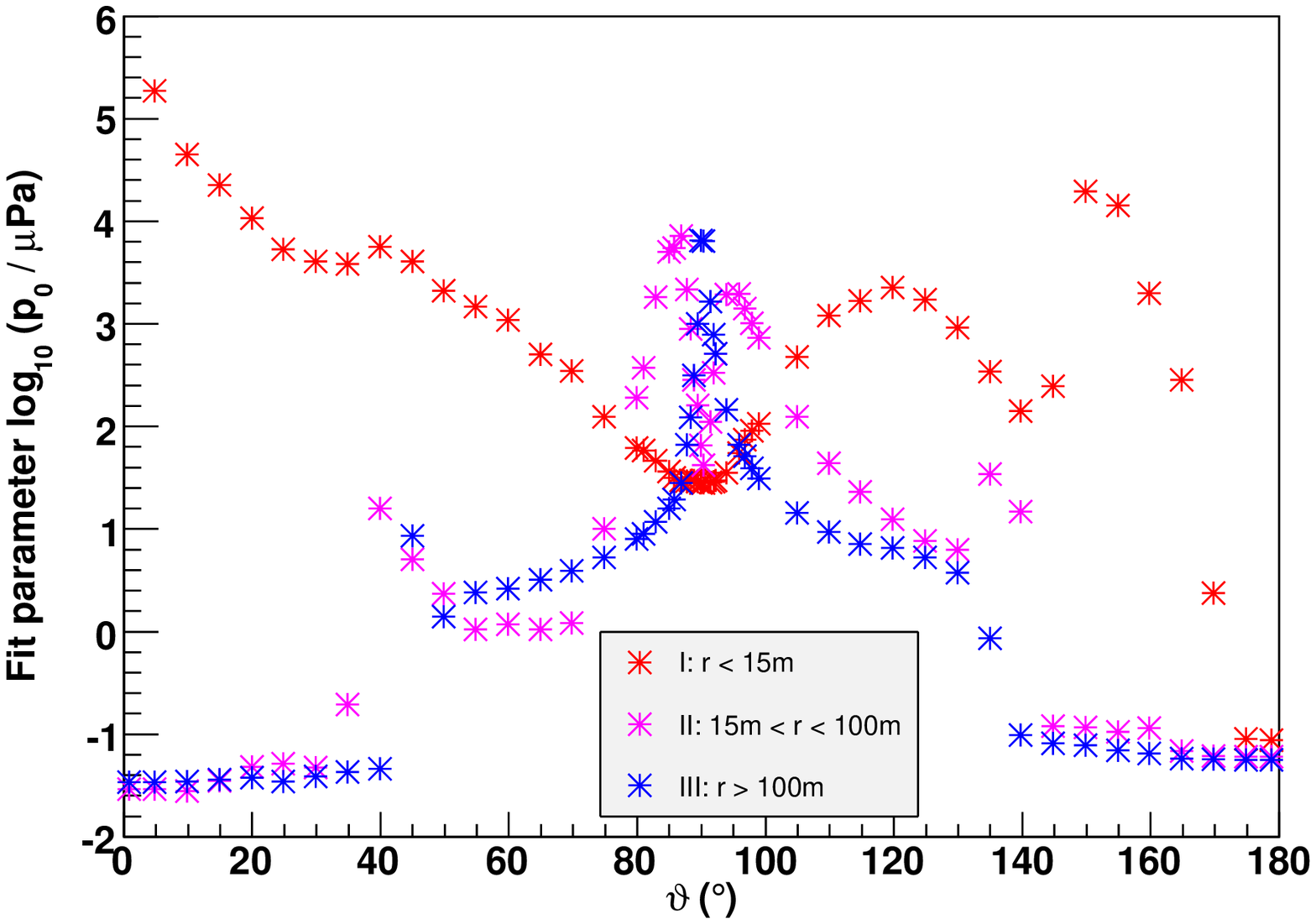}}
  \caption[Parameter set to describe the pressure field]{Parameter set
    to describe the pressure field. The values are obtained from fits
    $p_\mathrm{max} (r) = p_0 \, (r / 1 \, \mathrm{m})^a$ at fixed
    angle to the data shown in extracts in Fig.~\ref{fig:pressure}.}
  \label{fig:pressure_fits}
\end{figure}

We obtain the parameters $p_{0,i} (\vartheta)$ and $a_i (\vartheta)$
from linear fits in the double logarithmic plot for the three regions
discussed.  The fit results are presented in
Fig.~\ref{fig:pressure_fits}.

In the far field (region III, blue symbols), where the energy
distribution can be assumed as point-like, the amplitude drops like $1
/ r$ ($a = -1$) for nearly all angles, whereas the scaling factor
$p_0$ decreases strongly in the forward and backward directions where
only very small signals are expected for a line-like source. In a
narrow window around $\vartheta = 90^\circ$ the signal drops more
steeply, but $p_0$ is at least three orders of magnitude higher
compared to other directions.  Thus the main signal contribution
propagates perpendicular to the shower axis, but the steeper decrease
around $\vartheta = 90^\circ$ leads to a widening of the sonic disc
for large distances.

It is noteworthy, that for intermediate distances (region II, magenta
symbols) the amplitude drops like $1 / r$ expect for the narrow window
around $\vartheta = 90^\circ$ where it decreases like $1 / \sqrt{r}$
($a = -1 / 2$) as expected for a line source. This leads to the very
pronounced disc-shape of the sonic field.

In the forward and backward directions we expect no signal from the
disc-shape model. In the near field (region I, red symbols) we get
high $p_0$ in the forward direction, but the signal decreases like
$r^{-7}$ which leads to a strong suppression of the signal. In the
backward direction the signal decreases only like $1 / r$, but $p_0$ is
suppressed by at least three orders of magnitude compared to all other
directions, so that also in the backward direction no significant
signal contribution is expected.

Despite of apparent fluctuations in the fit parameters, the variation
with the angle $\vartheta$ is smooth around $\vartheta = 90^\circ$
where the signal is highest, which leads to reliable predictions of
the signal amplitude.

\begin{figure}[ht]
  \centering
  \includegraphics[width=0.9\textwidth]{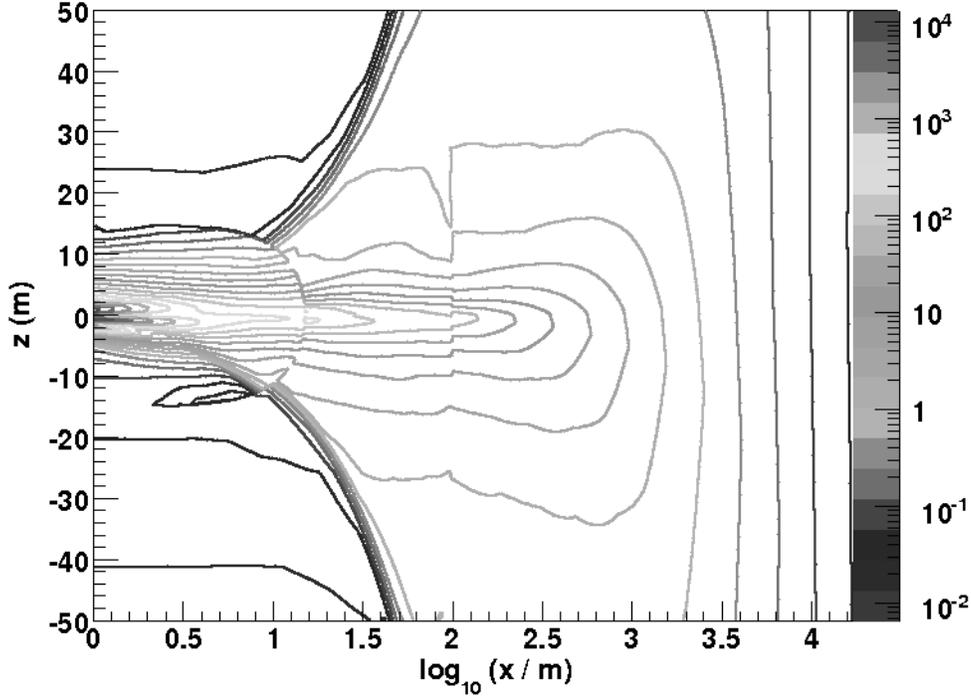}
  \caption[Parameterisation of the pressure
  amplitude]{Parameterisation of the pressure amplitude
    $p_\mathrm{max}$ produced by a hadronic cascade. The shower is
    centred at the origin and develops in positive $z$ direction. The
    colour scale gives the pressure in mPa / $10^9$\,GeV. The
    discontinuities at $r = 15$\,m and $r = 100$\,m result from the
    parameterisation method described in the text.}
  \label{fig:amplitude}
\end{figure}

We can now calculate $p_\mathrm{max}$ for every sensor position from
these 312 parameters\footnote{52 angles times three regions times two
  parameters ($p_0$ and $a$). The angular resolution of the
  lookup-table varies between 5$^\circ$ in the forward and backward
  directions and 0.5$^\circ$ perpendicular to the shower, where the
  main signal contribution is expected.} stored in a lookup-table. For
a given AM position we determine the distance $r$ to the shower centre
which sets the required region. Then, the angle $\vartheta$ is
determined, and the corresponding parameters $p_0$ and $a$ are
calculated by linear interpolation in the $\vartheta$ interval the
required angle is included. In a last step the absolute amplitude of
the signal is calculated from the energy of the cascade, since we
assume, as was discussed in Sec.~\ref{sec:hadronic_cascades}, that the
amplitude scales linearly with the cascade energy. The resulting
pressure field can be seen in Fig.~\ref{fig:amplitude}.

The plot shows lines of constant pressure produced by a hadronic
cascade centred at the origin and developing in the positive $z$
direction. The lobes perpendicular to the cascade axis, which form the
characteristic disc-shape since the plot is rotationally symmetric
around the $z$ axis, are clearly visible. Also the widening of the
disc in the far field which was discussed previously can be seen
nicely.

\medskip The second parameter of importance is the arrival time
$t_\mathrm{max}$ of the pressure peak at the acoustic module. It was
calculated together with $p_\mathrm{max}$ for all distances $r$ and
angles $\vartheta$, and it turns out that the spatial extension of the
energy deposition region is not important for the arrival times, which
have to be determined with a precision of 10\,$\mu \mathrm{s}$. The
time can simply be calculated from:

\begin{equation}
  \label{eq:parameterisation_t}
  t_\mathrm{max} (r, \vartheta) = \frac{r}{c}
\end{equation}

\noindent where $c$ is the speed of sound in water.


\chapter{Simulation study of an acoustic neutrino telescope}
\label{chap:detectors}
\minitoc

\bigskip In this chapter, the knowledge of acoustic pulses produced
by ultra high energy neutrinos gathered in the previous sections is
used to study the properties of an acoustic neutrino telescope. First,
the detector simulation is presented, then a reconstruction algorithm
for the direction and energy of the neutrino induced cascades is
discussed. In the last section the sensitivity of different detector
setups to an diffuse neutrino flux is derived.

\section{Detector simulation}

\begin{figure}[ht]
  \centering
  \includegraphics[width=0.9\textwidth]{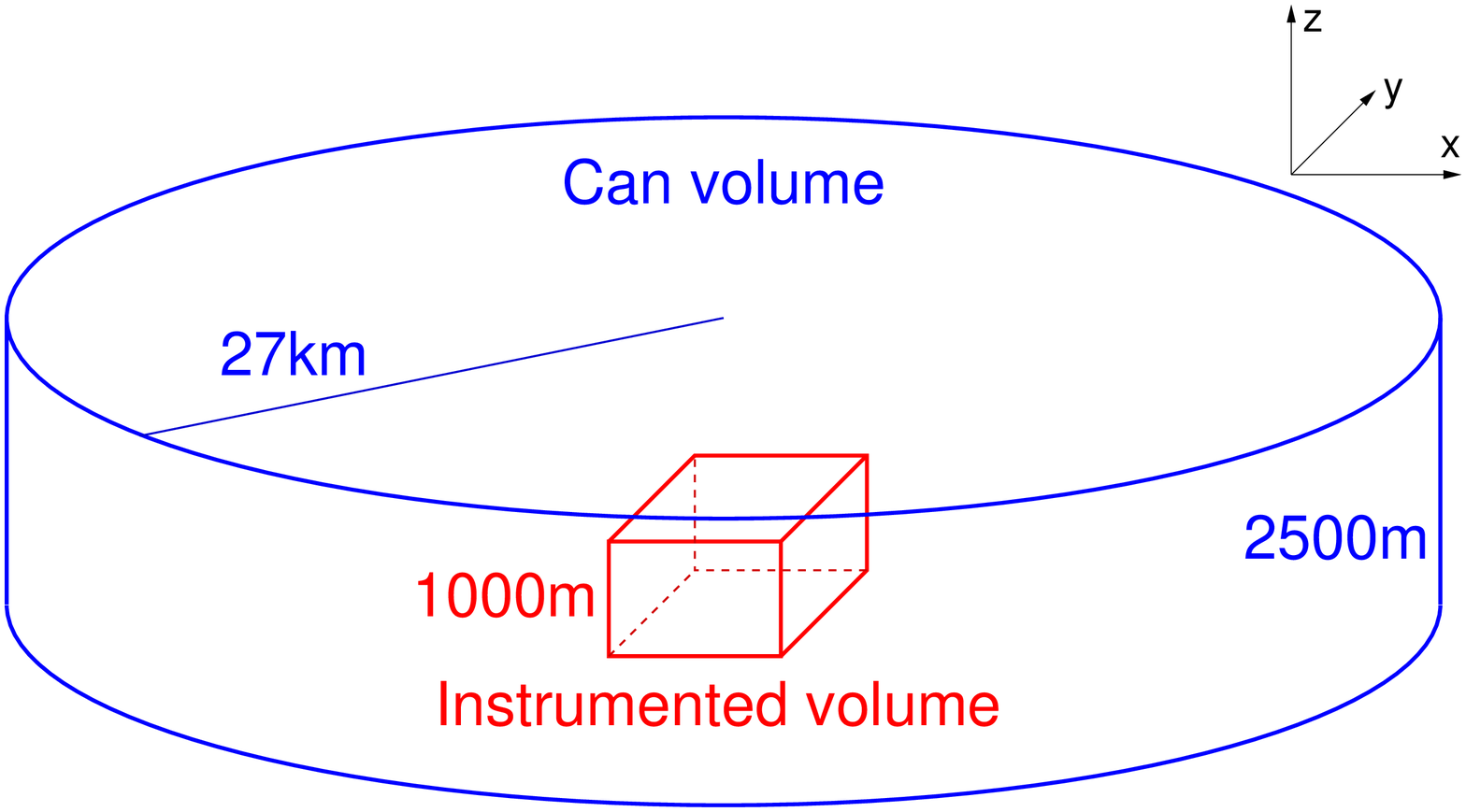}
  \caption[Schematic of the detector simulation setup]{Schematic of
    the detector simulation setup.}
  \label{fig:detector_can}
\end{figure}

Figure~\ref{fig:detector_can} shows a schematic view of the detector
simulation setup used for this study.

The simulation consists of a cylindrical can volume and a detection
volume instrumented with acoustic modules (AMs, cf.
Sec.~\ref{sec:noise}). In the first three sections of this chapter an
instrumented volume of $1 \times 1 \times 1 \, \mathrm{km}^3$ is
chosen. To avoid any bias on the simulation from the configuration of
the AMs inside the detection volume, we use 200 AMs distributed
homogeneously, but randomly over the instrumented volume. The
influence of the size of the instrumented volume and of the density of
the acoustic modules on the sensitivity of the neutrino telescope will
be discussed in Sec.~\ref{sec:sensitivity}.

An isotropic flux of downward going ultra high energy neutrinos
interacting inside the can volume is generated. The can has a height
of 2500\,m which is a typical depth in the Mediterranean Sea, and, for
the 1\,km$^3$ size detector, a radius of 27\,km since we assume that
no signal from neutrino interactions farther away can reach the
detector due to refraction in the water
(cf.~Sec.~\ref{sec:refraction}). The detection volume is horizontally
centred inside the can volume, 100\,m to 1100\,m above the sea floor.
Later, when larger instrumented volumes will be considered, only the
area of the instrumented volume will be increased at a constant height
of 1\,km, since we know from the water \v{C}erenkov telescopes being
built in the Mediterranean Sea that it is difficult from an
engineering point of view to deploy structures higher than about
1\,km.

A complete simulation code was developed during this work in C++,
which generates events in an acoustic neutrino telescope from the
following input:

\begin{itemize}
\item Detector description (A file containing the positions of the
  AMs; the AMs are assumed to have isotropic sensitivity, so no
  orientation of the AMs needs to be stored).
\item Height and radius of the can volume, and position of the
  detection volume relative to the can volume.
\item Number $N$ of neutrino interactions to be generated.
\item Neutrino energy range $[ E_\mathrm{min}, E_\mathrm{max} ]$.
\item Detection threshold $p_\mathrm{th}$ of the individual AMs.
\item Amplitude resolution $\sigma_p$, time resolution $\sigma_t$, and
  position resolution for the AMs $\sigma_\mathrm{AM}$ (see below).
\item Trigger threshold $N_t$ of the detector (see below).
\end{itemize}

\begin{figure}[ht]
  \centering
  \includegraphics[width=0.9\textwidth]{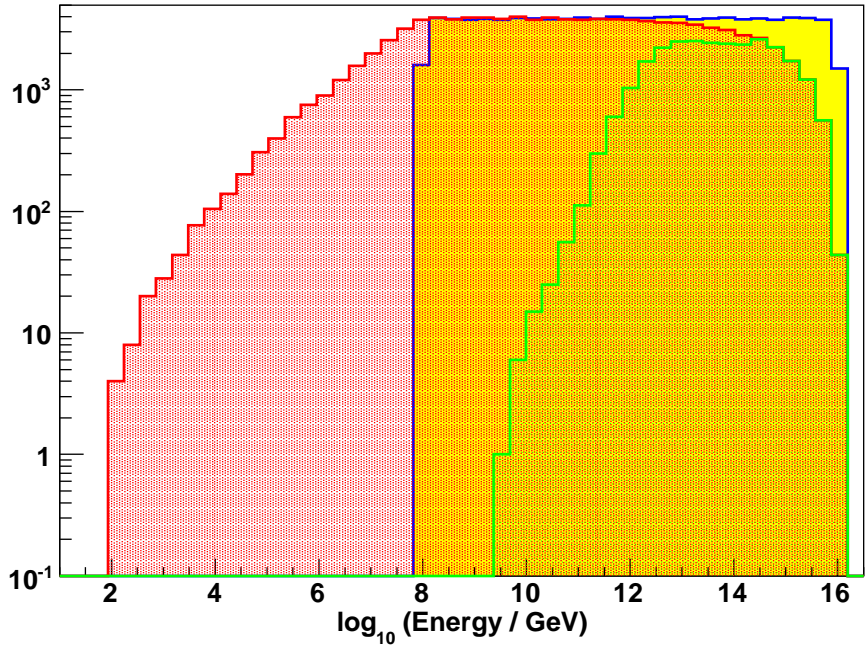}
  \caption[Simulated energy spectra]{Simulated energy spectra of the
    10$^5$ neutrinos injected into the can (yellow), the hadronic
    cascades produced (red), and the events fulfilling the trigger
    condition (green).}
  \label{fig:energy_nu_hadron}
\end{figure}

For each of the $N$ neutrino interactions the following steps are
performed during the simulation:

\begin{enumerate}
\item The energy $E_\nu$ of the neutrino is determined randomly in the
  predefined energy range $[ E_\mathrm{min}, E_\mathrm{max} ]$ from an
  energy distribution which is flat in $\log \, E_\nu$
  (cf.~Fig.~\ref{fig:energy_nu_hadron}):
  \begin{displaymath}
    \frac{\mathrm{d} N}{\mathrm{d} (\log E_\nu)} = \mathrm{const.}
    \quad \Leftrightarrow \quad \frac{\mathrm{d} N}{\mathrm{d} E_\nu}
    \propto E_\nu^{-1}
  \end{displaymath}
\item The position of the neutrino interaction vertex is chosen
  randomly such that the interactions are distributed homogeneously
  over the can volume.
\item The direction of the hadronic cascade, which is assumed to
  coincide with the direction of the neutrino and thus is down-going
  (cf.~Sec.~\ref{sec:nu_propagation}), is determined randomly to
  guarantee an isotropic flux of down-going neutrinos (i.e.~flat
  distribution of the azimuth $\varphi \in [0, 2 \pi [$, and flat
  distribution of the cosine of the zenith $\cos \, \theta \in [0,
  1]$).
\item The energy $E_\mathrm{had}$ of the hadronic cascade produced in
  the interaction is drawn randomly using $E_\nu$ and the distribution
  of the kinematic energy variable $y$ shown in
  Fig.~\ref{subfig:y_dist_a} which describes the energy transfer to
  the hadronic system.
\item For all acoustic modules in the detector description:
  \begin{enumerate}
  \item Each of the three position coordinates of the AM is smeared
    out using a Gaussian probability density function with a width
    given by the position resolution $\sigma_\mathrm{AM}$ of the
    detector. This accommodates the fact that for a submarine
    structure extending over several cubic kilometres the position of
    the individual components can only be fixed with a precision of
    approximately $\sigma_\mathrm{AM}$. We use $\sigma_\mathrm{AM} =
    10$\,cm for our simulation which is the value the ANTARES neutrino
    telescope will achieve for the positioning precision of its
    photomultiplier tubes.
  \item For the new AM position the amplitude $p_\mathrm{max}$ and
    arrival time $t_\mathrm{max}$ of the acoustic signal are
    calculated according to the parameterisation presented in the
    previous section.
  \item The amplitude and arrival time are smeared by the amplitude
    resolution $\sigma_p$ and the time resolution $\sigma_t$ of the
    acoustic module. We use $\sigma_p = 2$\,mPa and $\sigma_t = 10 \,
    \mathrm{\mu s}$ which corresponds to a sampling frequency of
    100\,kHz in the AM.
  \item If the smeared amplitude is larger than the detection
    threshold $p_\mathrm{th}$ of the AM, amplitude and arrival time
    are stored together with the ID of the acoustic module as a {\em
      hit}.
  \end{enumerate}
\item If the number of hits is larger than or equal to the trigger
  threshold $N_t$ of the detector the set of hits is written to disc
  as an {\em event}.
\end{enumerate}

Figure~\ref{fig:energy_nu_hadron} shows the energy spectra of the
neutrinos (yellow), the hadronic cascades (red), and the triggered
events (green) for the 1\,km$^3$ detector with 200 AMs, and
$p_\mathrm{th} = 5$\,mPa and $N_t = 5$ which is the minimum number of
hits to reconstruct the position and absolute time of the neutrino
interaction vertex.

We generated 10$^5$ neutrinos with energies between 10$^8$\,GeV and
10$^{16}$\,GeV. The energy spectrum of the hadronic cascades produced
in the interactions is shifted to lower energies since a significant
part of the energy is taken away by the lepton. 27\% of all showers
fulfilled the condition to produce a pressure amplitude $> 5$\,mPa in
$\ge 5$ AMs. It can be seen easily, that the lower energy threshold of
an acoustic neutrino telescope will be around 10$^9$\,GeV.

In the next section we will present an algorithm to reconstruct the
cascade direction and energy from a set of hits.

\section{Event reconstruction and selection cuts}

A two step reconstruction algorithm to determine the cascade direction
and energy was developed for this thesis. In the first step, the
position $\vec{r}_0$ and absolute time $t_0$ of the interaction vertex
are estimated, in a second step the direction $(\theta, \varphi)$ and
the energy $E_\mathrm{had}$ of the shower are fitted.

For the determination of $\vec{r}_0$ and $t_0$ we assume that the
cascade is an isotropic point source, and minimise the sum $\Delta_t$
of the residuals of the arrival times. If $N_h$ is the number of hits
in an event, $t_i$ is the arrival time measured at AM $i$, and
$\vec{r}_i$ is the position of this acoustic module, then

\begin{equation}
  \label{eq:delta_pos}
  \Delta_t (\vec{r}_0, t_0) = \sum_{i = 1}^{N_h} \, \frac{\left( \vert
    \vec{r}_i - \vec{r}_0 \vert / c - (t_i - t_0)
  \right)^2}{\sigma_t^2}
\end{equation}

\noindent where $c$ is the sound velocity in water. The function
$\Delta_t (\vec{r}_0, t_0)$ is not a $\chi^2$ function as known from
statistics because the errors are systematically underestimated. To
come to a real $\chi^2$ fit one would also have to include the error
introduced by the positioning uncertainty $\sigma_\mathrm{AM}$ into
the measurement of the arrival times.

For the minimisation we use the C++ implementation of the
Minuit\footnote{http://wwwasdoc.web.cern.ch/wwwasdoc/minuit/minmain.html}
minimisation package that comes with the
ROOT\footnote{http://root.cern.ch/} data analysis framework. As a
starting point for the minimisation we initialise the variable
$\vec{r}_0$ with the centre of gravity of all AMs that have been hit
and $t_0$ with the time of the first hit.

After the position of the vertex has been found we reconstruct the
direction and energy of the cascade by minimising the sum $\Delta_p$
of the residuals between the measured amplitudes $p_i$ and the
amplitudes $p_\mathrm{exp}$ expected for a given set of $(\theta,
\phi, E_\mathrm{had})$ at the position $\vec{r}_0$ which is now fixed.

\begin{equation}
  \label{eq:delta_dir}
  \Delta_p (\theta, \phi, \ln \, E_\mathrm{had}) = \ln \, \left( \ln
    \, \sum_{i = 1}^{N_h} \, \frac{\left( p_i - p_\mathrm{exp}
        (\vec{r}_i, \vec{r}_0, \theta, \phi, E_\mathrm{had})
      \right)^2}{\sigma_p^2} \right)
\end{equation}

The expected amplitude $p_\mathrm{exp}$ is calculated by using the
parameterisation of the pressure field presented in
Sec.~\ref{sec:signal_parameterisation}. Taking the logarithm of the
sum twice is necessary for numerical reasons, since the values for
$p_\mathrm{exp}$ fluctuate by several orders of magnitude depending
whether the considered AM is inside or outside the sonic disc when
varying the direction, and thus the sum over the residuals becomes
very large. The minimisation is again carried out by using Minuit,
whereby $\ln E_\mathrm{had}$ is used instead of $E_\mathrm{had}$ as
variable in the minimisation algorithm.  The latter one would produce
too large numerical uncertainties since it can vary over at least ten
orders of magnitude.

We initialise the direction variables to start the minimisation with
the normal of a plane through the three AMs that measured the highest
amplitude and do not lie on a straight line, since they are expected
to lie inside the sonic disc. The energy variable is initialised by a
simple energy estimator assuming a $1 / \sqrt{r}$ decrease of the
amplitude in all directions from the fixed position $\vec{r}_0$.

\begin{figure}[p]
  \centering
  \includegraphics[width=\textwidth]{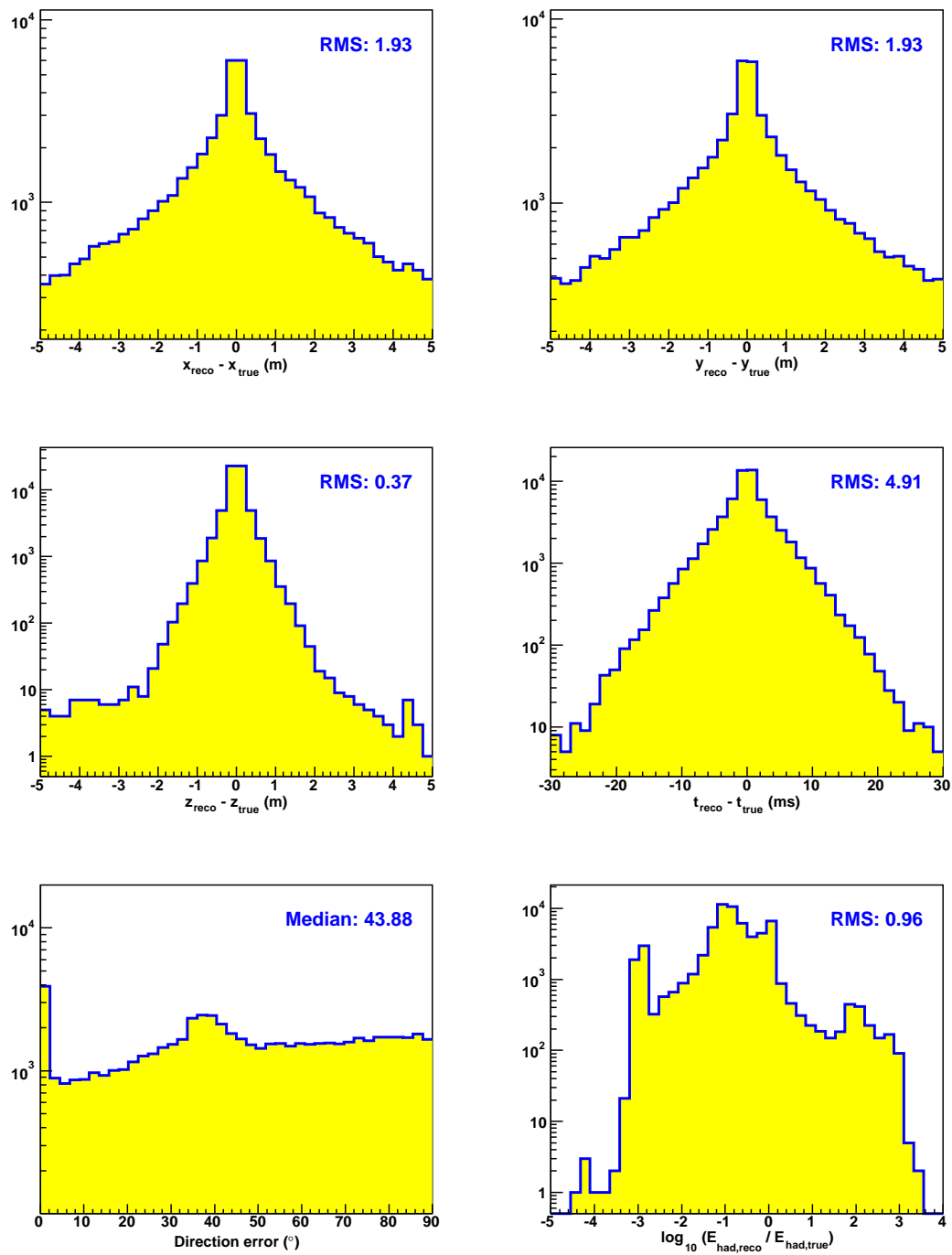}
  \caption[Reconstruction errors for hadronic cascades]{Reconstruction
    errors for hadronic cascades.}
  \label{fig:reco_errors_wo_cuts}
\end{figure}

For 78\% of all events which have produced a hit with an amplitude $>
5$\,mPa in at least 5 AMs (21\% of all hadronic cascades) the
reconstruction algorithm finds a minimum. The errors of the
reconstructed values compared to the Monte Carlo truth are shown in
Fig.~\ref{fig:reco_errors_wo_cuts}.

The reconstruction of the position and the time (upper two rows in
Fig.~\ref{fig:reco_errors_wo_cuts}) are very accurate. This is a
prerequisite for the determination of the direction and the energy for
which this information is required.  In the horizontal plane, where
the distance from the vertex to the detector extends up to 27\,km, the
vertex position can be reconstructed with an accuracy of 1.93\,m. In
the vertical direction, where the distances to the detector are much
smaller -- the can height is only 2.5\,km -- the position can be fixed
much more precisely, up to 37\,cm.

The direction error is the angle between the true direction of the
hadronic cascade and the reconstructed direction. Due to the
pancake-like shape of the pressure field, there is a 180$^\circ$
ambiguity in the reconstructed direction, i.e.~if we look at the
signature of some hadronic cascade in the detector, a hadronic cascade
at the same position, but developing in the exact opposite direction,
would produce the same hit pattern in the detector. Further, we do not
expect any neutrinos from below the horizon. So we define the
direction error $\Delta \alpha$ to be $180^\circ - \Delta \alpha$ for
$\Delta \alpha > 90^\circ$.

There are events for which the direction is reconstructed very well,
superimposed by a nearly flat distribution of events for which the
direction reconstruction seems to fail completely. The mean error in
$\log (E_\mathrm{had,reco} / E_\mathrm{had,true})$ is about -1,
i.e.~the algorithm tends to systematically underestimate the energy of
the hadronic system by a factor of 10, which can be corrected for
afterwards. The width of the distribution is 0.96. That means, that
the energy of the hadronic cascade can on average be determined up to
a factor of $10^{0.96} = 9.1$. In addition, there is a peak at -3,
which indicates that there are events for which the energy is
underestimated by a factor of 1000.

\begin{figure}[ht]
  \centering
  \subfigure[Direction reconstruction.]{
    \label{subfig:errors_theta_a}
    \includegraphics[width=0.48\textwidth]{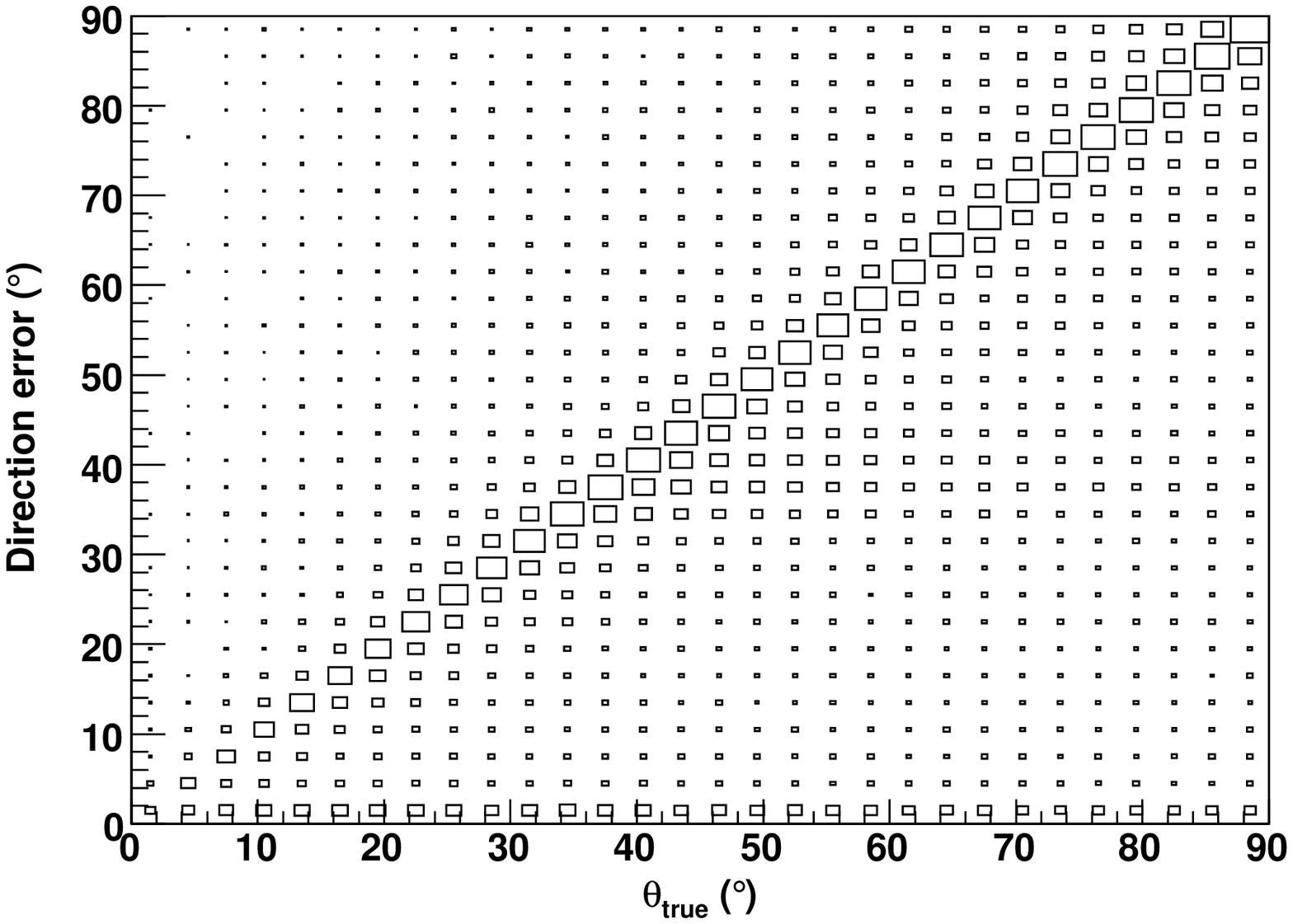}}
  \subfigure[Energy reconstruction.]{
    \label{subfig:errors_theta_b}
    \includegraphics[width=0.48\textwidth]{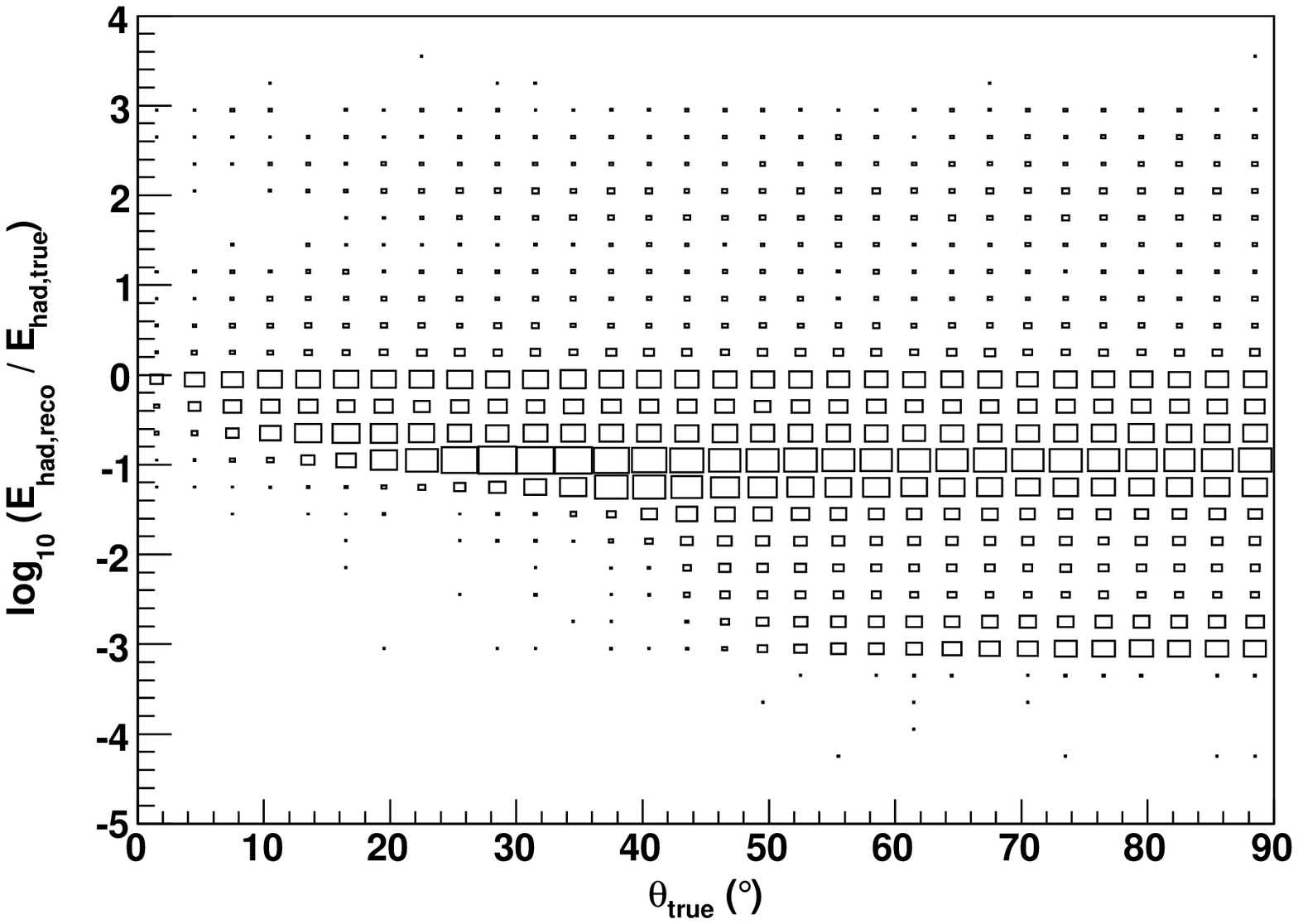}}
  \caption[Reconstruction errors as a function of the zenith
  angle]{Reconstruction errors as a function of the true zenith angle
    of the cascade.}
  \label{fig:errors_theta}
\end{figure}

Further analysis of the misreconstructed events revealed, that mostly
events produced from cascades with large zenith angles account for the
badly reconstructed events, which is not very surprising because for
an event with large $\theta_\mathrm{true}$ produced far from the
detector the core of the sonic disc will not transverse the detector
and hits are only produced from the remainder of the pressure field
which has only low direction characteristics.
Figure~\ref{fig:errors_theta} shows the errors in the direction and
energy reconstruction as a function of the zenith angle of the cascade
producing the event.

Obviously, the direction reconstruction becomes worse for the reasons
discussed above the more inclined the shower is. It is further
visible, that for cascades with zenith angles larger than about
$20^\circ$, where only small parts of the core of the sonic disc are
inside the detector, the reconstruction algorithm starts to
systematically underestimate the cascade energy. For very inclined
showers ($\theta_\mathrm{true} > 50^\circ$) the reconstruction
algorithm starts to underestimate the energy by several orders of
magnitude. Close inspection of those events reveals that they are
produced by cascades with energies above approximately $10^{13}$\,GeV,
where even the small pressure pulses emitted in the forward and
backward directions start to contribute to the measured hits.

\begin{figure}[p]
  \centering
  \includegraphics[width=\textwidth]{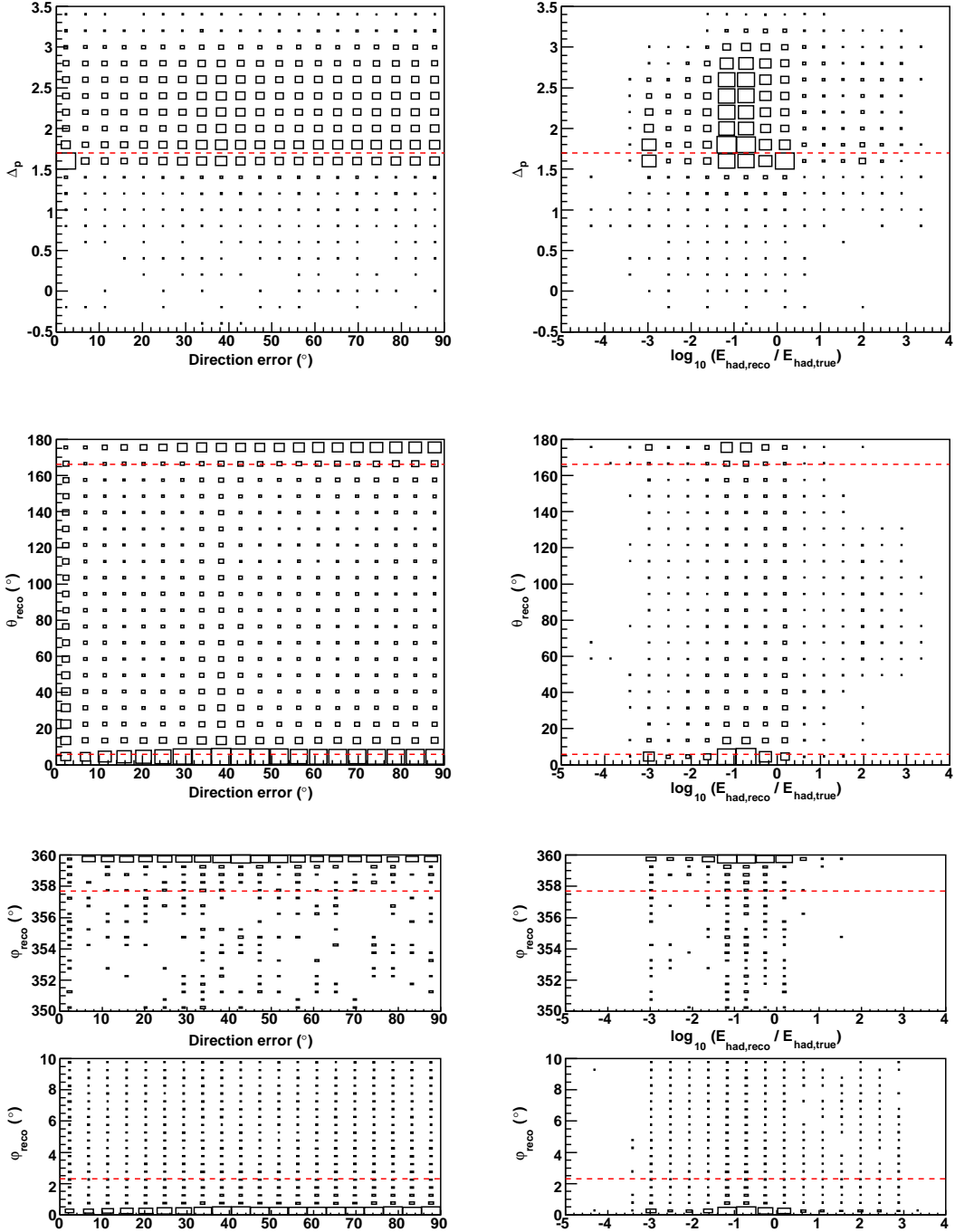}
  \caption[Reconstruction errors as a function of $\Delta_p$,
  $\theta_\mathrm{reco}$, and $\varphi_\mathrm{reco}$]{Errors in the
    direction and energy reconstruction as a function of the final
    value of the minimisation function $\Delta_p$, of the
    reconstructed zenith angle $\theta_\mathrm{reco}$, and of the
    reconstructed azimuth $\varphi_\mathrm{reco}$ (For the latter only
    the boundaries around 0 and 360$^\circ$ are shown for better
    visibility of the effects of the cut). The dashed lines indicate
    possible selection cuts.}
  \label{fig:delta_d_dist}
\end{figure}

\medskip The large uncertainty in the direction reconstruction poses
no problem when only a diffuse neutrino flux is measured, which is
expected for many of the cosmological models presented in
Chap.~\ref{chap:sources}. They predict a large number of similar
sources at cosmological distances, so that only a few neutrinos from
each individual source will be detected. Then, the distribution of the
incident neutrinos will be isotropic and the direction of individual
neutrinos is not of particular interest since we do not expect the
discovery of any point sources of neutrinos. In this case only the
energy resolution is of interest since one usually wants to measure
the neutrino flux as a function of energy. Nevertheless, we
investigated whether the resolution of an acoustic neutrino telescope
can be improved by using selection cuts\footnote{A selection cut is a
  criteria to decide from quantities which are {\em measured} or
  calculated from measured quantities only, whether a particular event
  was well reconstructed.}. It turns out that there are three
quantities which are suitable to cut on.
Figure~\ref{fig:delta_d_dist} shows in the left column the direction
error and in the right column the error in the energy reconstruction
as a function of the final value of the minimisation function
$\Delta_p$, of the reconstructed zenith angle $\theta_\mathrm{reco}$,
and of the reconstructed azimuth $\varphi_\mathrm{reco}$.

Badly reconstructed events have a tendency to have large values of
$\Delta_p$, which is related to the statistical $\chi^2$ of the
direction and energy fit. A good cut value for keeping well
reconstructed events is $\Delta_p < 1.7$, where we tried to find a
compromise between improving the resolution and reducing the gathered
set of events not too much.

Further, the reconstruction algorithm tends to drive the angles
towards the boundary values 0 and $180^\circ$ or $360^\circ$
respectively, so that badly reconstructed events have values of
$\theta_\mathrm{reco}$ and $\varphi_\mathrm{reco}$ near these
boundaries and can be conveniently cut. The only delicate cut is at
the lower bound for $\theta_\mathrm{reco}$, since we have seen that
down going cascades ($\theta_\mathrm{true} = 0$) contribute most to
the well reconstructed events. Convenient cut values are $5.7^\circ <
\theta_\mathrm{reco} < 166^\circ$, which does not eliminate too many
good events since the solid angle with $\theta < 5.7^\circ$ is very
small ($31 \cdot 10^{-3}$\,sr), and $2.3^\circ < \varphi_\mathrm{reco}
< 357.7^\circ$. The influence of these cuts on the reconstruction
accuracy is summarised in Tab.~\ref{tab:cuts}.

\begin{table}[hbt]
  \centering
  \caption[Influence of the selection cuts]{Influence of the different
    selection cuts on the number of events, the median of the
    direction error $\Delta \alpha_\mathrm{med}$, and the energy
    resolution.}
  \label{tab:cuts}
  \begin{tabular}{lrrrr}
    \hline
    Cut & \multicolumn{2}{c}{\# Events} & $\Delta \alpha_\mathrm{med}$
    & $\log \frac{E_\mathrm{reco}}{E_\mathrm{true}}$ \\ \hline
    none & 63049 & (100\%) & 43.88$^\circ$ & 0.96 \\
    $\Delta_p$ & 13585 & (22\%) & 37.12$^\circ$ & 1.13 \\
    $\theta_\mathrm{reco}$ & 29952 & (48\%) & 37.12$^\circ$ & 1.09 \\
    $\varphi_\mathrm{reco}$ & 40257 & (64\%) & 39.38$^\circ$ & 0.94 \\
    all & 5982 & (9.5\%) & 12.38$^\circ$ & 1.04 \\ \hline
  \end{tabular}
\end{table}

\begin{figure}[p]
  \centering
  \includegraphics[width=\textwidth]{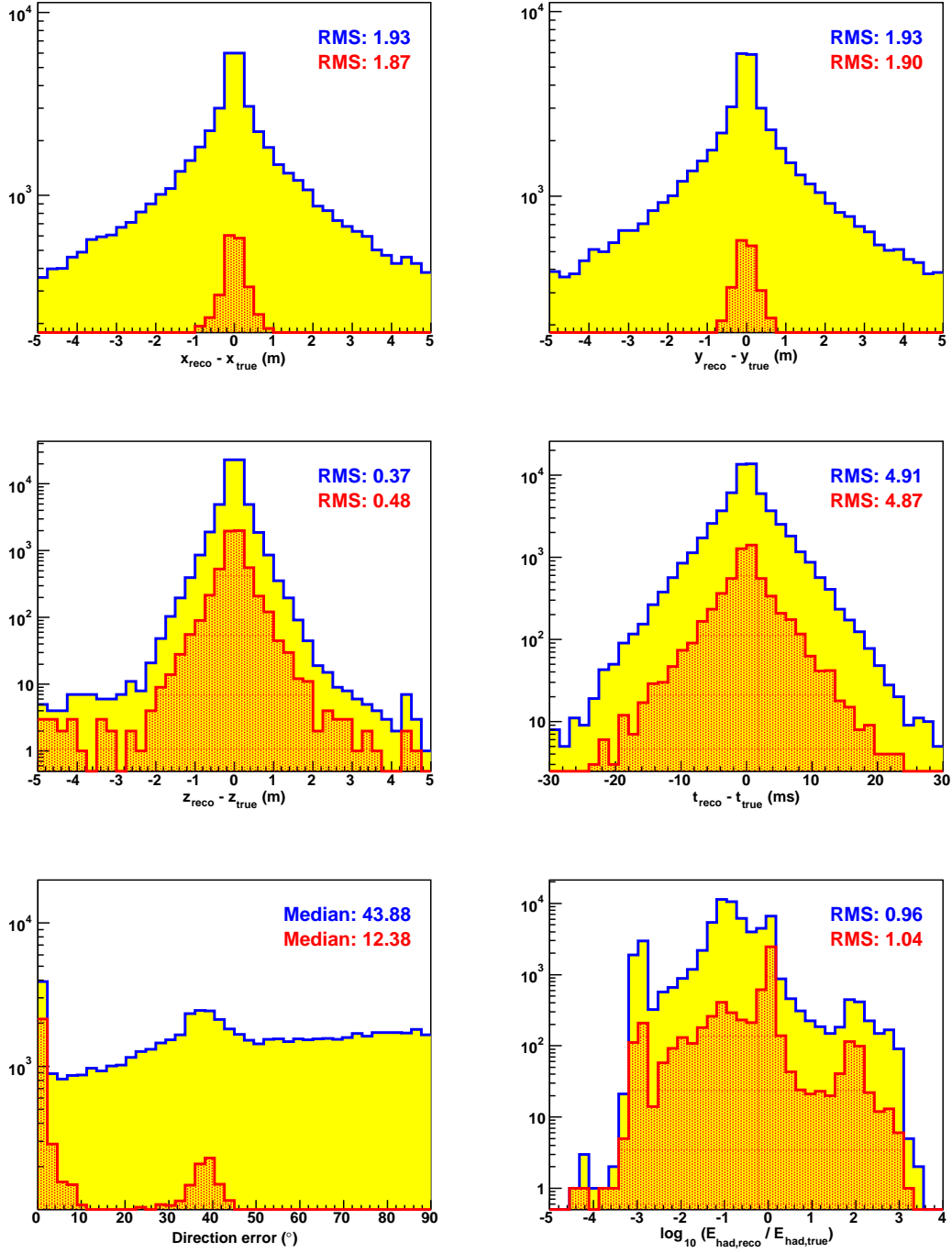}
  \caption[Reconstruction errors for hadronic cascades after selection
  cuts]{Reconstruction errors for hadronic cascades before (yellow)
    and after (red) the selection cuts are applied.}
  \label{fig:reco_errors}
\end{figure}

\begin{figure}[ht]
  \centering
  \includegraphics[width=0.9\textwidth]{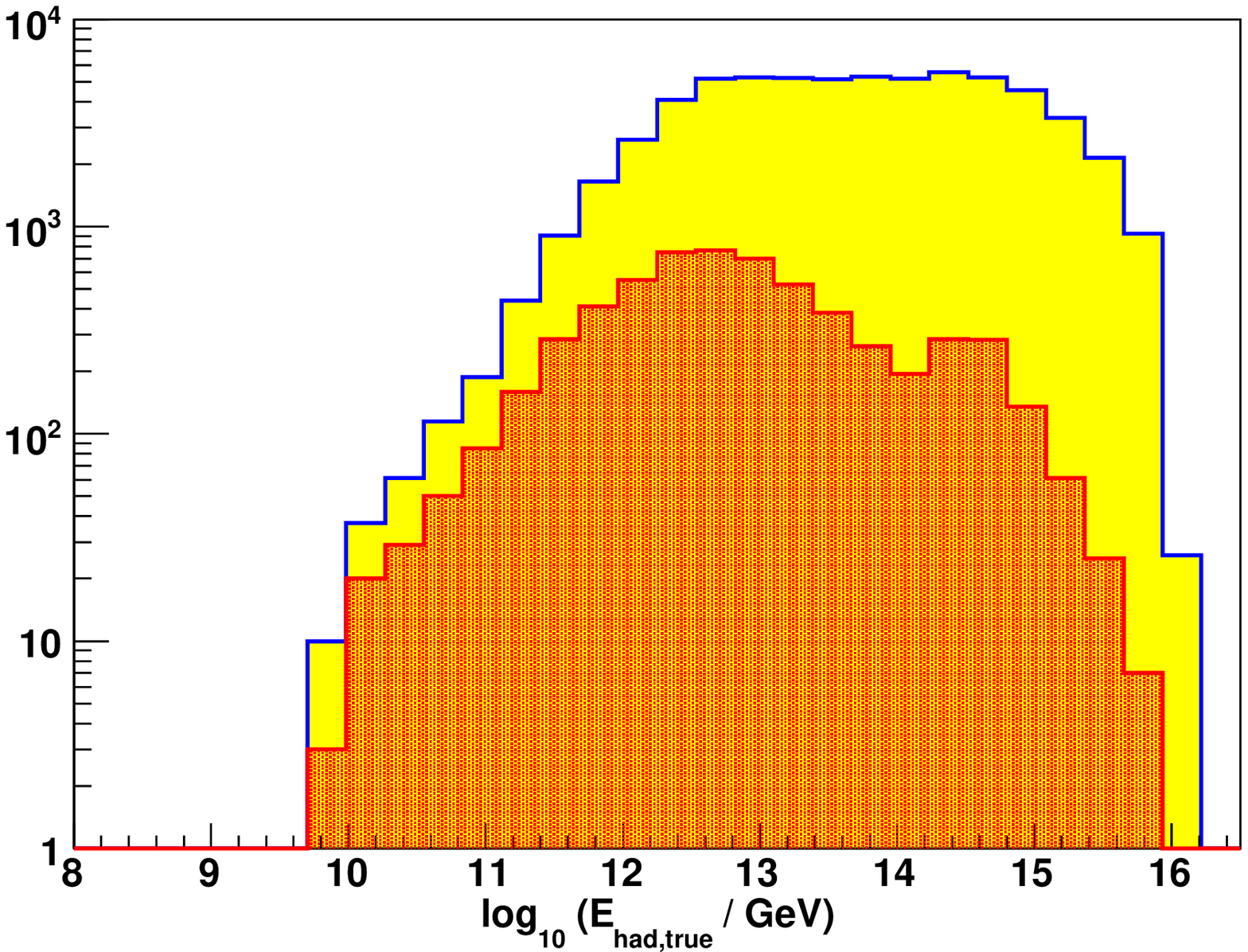}
  \caption[Energy spectrum of reconstructed events]{Energy spectrum of
    the events for which the reconstruction algorithm converged
    (yellow) and for the events which passed the selection cuts (red).}
  \label{fig:energy_recos}
\end{figure}

Obviously, only all three cuts together lead to a significant
improvement of the direction resolution, whereas the energy resolution
remains nearly unaffected. Figure~\ref{fig:reco_errors} shows the
distribution of the errors after the cuts are applied. The resulting energy
spectrum of the events can be seen in Fig.~\ref{fig:energy_recos}.

The selection cuts mostly discard events with highest energies above
$10^{13}$\,GeV for which the reconstruction fails for the reasons
elaborated before, namely that at these energies inclined showers
produced far from the detector emit pressure amplitudes in the forward
and backward directions sufficiently high to trigger the detector.

In summary we can say, that it is possible with an acoustic neutrino
telescope to reconstruct the energy of neutrino induced hadronic
cascade up to a factor of ten, which is sufficient to scrutinise
cosmological models which predict neutrino fluxes extending over
several orders of magnitude in energy. Since these hypothetical
sources are expected to be distributed isotropically over the sky at
cosmological distances, only a diffuse flux of neutrinos is expected
at Earth, so that the low direction resolution of the telescope is
acceptable. The direction resolution can be significantly improved up
to 12$^\circ$ in the lower energy range (10$^9$\,GeV to
$10^{13}$\,GeV) at the expense of a reduced statistics by using
selection cuts.

\section{Separation of background}

Another important topic which needs to be studied towards an acoustic
neutrino telescope is the existence of background sources, i.e.~of
sound sources which might wrongly be identified as neutrinos. It is
completely unknown whether there are any natural or anthropogenic
sources which emit bipolar acoustic pulses that can be detected
coherently over large distances. The Erlangen ANTARES group works
towards instrumenting several structures of the ANTARES neutrino
telescope with acoustic sensors \cite{Lahmann:2005} to measure the
rate with which such background events occur.

In this section we will discuss the consequences of the existence of
such hypothetical background sources for the development of an
acoustic neutrino telescope. We implemented point sources which emit
spherical waves into the simulation code. Their signal amplitude
decreases like $1 / r$, where $r$ is the distance from the source.
Absorption was neglected completely for the point source study. To
have some energy measure for the point sources for comparison with
hadronic cascades, we introduce an effective energy for the point
sources: A point source is defined to have energy $E$ if it produces,
in a distance of 400\,m from the source, a pressure amplitude equal to
the amplitude produced by a hadronic cascade of energy $E$ at this
distance perpendicular to the cascade axis.

We simulated $3 \cdot 10^5$ point sources with the same energy
spectrum used for the neutrino-like events
(Fig.~\ref{fig:energy_nu_hadron}). From these $3 \cdot 10^5$ point
sources 52\% (154965) fulfilled the trigger condition to produce an
amplitude $\ge 5$\,mPa in five or more acoustic modules (neutrino-like
events: 27\%). The reconstruction algorithm converged for 44\%
(132279) of all point sources generated (neutrino-like events: 21\%),
whereby the vertex position and energy of the point sources are
reconstructed with the same accuracy as for neutrino-like events, and
the direction reconstruction naturally fails completely. The selection
cuts presented in the previous section are survived by 1.6\% (4811) of
the point sources (neutrino-like events: 2.0\%).

\begin{figure}[ht]
  \centering
  \subfigure[$f_{10}$.]{
    \label{subfig:dist_Nhits10MaxAmplitude}
    \includegraphics[width=0.48\textwidth]{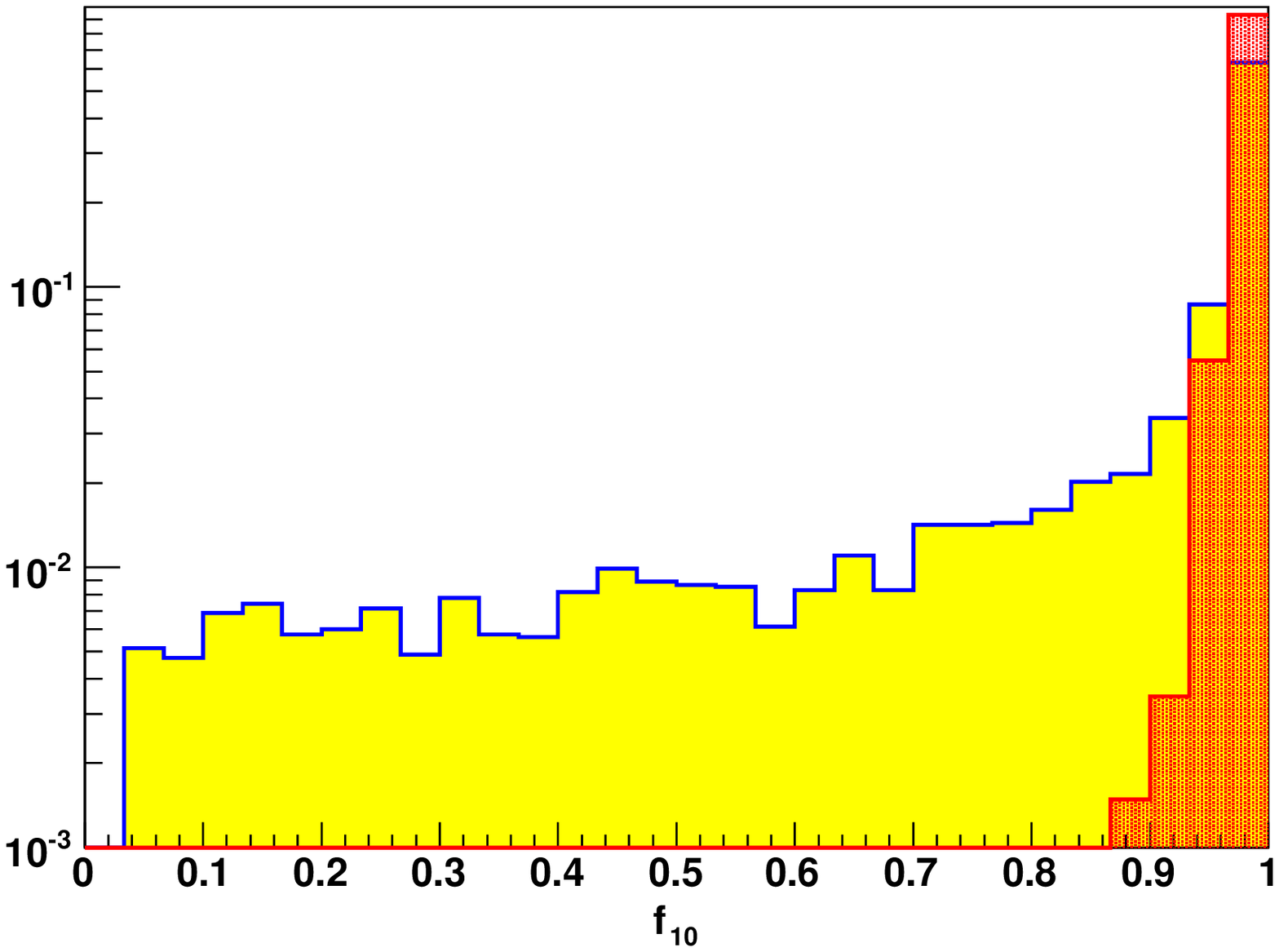}}

  \subfigure[$f_{50}$.]{
    \label{subfig:dist_Nhits50MaxAmplitude}
    \includegraphics[width=0.48\textwidth]{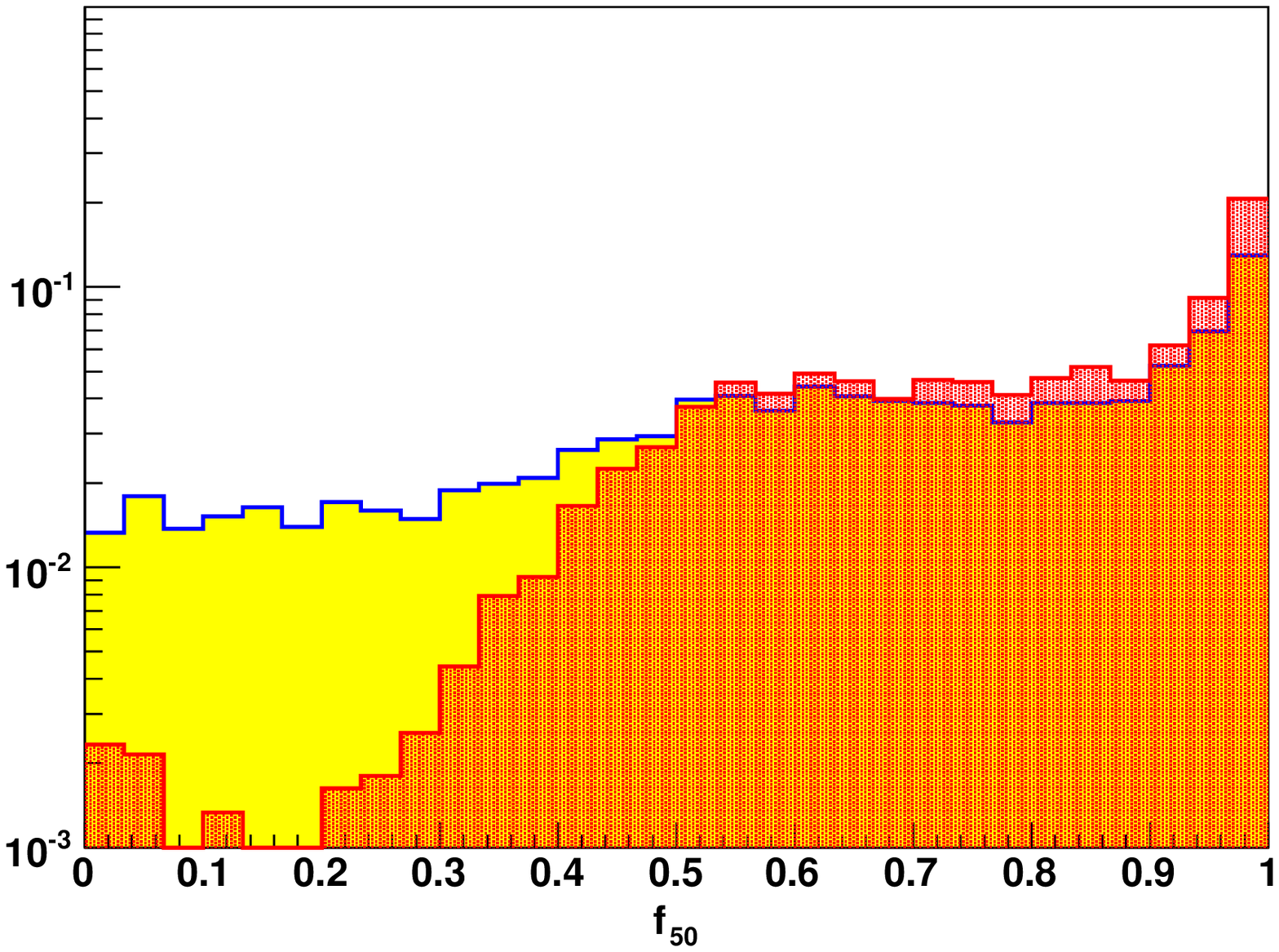}}
  \subfigure[$f_{90}$.]{
    \label{subfig:dist_Nhits90MaxAmplitude}
    \includegraphics[width=0.48\textwidth]{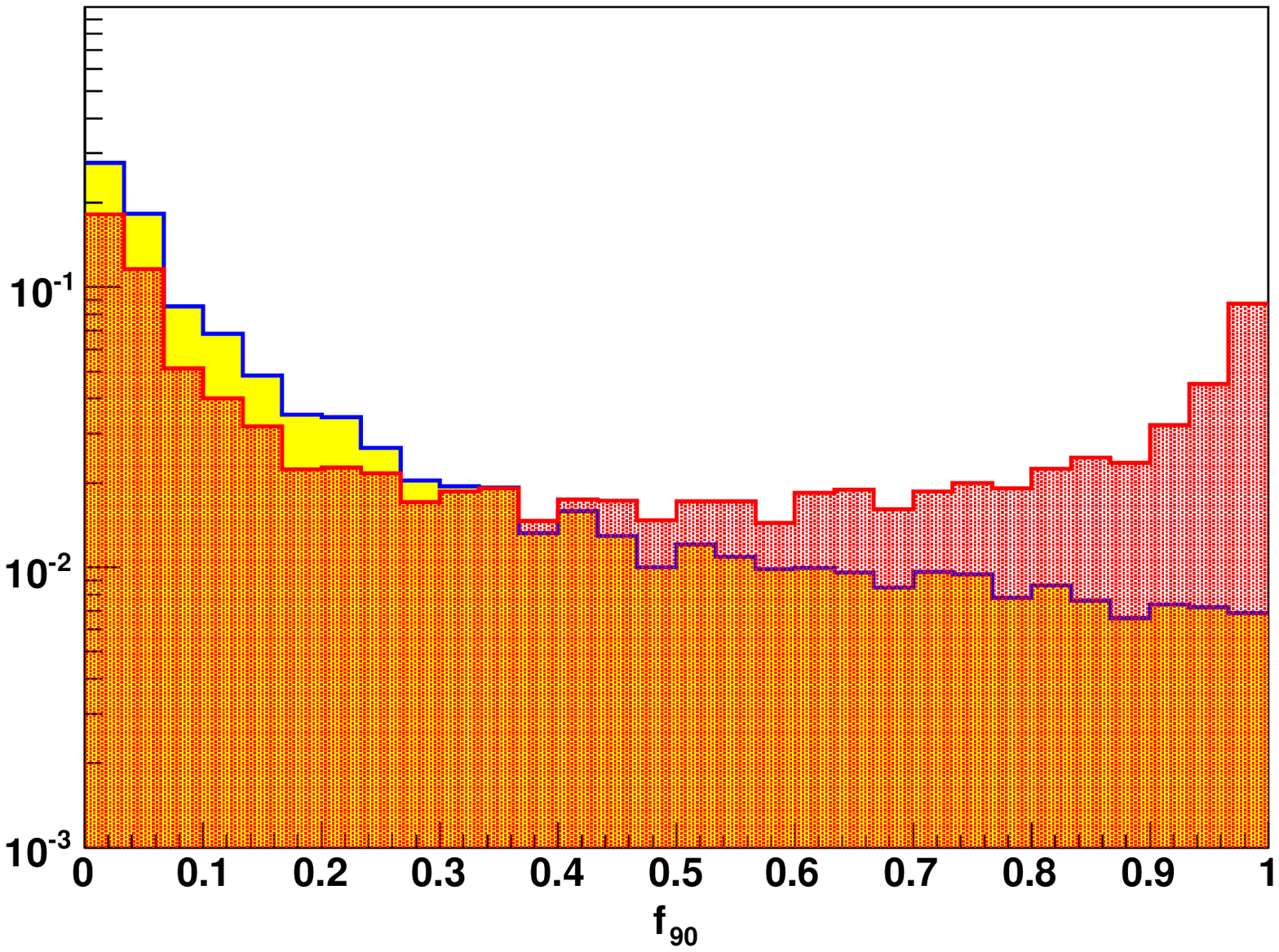}}
  \caption[Distribution of the variables $f_{10}$, $f_{50}$, and
  $f_{90}$]{Distribution of the variables $f_{10}$, $f_{50}$, and
    $f_{90}$ for neutrino-like events (yellow) and point sources (red)
    which pass the reconstruction algorithm, but without any selection
    cuts applied.}
  \label{fig:dist_NhitsXxMaxAmplitude}
\end{figure}

This means, that the presented reconstruction algorithm is not capable
to distinguish between point sources and the more disc-like neutrino
induced events. Nonetheless, the two event classes produce different
hit patterns in the detector which, if it should prove necessary,
could be used to distinguish between them. To see this, we define the
variable $f_n$ to be the fraction of hits in an event which have an
amplitude of $n / 100$ or more of the maximum amplitude measured in an
event. For neutrino-like events we expect only few high amplitude hits
around the sonic disc, whereas for point sources the wave front is
expected to propagate through the entire detector, producing similar
amplitudes in large parts of it. The distribution of $f_{10}$,
$f_{50}$, and $f_{90}$ for neutrino-like events and point sources is
shown in Fig.~\ref{fig:dist_NhitsXxMaxAmplitude}.

The distributions of $f_{10}$ and $f_{50}$ show, that for point
sources, in contrary to neutrino induced events, there are practically
no hits which have an amplitude lower than 10\% of the maximum
amplitude and only a few hits with an amplitude lower than 50\% of
the maximum amplitude. Further, neutrino-like events contain only a
few fits with an amplitude larger than 90\% of the maximum amplitude
measured in the event.

These distributions might allow to separate point sources from
neutrino like events if the rate of point sources is lower or of the
same order of magnitude as the rate of neutrino induced events. Thus,
it is crucial to measure the rate and properties of acoustic
background with long correlation length, as we plan to do within the
ANTARES experiment, before deducing the feasibility of building an
acoustic neutrino telescope.  For the remainder of this work we will
assume, that there is no such background.

\section{Sensitivity of an acoustic neutrino telescope}
\label{sec:sensitivity}

In this final section we will derive the sensitivity of an acoustic
neutrino telescope to a diffuse flux of ultra high energy neutrinos. A
general introduction to the calculation of flux limits is given in
Appendix~\ref{chap:limits}.

For the simulated acoustic neutrino telescope we define the effective
volume $V_\mathrm{eff}$ as

\begin{equation}
  \label{eq:v_eff}
  V_\mathrm{eff} = \frac{N_\mathrm{det}}{N_\mathrm{gen}} \, V_\mathrm{gen}
\end{equation}

\noindent where $N_\mathrm{det}$ is the number of detected neutrinos,
i.e. the number of cascades which produce a trigger and for which the
reconstruction algorithm converges, whereby the selection cuts can be
applied or not. $N_\mathrm{gen}$ is the number of cascades produced in
the generation volume $V_\mathrm{gen}$, which in our case is the can
volume. The aperture (\ref{eq:aperture}) is then given by

\begin{equation}
  \label{eq:aperture_veff}
  \mathcal{A} (E) = \int \mathrm{d} \Omega \, \sigma_\mathrm{tot} (E) N_t
  (V_\mathrm{eff}) P (E, \Omega)
\end{equation}

\noindent where $\sigma_\mathrm{tot}$ is the total neutrino nucleon
cross section (\ref{eq:cross_section}), $N_t$ is the number of target
nucleons inside the effective volume $V_\mathrm{eff}$, and $P (E,
\Omega)$ is the probability for a neutrino of energy $E$ and direction
$\Omega$ to propagate to the can level. As was laid out in
Sec.~\ref{sec:nu_propagation}, we assume that all neutrinos coming
from above the horizon can propagate unperturbed to the can level,
whereas all neutrinos from below are absorbed inside the Earth:

\begin{displaymath}
  P (E, \Omega) = \left\{
    \begin{array}{ll}
      1 & \mathrm{for} \quad \theta < 90^\circ \\
      0 & \mathrm{for} \quad \theta \ge 90^\circ
    \end{array}
  \right.
\end{displaymath}

\noindent and thus

\begin{equation}
  \mathcal{A} (E) = 2 \pi \sigma_\mathrm{tot} (E) N_t
  (V_\mathrm{eff})
\end{equation}

\begin{figure}[ht]
  \centering
  \includegraphics[width=0.9\textwidth]{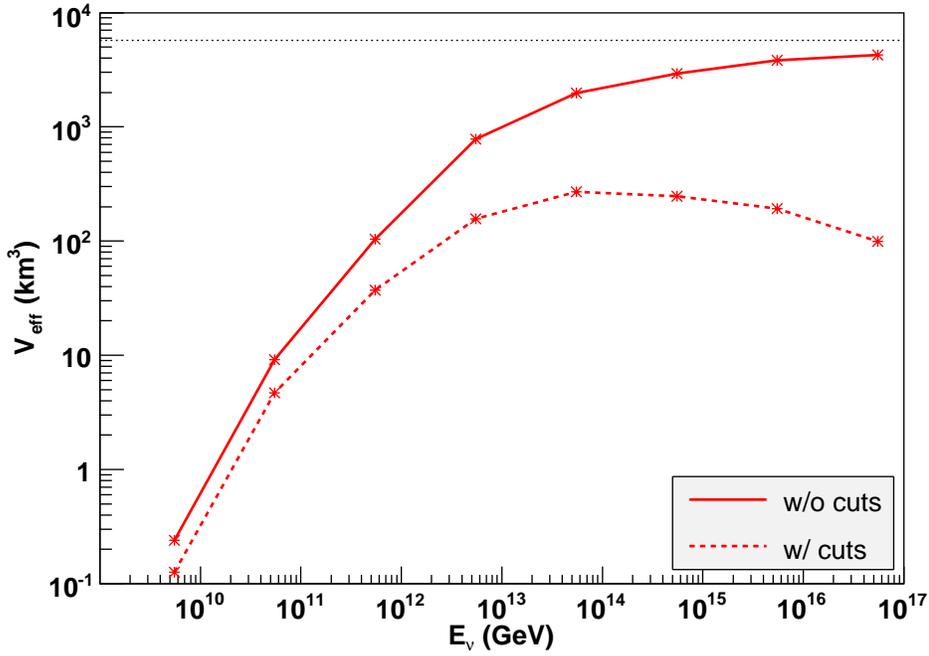}
  \caption[Effective volume as a function of energy]{Effective volume
    of the 1\,km$^3$ detector as a function of the neutrino energy
    without selection cuts, and with selection cuts applied. The
    dotted line indicates the can volume.}
  \label{fig:v_eff_rand200_0001km3_05mPa}
\end{figure}

The effective volume of the neutrino telescope presented in the
previous sections as a function of the neutrino energy is shown in
Fig.~\ref{fig:v_eff_rand200_0001km3_05mPa}.

Without the selection cuts applied the effective volume approaches the
can volume of 5726\,km$^3$ asymptotically, which is a natural bound
for the effective volume since we have shown that no signals from
beyond the can volume reach the detector. When we apply the selection
cuts, the effective volume starts to decrease again above
approximately $10^{14}$\,GeV due to the fact, that the selection cuts
mainly discard highest energy events.

\begin{figure}[ht]
  \centering
  \subfigure[Without selection cuts.]{
    \label{subfig:v_eff_vs_p_th_a}
    \includegraphics[width=0.48\textwidth]{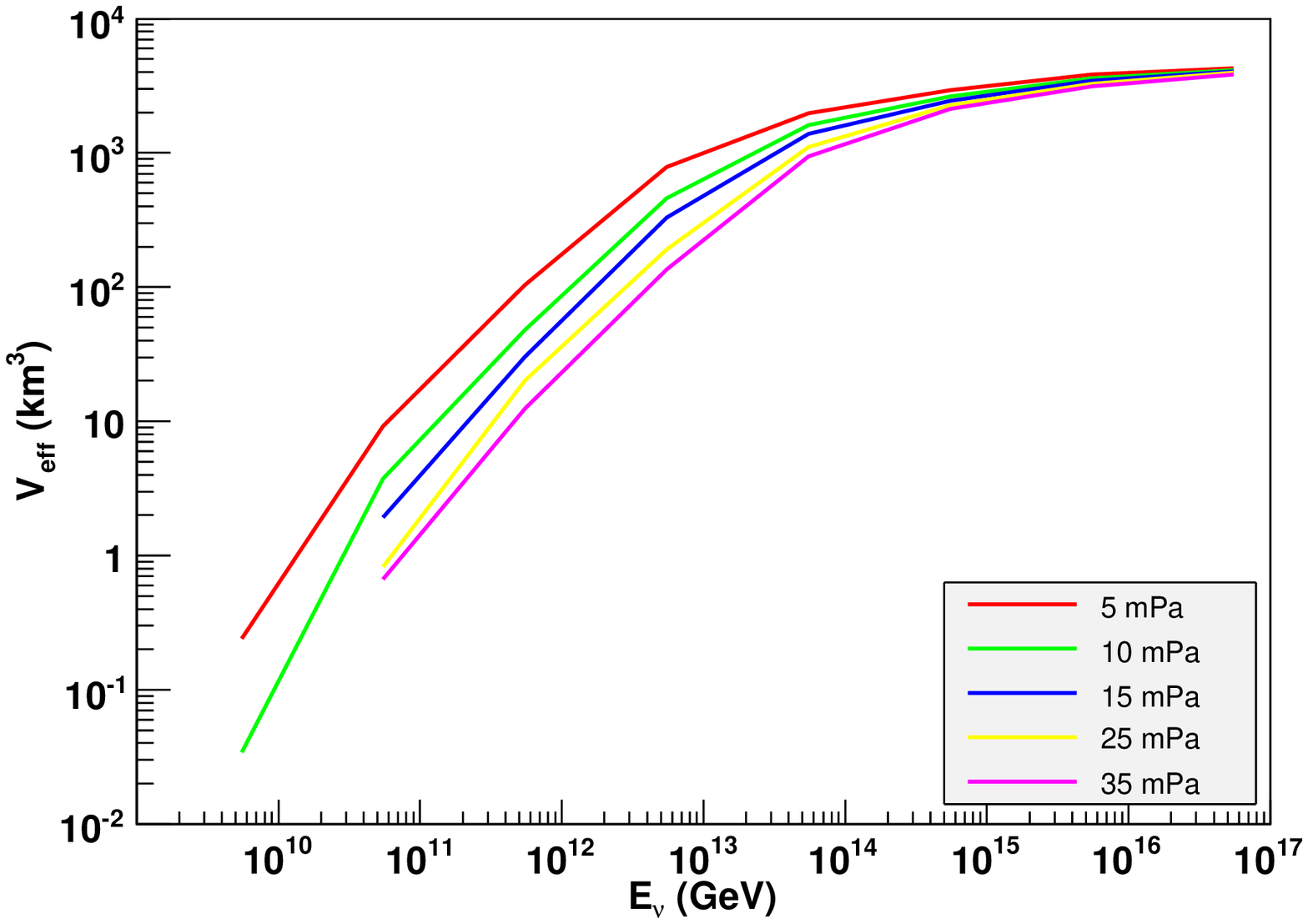}}
  \subfigure[With selection cuts applied.]{
    \label{subfig:v_eff_vs_p_th_b}
    \includegraphics[width=0.48\textwidth]{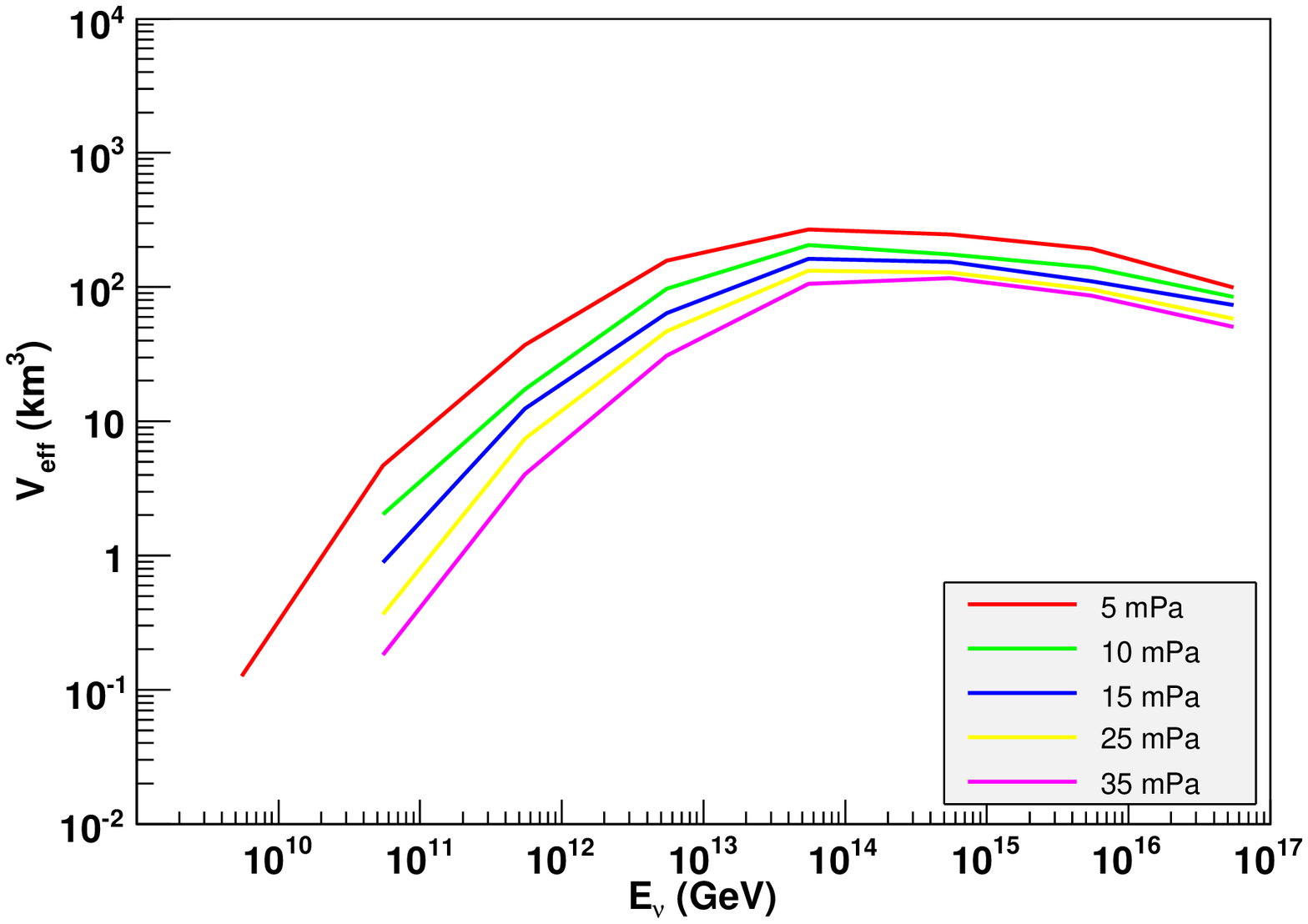}}
  \caption[Effective volume as a function of the detection threshold
  $p_\mathrm{th}$]{Effective volume as a function of the detection
    threshold $p_\mathrm{th}$. The detector consists of 200 AMs
    distributed over 1\,km$^3$.}
  \label{fig:v_eff_vs_p_th}
\end{figure}

\medskip We will now investigate the influence of the detector
parameters on the effective volume. Figure~\ref{fig:v_eff_vs_p_th}
shows the dependence of the effective volume on the detection
threshold $p_\mathrm{th}$ of the individual acoustic modules.

As we have seen, this detection threshold is set by the ambient
background noise in the sea. High-level background noise, which would
force an increase of the detection threshold, mainly affects the
sensitivity for low energies. Whereas for the highest energies the
effective volume is nearly independent of $p_\mathrm{th}$, it
decreases by about a factor of ten at $10^{10}$\,GeV when
$p_\mathrm{th}$ is increased from 5\,mPa to 35\,mPa. Thus, the
optimisation of filtering algorithms to achieve a $p_\mathrm{th}$ as
low as possible is crucial for an acoustic neutrino telescope.

\begin{figure}[ht]
  \centering
  \includegraphics[width=0.9\textwidth]{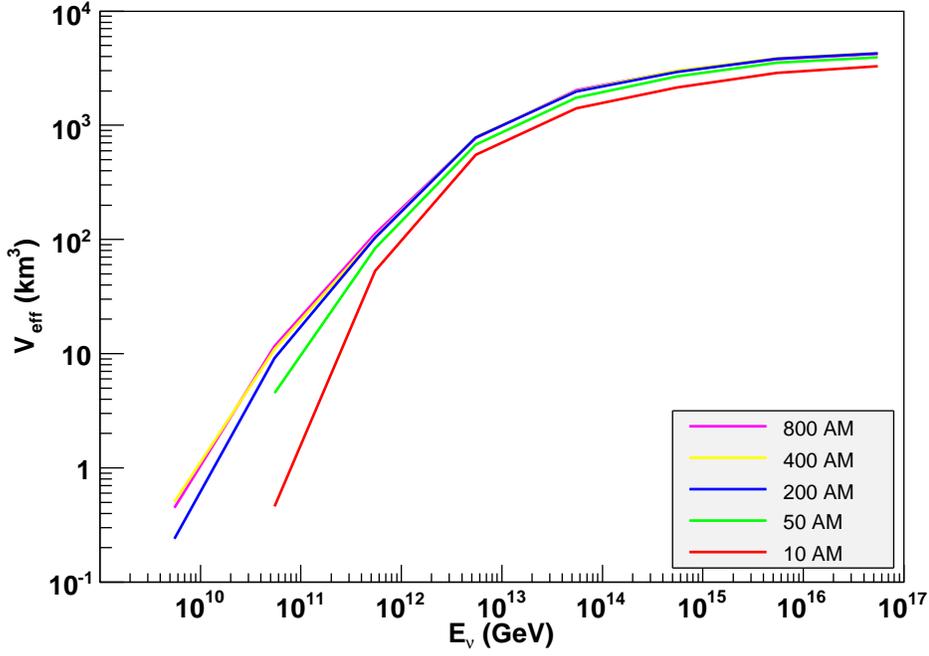}
  \caption[Effective volume as a function of the sensor
  density]{Effective volume as a function of the AM density. The
    instrumented volume is 1\,km$^3$ with $p_\mathrm{th} = 5$\,mPa for
    each detector.}
  \label{fig:v_eff_vs_am_density}
\end{figure}

The dependence of the effective volume on the density of acoustic
modules is shown in Fig.~\ref{fig:v_eff_vs_am_density}.

It can be seen that an increase of the instrumentation density above
$200 \, \mathrm{AM} / \mathrm{km}^3$ does not lead to a significant
increase in sensitivity. If one is only interested in neutrinos at
highest energies $E_\nu \gtrsim 10^{13}$\,GeV one can even build a
detector with as few as $10 \, \mathrm{AM} / \mathrm{km}^3$ utilising
the fact, that acoustic signals emitted in all directions from the
cascade can be detected at these energies. We cannot show the
dependence of the effective volume on the instrumentation density with
selection cuts applied, since the selection cuts are sensitive to the
detector geometry, and would have to be redeveloped for each detector.

At last we will investigate the dependence of $V_\mathrm{eff}$ on the
instrumented volume $V_\mathrm{inst}$. We assume that the instrumented
volume is set on the sea bed at a depth of 2500\,m and has cylindrical
shape with a height of 1000\,m, and a radius of $r_\mathrm{inst} =
\sqrt{V_\mathrm{inst} / 1 \, \mathrm{km} / \pi}$. Since, for a
constant instrumentation density, the computing time of the simulation
rises linearly with the instrumented volume, simulations for very
large volume detectors are impractical and we will estimate this
dependence from the simulations of the 1\,km$^3$ detector.

If the effective volume is smaller than the instrumented volume, we
can assume that the effective volume increases linearly with the
instrumented volume:

\begin{equation}
  \label{eq:v_eff_1}
  V_\mathrm{eff} = \frac{V_\mathrm{eff, 1 \, \mathrm{km}^3}}{1 \,
    \mathrm{km}^3} \, V_\mathrm{inst}
\end{equation}

\begin{figure}[ht]
  \centering
  \subfigure[Without selection cuts.]{
    \label{subfig:v_inst_vs_v_eff_a}
    \includegraphics[width=0.48\textwidth]{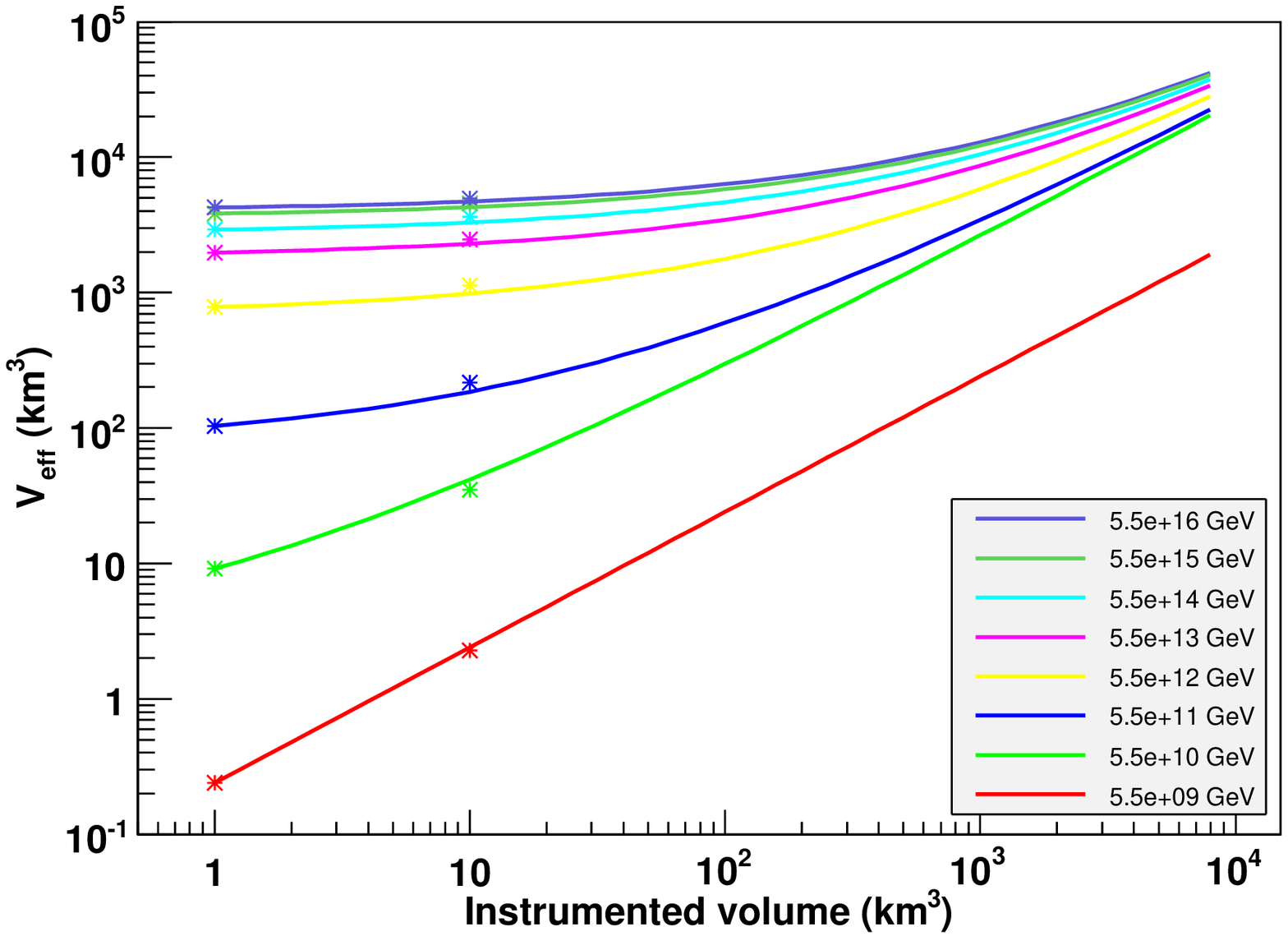}}
  \subfigure[With selection cuts applied.]{
    \label{subfig:v_inst_vs_v_eff_b}
    \includegraphics[width=0.48\textwidth]{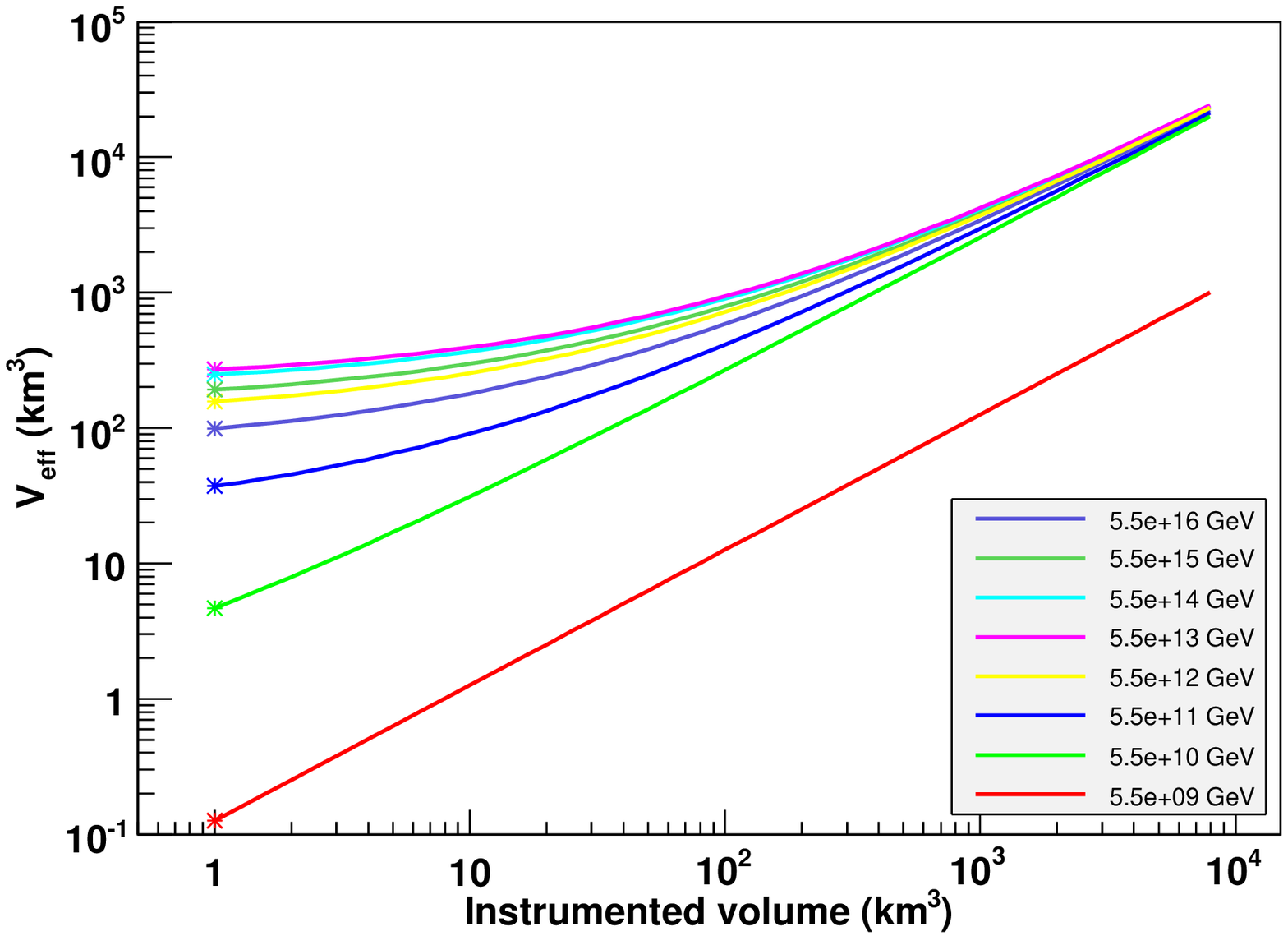}}
  \caption[Effective volume as a function of the instrumented
  volume]{Effective volume as a function of the instrumented volume
    for different neutrino energies. The solid lines are calculated
    from (\ref{eq:v_eff_1}) and (\ref{eq:v_eff_2}), where $b$ was
    determined from $V_\mathrm{inst}$ at 1\,km$^3$. The simulation
    results (data points) for $V_\mathrm{inst} = 10 \, \mathrm{km}^3$
    agree well with the model. The selection cuts cannot be applied to
    the events from larger detectors, so no simulation results for
    larger detectors are shown here.}
  \label{fig:v_inst_vs_v_eff}
\end{figure}

If the effective volume is larger than the instrumented volume, we
assume that the complete instrumented volume contributes to the
effective volume, and that it further extends up to the sea surface
and some distance $b$, which is independent of $V_\mathrm{inst}$,
horizontally beyond the instrumented volume:

\begin{equation}
  \label{eq:v_eff_2}
  V_\mathrm{eff} = \pi (r_\mathrm{inst} + b)^2 \cdot 2.5 \, \mathrm{km}
\end{equation}

\noindent where $b$ can be calculated from the simulation of the
1\,km$^3$ detector. The resulting dependence of the effective volume
on the instrumented volume is shown in Fig.~\ref{fig:v_inst_vs_v_eff}
for different neutrino energies.

At low energies the effective volume is proportional to the
instrumented volume, whereas at highest energies $V_\mathrm{eff}$
remains nearly constant over a wide range of $V_\mathrm{inst}$, and
starts to increase only when $V_\mathrm{inst}$ becomes larger than the
initial effective volume for small detectors.

\medskip This now allows us to estimate the sensitivity of an acoustic
neutrino telescope to the neutrino fluxes expected to be produced by
the different cosmological models presented in
Chap.~\ref{chap:sources}, and to compare the acoustic detection
technique to the other experiments discussed in
Chap.~\ref{chap:experiments}. We will restrict the discussion to a
detector with $200 \, \mathrm{AM} / \mathrm{km}^3$, which was shown
above to be sufficient, and a detection threshold of the acoustic
modules of $p_\mathrm{th} = 5$\,mPa.

\begin{figure}[ht]
  \centering
  \includegraphics[width=0.9\textwidth]
  {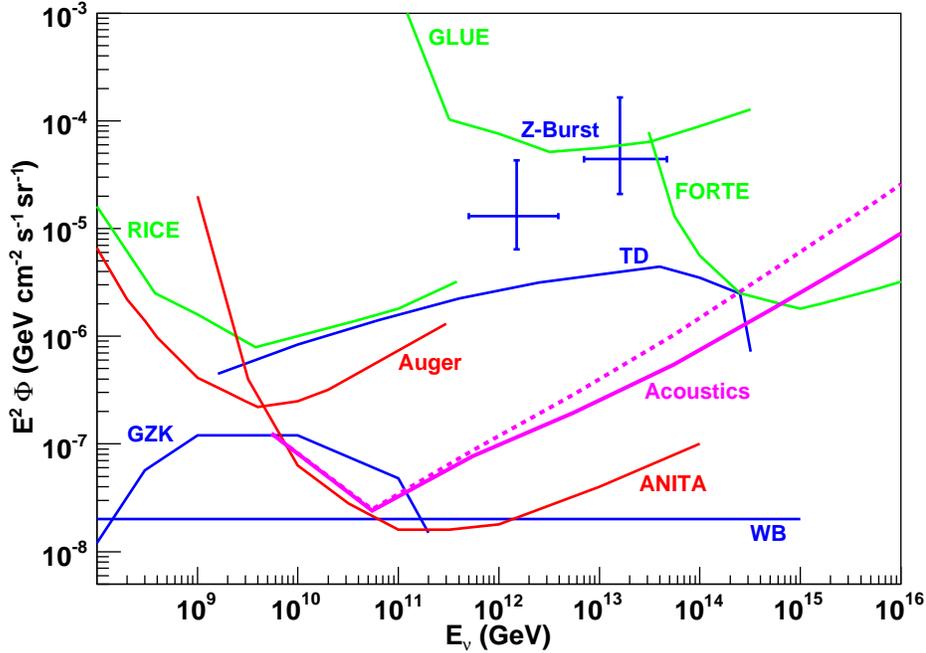}
  \caption[Model independent sensitivity of an acoustic neutrino
  telescope]{Model independent sensitivity of a 1500\,km$^3$ acoustic
    neutrino telescope after 5 years measurement time without
    selection cuts (solid line) and with selection cuts applied
    (dashed line) (RICE from \cite{Miocinovic:2005jh}, Auger from
    \cite{Bertou:2001vm}, GLUE from \cite{Gorham:2003da}, FORTE from
    \cite{Lehtinen:2003xv}, ANITA from \cite{Miocinovic:2005jh}, WB
    from \cite{Bahcall:1999yr}, GZK from \cite{Engel:2001hd}, Z-Bursts
    from \cite{Fodor:2001qy}, TD from \cite{Yoshida:1996ie}).}
  \label{fig:flux_limit_rand200_1500km3_05mPa_5y}
\end{figure}

First we will present a model independent flux limit at 90\%
confidence level (C.L.), calculated as described in
Appendix~\ref{chap:limits}, and compare it to the results shown in
Fig.~\ref{fig:models_vs_limits}. It turns out, that an instrumented
volume of greater than thousand cubic kilometres is required to obtain
competitive results.
Figure~\ref{fig:flux_limit_rand200_1500km3_05mPa_5y} shows the
resulting flux limit for an instrumented volume of 1500\,km$^3$
(diameter: 44\,km, height: 1\,km) after a measurement time of 5 years.

Such a detector would have a sensitivity comparable to the sensitivity of
the ANITA radio \v{C}erenkov detector. This would allow to test
theoretical source models with two experimental approaches which are
completely independent, and thus have completely different sources of
systematic errors. This is of utter importance when measuring in an
energy region where no calibration sources are available. A test of
all the source models shown as blue curves in
Fig.~\ref{fig:flux_limit_rand200_1500km3_05mPa_5y} would be
possible.

\begin{figure}[ht]
  \centering
  \includegraphics[width=0.9\textwidth]
  {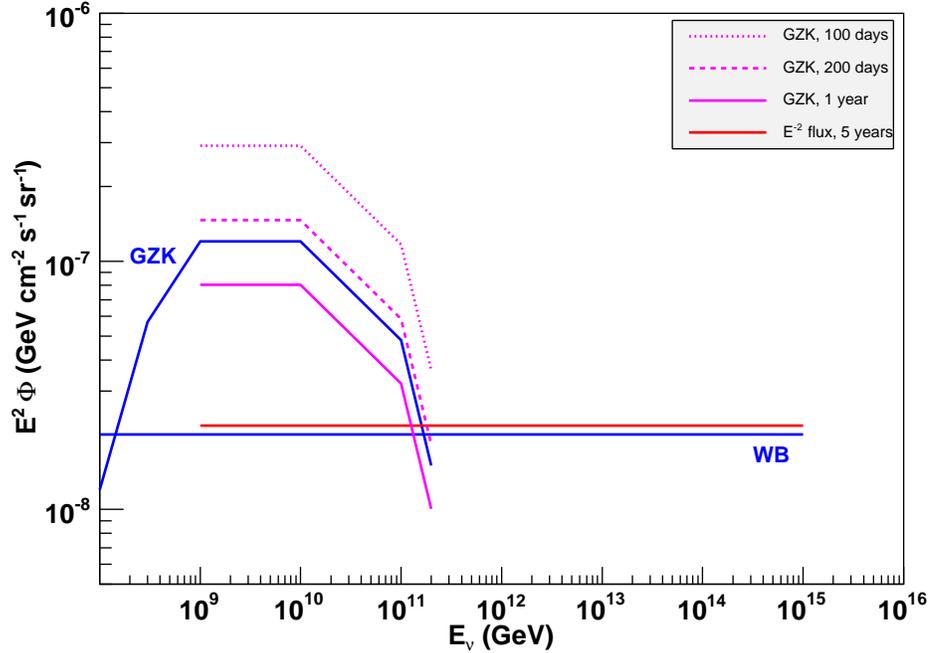}
  \caption[Sensitivity of an acoustic neutrino telescope to GZK
  neutrinos and to the WB limit]{Sensitivity of a 1500\,km$^3$
    acoustic neutrino telescope to GZK neutrinos and the Waxman
    Bahcall upper bound (WB from \cite{Bahcall:1999yr}, GZK from
    \cite{Engel:2001hd}).}
  \label{fig:flux_limit_rand200_1500km3_05mPa_models}
\end{figure}

The sensitivity to the small flux of GZK neutrinos and the Waxman
Bahcall upper bound is examined more closely in
Fig.~\ref{fig:flux_limit_rand200_1500km3_05mPa_models}, which shows
the 90\% C.L. upper limit on these fluxes.

With the presented detector a non-detection of GZK neutrinos after 245
days would rule out the presented model on a 90\% C.L. The Waxman
Bahcall upper bound could be reached after about 5 years of
measurement time.

\medskip Despite of the large instrumented target mass the proposed
detector requires only\footnote{For comparison: The ATLAS detector at
  the LHC is expected to have more than 10$^8$ readout channels} $3
\cdot 10^5$ DAQ channels\footnote{$200 \, \mathrm{AM} / \mathrm{km}^3
  \times 1500 \, \mathrm{km}^3$} to be read out. Further, data is
taken with a frequency of 100\,kHz only, and will only be transmitted
from the acoustic module to a central DAQ system when the AM
identifies a candidate bipolar pulse. Thus, from an engineering point
of view, the design of such a detector should be feasible. Further, the
sensitivity should be improved by the inclusion of electromagnetic
cascades, which also produce utilisable acoustic signals, but had to
be neglected completely in this work.

\section{Comparison with other simulations}
\label{sec:comparison}

There are other groups which are also studying acoustic neutrino
telescopes in water \cite{Perkin:2005, Niess:2005PhD} or ice \cite{Besson:2005},
where the latter one has published sensitivity estimates for a very
large volume detector which can be compared with our results.

The speed of sound in ice ($c_\mathrm{ice} \approx 3500 \, \mathrm{m}
/ \mathrm{s}$) is about 2.3 times higher than in water. Since hadronic
cascades are expected to develop equally in water and ice and the
amplitude of the acoustic signal is proportional to $c^2$
(\ref{eq:kirchhoff_short}) acoustic pulses in ice are expected to have
a signal amplitude 5.4 times larger than in water for equal energies
of the hadronic system. Thus, an acoustic neutrino telescope in ice is
expected to be more sensitive than in water, especially at low
energies.

In \cite{Besson:2005} the simulation of an acoustic neutrino telescope
with an instrumented volume $V_\mathrm{inst} = 98 \, \mathrm{km}^3$
and 27300 AMs ($280 \, \mathrm{AM} / \mathrm{km}^3$) arranged on 91
vertical strings submerged in ice is presented. For the simulation of
the ice detector an AM threshold $p_\mathrm{th} = 9$\,mPa is used
yielding a lower energy threshold of about $5 \cdot 10^9$\,GeV
comparable to the lower energy threshold of our detector obtained with
$p_\mathrm{th} = 5$\,mPa.  Table~\ref{tab:comparison_ice_water}
compares the effective volume of the ice and water acoustic telescopes
in the neutrino energy range where the two simulations overlap.

\begin{table}[hbt]
  \centering
  \caption[Comparison of $V_\mathrm{eff}$ for the ice and water
  acoustic neutrino telescopes]{Comparison of the effective volume
    $V_\mathrm{eff}$ for the ice and water acoustic neutrino
    telescopes (Values for $V_\mathrm{eff,ice}$ from \cite{Besson:2005}).}
  \label{tab:comparison_ice_water}
  \begin{tabular}{rrr}
    \hline
    \multicolumn{1}{c}{$E_\nu$} &
    \multicolumn{1}{c}{$V_\mathrm{eff,ice}$} &
    \multicolumn{1}{c}{$V_\mathrm{eff,water}$} \\
    \multicolumn{1}{c}{(GeV)} & \multicolumn{1}{c}{(km$^3$)} &
    \multicolumn{1}{c}{(km$^3$)} \\ \hline
    $5.5 \cdot 10^9$ & 175 & 25 \\
    $5.5 \cdot 10^{10}$ & 445 & 300 \\ \hline
  \end{tabular}
\end{table}

As expected, the effective volume, and thus the sensitivity, of the
ice detector is larger than of the water detector. Apart from the
higher sound velocity in ice, there are three other reasons leading to
an increase of the effective volume for the ice detector: For the ice
simulation it is assumed that a constant fraction $y = 0.2$ of the
neutrino energy is transferred to the hadronic system, whereas our
simulation uses the full $y$ distribution (Fig.~\ref{fig:y_dist}) with
a median of 0.08 which leads, in comparison, to a decrease of the
sensitivity of the water detector. Further,
Fig.~\ref{fig:v_eff_vs_am_density} shows that for the lowest neutrino
energies the difference between $280 \, \mathrm{AM} / \mathrm{km}^3$
used for the ice detector and $200 \, \mathrm{AM} / \mathrm{km}^3$ for
the water detector induces about a factor of two in effective volume,
whereas nearly no difference is expected at $5.5 \cdot 10^{10}$\,GeV.
For the simulation of the ice detector a reconstruction efficiency of
unity is assumed for all down-going cascades. The reconstruction
algorithm we use for the water telescope has an efficiency of about
50\% at $5.5 \cdot 10^9$\,GeV and of 78\% at $5.5 \cdot 10^{10}$\,GeV.

Taking the higher density of acoustic modules and the reconstruction
efficiency into account leads to a ratio between the effective volumes
gained with the two simulations of the order of 1.5, which can be
accounted for by the increased sound velocity in ice and the different
$y$ distributions used. Thus, the results obtained by the two
completely independent simulations are compatible, whereby an acoustic
neutrino telescope in ice has a slightly increased sensitivity
compared to water as detection medium due to the increased sound
velocity in ice.


\chapter{Summary}

We have presented a simulation study on a new technique for the
detection of ultra high energy neutrinos from cosmological sources:
Listening for thermoacoustic sound pulses in water which are emitted
from particle cascades produced in neutrino interactions.

Whereas we think today, that we pretty well understand the processes
which produce energy in ``ordinary'' stars, there is a plethora of
theoretical models describing the more exotic objects in which
particles can be accelerated to energies of 10$^{20}$\,eV and beyond.
Until today, only a few particles, presumably protons, could be
measured at these highest energies. But an accelerator which is
capable to accelerate hadrons should also produce neutrinos in
interactions of these hadrons with the surrounding shell, interstellar
matter, or radiation. These neutrinos will propagate unperturbed to
the Earth, conserving the direction and energy information. This is
the big advantage compared to the detection of the hadrons, which
reach us only after being deflected in the magnetic fields between the
galaxies, and which might have undergone several interactions on their
way, transferring some part of their energy to secondary particles.
Measuring the ultra high energy neutrinos would allow us to better
understand the processes taking place in these accelerators.

Since the neutrino flux decreases significantly with increasing
energy, its detection requires ever larger target masses. Today, water
\v{C}erenkov detectors in the gigaton range (1\,km$^3$ of water or
ice) are being constructed, which will allow to detect neutrinos with
energies up to about 100\,TeV. The detection of neutrinos with
energies up to 100\,ZeV and above will require a teraton detector
(1000\,km$^3$ of water or ice). There are several experimental
approaches to observe these big target masses, and it would be
desirable to implement as many of them as possible, since there are no
known calibration sources in this energy range, and thus cross
calibration between different experiments, yielding different
systematic errors, is essential.

In our study we investigate the thermoacoustic detection method,
already proposed in 1957 by G.A.~Askarian. An ultra high energy
neutrino which interacts in a fluid, in our case sea water, produces a
particle cascade which dissipates its energy in a narrow cylinder in
the medium by heating this region. This leads to a rapid expansion of
the fluid which propagates as a bipolar sonic pulse of a few
microseconds duration perpendicular to the shower axis. The disc-like
pressure field can be described by an inhomogeneous wave equation
(\ref{eq:thermoacoustic_model}). We have shown the validity of the
model by experiments in which a 177\,MeV proton beam and an infrared
Nd:YAG laser beam were dumped into water tanks, and the acoustic
signals produced were measured. The measured acoustic pulses agree
with simulations based on the thermoacoustic model very well. Thus, we
have proceeded to study the application of the thermoacoustic model to
the detection of ultra high energy neutrinos, since the absorption
length of sound in water is about ten times larger than the absorption
length of light. These lengths determine the required spacing of the
sensors in a detector. Therefore, using acoustic techniques, much
larger target masses can be instrumented with the same number of
sensors.

We envision an acoustic neutrino telescope set on the sea bed of, for
example, the Mediterranean Sea. Since the total neutrino nucleon cross
section (\ref{eq:cross_section}) increases with energy, only neutrinos
coming from above the horizon will be able to propagate to the
detector. Neutrinos coming from below will be absorbed in the Earth.
When the neutrino interacts, most of its energy is transfered to the
lepton (a neutrino in the case of an neutral current interaction; a
charged lepton in the case of a charged current interaction). This
part of the total energy cannot be detected with an acoustic detector,
since neutrinos deposit no energy in the medium and the energy
deposition of muons and electrons is too sparse to produce a
significant acoustic pulse, for the electrons due to the LPM effect
occurring at these highest energies. The energy deposition of tau
leptons, which produce hadronic cascades in the tau decay, was not
investigated yet. The remaining energy is deposited by a hadronic
cascade in a narrow volume of about 5\,m in length and 2\,cm in
diameter, which is energy independent. From this energy deposition
region bipolar acoustic pulses are emitted whose amplitude is
proportional to the energy in the hadronic system.

These bipolar pulses propagate mainly perpendicular to the cascade
axis. For their detection over large distances sound absorption in the
water has to be taken into account. The absorption length is strongly
frequency dependent, and is about 1\,km for the central frequency of
20\,kHz of the pulses. Based on Monte Carlo simulations of hadronic
cascades, on the thermoacoustic model, and on the frequency dependent
absorption length a parameterisation of the amplitude of the bipolar
pulses was derived (Fig.~\ref{fig:amplitude}). For the detection of
the neutrino induced acoustic signals we use acoustic modules (AMs):
Hypothetical devices which are able to detect bipolar pulses with an
amplitude larger than some threshold pressure $p_\mathrm{th}$ in the
ambient noise always present in the sea and record their amplitude and
arrival time. The ambient noise is well known and fortunately has a
minimum between 30\,kHz and 200\,kHz, depending on the weather
conditions, the frequency range where main contributions of the signal
are. Further it was shown that due to refraction no signals produced
further away than 27\,km from an AM can be detected, which sets a
natural limit to the target mass that can be observed by a single AM.

Based on the acoustic modules a complete detector simulation code was
developed and tested with an exemplary detector of 1\,km$^3$ size
containing 200 AMs. We have shown that a significant part (27\%) of all
neutrinos with energies between 10$^8$\,GeV and 10$^{16}$\,GeV which
interact in the surroundings of the detector generate a trigger, and
that 79\% of these triggered events can be reconstructed with a simple
minimisation algorithm. The energy of the hadronic cascade can be
determined up to a factor of 10, which is sufficient since all
predicted neutrino fluxes extend over several orders of magnitude in
energy. The direction resolution can be improved up to 12$^\circ$
with rigid selection cuts, but it is of only minor importance for
diffuse neutrino fluxes. An important factor towards the realisation
of an acoustic neutrino telescope will be the rate of background
events with long correlation lengths, i.e. of natural or anthropogenic
sources which produce causally correlated bipolar acoustic pulses that
could be identified wrongly as neutrinos. So far, this rate is
completely unknown, and the Erlangen ANTARES group has proposed an
experiment to measure it within the ANTARES experimental setup.

We could show that using an underwater acoustic neutrino telescope
with an instrumented volume of 1500\,km$^3$, where only $200 \,
\mathrm{AM} / \mathrm{km}^3$ would be required, many cosmological
models predicting ultra high energy neutrino fluxes could be tested
within 5 years of measurement time
(Figs.~\ref{fig:flux_limit_rand200_1500km3_05mPa_5y} and
\ref{fig:flux_limit_rand200_1500km3_05mPa_models}). Such a detector
would be a possible approach to the detection of ultra high energy
cosmological neutrinos, which is complementary to the radio
\v{C}erenkov experiments currently under construction.


\chapter{Zusammenfassung}

\selectlanguage{ngerman}Die vorliegende Arbeit beschreibt eine
Simulationsstudie zu einer neuen Nachweistechnik f"ur
h"ochstenergetische Neutrinos aus kosmischen Quellen: Der Nachweis
thermoakustischer Schallpulse, die im Meer von neutrinoinduzierten
Teilchenschauern erzeugt werden.

Physiker nehmen heutzutage an, die Energieerzeugung in \glqq
normalen\grqq\ Sternen sehr gut verstanden zu haben. Es gibt jedoch
eine Vielzahl theoretischer Modelle, die exotische kosmische Objekte
beschreiben, die in der Lage sind Teilchen auf Energien von
10$^{20}$\,eV oder h"oher zu beschleunigen. Bis heute konnten nur
wenige dieser Teilchen --- man geht davon aus, dass es sich um
Protonen handelt --- auf der Erde nachgewiesen werden. Ein
Beschleuniger, der Hadronen auf solch hohe Energien beschleunigt, muss
auch eine Quelle h"ochstenergetischer Neutrinos sein, die bei der
Wechselwirkung der Protonen mit der umgebenden Materie oder Strahlung
entstehen. Diese Neutrinos k"onnen als nur schwach wechselwirkende
Teilchen, nachdem sie den Beschleuniger verlassen haben, die Erde
ungest"ort erreichen. Neutrinos tragen, im Gegensatz zu Protonen, die
in den intergalaktischen Magnetfeldern abgelenkt werden und mit dem
Restgas wechselwirken k"onnen, wobei ein Teil ihrer Energie an
Sekund"arteilchen "ubertragen wird, eine Richtungs- und
Energieinformation. Der Nachweis dieser h"ochstenergetischen Neutrinos
w"urde einen tiefen Einblick in die Beschleunigungsprozesse, die in
kosmologischen Objekten stattfinden, erlauben.

Da der Neutrinofluss mit zunehmender Energie stark abnimmt, werden
immer gr"o"sere Detektormassen zu seinem Nachweis ben"otigt. Derzeit
geplante oder im Aufbau befindliche Wasser-\v{C}erenkov Detektoren mit
einer Masse von einer Gigatonne (1\,km$^3$ Wasser oder Eis)
erm"oglichen den Nachweis von Neutrinos mit einer Energie von bis zu
100\,TeV. Die Messung von Neutrinos mit Energien von 100\,ZeV und
dar"uber wird die Entwicklung von Detektoren mit einer Masse im
Bereich einer Teratonne (1000\,km$^3$ Wasser oder Eis) erfordern. Es
gibt verschiedenste experimentelle Ans"atze zur Instrumentierung solch
gro"ser Volumina, und es w"are sinnvoll in Zukunft m"oglichst viele
von ihnen zu realisieren, da in diesem Energiebereich keinerlei
Kalibrationsquellen bekannt sind. Somit ist ein Vergleich zwischen
Ergebnissen, die mit verschiedenen experimentellen Techniken gemessen
wurden und somit unterschiedliche systematische Fehler aufweisen,
unerl"asslich.

Diese Studie besch"aftigt sich mit der thermoakustischen
Nachweismethode, die erstmals 1957 von G.A.~Askarian vorgeschlagen
wurde. Ein h"ochstenergetisches Neutrino wechselwirkt in einem Fluid,
in diesem Fall Wasser, und es entsteht ein Teilchenschauer, dessen
Energie innerhalb eines schmalen Zylinders in Form von W"arme an das
Wasser abgegeben wird. Dies f"uhrt zu einer pl"otzlichen Ausdehnung
der Fl"ussigkeit, und es breitet sich ein bipolarer akustischer Puls
von wenigen Mikrosekunden L"ange senkrecht zur Schauerachse aus. Das
entstehende scheibenf"ormige Druckfeld kann durch die inhomogene
Wellengleichung (\ref{eq:thermoacoustic_model}) beschrieben werden.
Wir konnten dieses Modell in zwei Experimenten, in denen ein Strahl
von 177\,MeV Protonen, bzw. ein infraroter Nd:YAG Laser in einen mit
Wasser gef"ullten Tank geschossen wurden, und die dabei entstehenden
akustischen Pulse aufgezeichnet wurden, best"atigen. Die gemessenen
Signale stimmen mit Simulationsrechnungen, die unter der Annahme eines
thermoakustischen Schallerzeugungsmechanismus gemacht wurden, sehr gut
"uberein. Dieses Ergebnis ermutigte uns, die M"oglichkeit des
Nachweises ultrahochenergetischer Neutrinos mit einem akustischen
Detektor weiter zu studieren, insbesondere da die Abschw"achl"ange von
Schall in Wasser etwa zehn mal gr"o"ser ist als die Abschw"achl"ange
von Licht. Diese Abschw"achl"ange ist es, die den f"ur einen Detektor
zu w"ahlenden Abstand zwischen den einzelnen Sensoren bestimmt. Somit
kann mittels eines akustischen Detektors ein sehr viel gr"o"seres
Volumen mit der selben Anzahl von Sensoren best"uckt werden, als dies
in einem optischen Detektor m"oglich w"are.

Wir untersuchen ein m"ogliches akustisches Neutrinoteleskop auf dem
Meeresgrund, zum Beispiel auf dem Grund des Mittelmeers. Da der totale
Wirkungsquerschnitt (\ref{eq:cross_section}) f"ur die Wechselwirkung
von Neutrinos mit Nukleonen mit wachsender Neutrinoenergie ansteigt,
k"onnen bei h"ochsten Energien nur von oben kommende Neutrinos den
Detektor erreichen; von unten kommende Neutrinos werden in der Erde
absorbiert. Bei der Wechselwirkung wird ein Gro"steil der
Neutrinoenergie auf das entstehende Lepton "ubertragen (ein geladenes
Lepton im Falle einer Wechselwirkung "uber einen geladenen Strom; ein
Neutrino im Falle einer Neutralstromwechselwirkung). Dieser Teil ist
f"ur die Detektion mit einem akustischen Neutrinoteleskop verloren, da
Neutrinos keine, und Muonen und Elektronen nur sehr d"unn Energie im
Medium deponieren, letztere auf Grund des LPM Effekts. Die
Energiedeposition von Tau-Leptonen, bei deren Zerfall ein hadronischer
Schauer entstehen kann, wurde im Rahmen dieser Arbeit nicht
untersucht. Die verbleibende Energie wird in Form eines hadronischen
Teilchenschauers in einem zylindrischen Volumen von etwa 5\,m L"ange
und 2\,cm Durchmesser abgegeben. In diesem Volumen, dessen Abmessungen
energieunabh"angig sind, wird dann ein bipolarer akustischer Puls
ausgesendet, dessen Amplitude proportional zur Energie der
hadronischen Kaskade ist.

Das Signal breitet sich haupts"achlich senkrecht zur Achse des
Schauers aus. Beim Nachweis muss jedoch noch die Abschw"achung des
Signals im Wasser ber"ucksichtigt werden. Die stark frequenzabh"angige
Absorptionsl"ange betr"agt f"ur die 20\,kHz Schwerpunktsfrequenz des
Signals etwa 1\,km. Basierend auf Monte Carlo Simulationen
hadronischer Schauer, dem thermoakustischen Modell und der
frequenzabh"angigen Absorptionsl"ange wurde eine Parametrisierung der
Amplitude des akustischen Pulses f"ur beliebige Sensorpositionen
entwickelt (Abb.~\ref{fig:amplitude}). F"ur die Messung
neutrinoinduzierter Schallpulse definieren wir ein hypothetisches {\em
  akustisches Modul} (AM): Ein Ger"at, das bipolare akustische
Schallpulse, wie sie zum Beispiel von neutrinoinduzierten hadronischen
Schauern ausgesandt werden, mit einer Amplitude gr"o"ser einer
Schwelle $p_\mathrm{th}$ aus dem umgebenden Untergrundrauschen
herausfiltern kann, und ihre Ankunftszeit und Amplitude messen und
weitersenden kann. Das Umgebungsrauschen im Meer ist eine bekannte
Gr"o"se, und hat ein Minimum zwischen 30\,kHz und 200\,kHz je nach
Windst"arke. Dies ist der Frequenzbereich in dem neutrinoinduzierten
akustischen Pulse erwartet werden. Weiter konnte gezeigt werden, dass
auf Grund von Brechung keine Signale, die weiter als 27\,km von einem
Detektor erzeugt werden, diesen erreichen k"onnen.  Dadurch wird eine
nat"urliche obere Grenze f"ur das Volumen festgelegt, das mit einem
einzelnen Sensor beobachtet werden kann.

Aufbauend auf akustischen Modulen wurde eine vollst"andige
Detektor-Si\-mu\-la\-tions\-kette entwickelt, und am Beispiel eines
1\,km$^3$ gro"sen Detektors bestehend aus 200 AMs untersucht. Wir
konnten zeigen, dass ein erheblicher Teil (27\%) aller in der Umgebung
des Detektors wechselwirkenden Neutrinos mit einer Energie zwischen
10$^8$\,GeV und 10$^{16}$\,GeV im Detektor Trigger ausl"osen, und dass
f"ur 79\% dieser getriggerten Ereignisse die Richtung und Energie des
hadronischen Schauers mit einem einfachen Minimierungs-Algorithmus
bestimmt werden k"onnen. Die Energie kann bis auf einen Faktor 10
genau bestimmt werden, was noch ausreichend ist, da die zu
untersuchenden kosmologischen Modelle ein Neutrinospektrum
voraussagen, das sich "uber mehrere Zehnerpotenzen in der Energie
erstreckt. Die Richtungsrekonstruktion kann mittels Schnitten bis auf
12$^\circ$ verbessert werden, was aber f"ur diffuse Neutrinofl"usse
nur eine untergeordnete Rolle spielt. Ein wichtiger Faktor bei der
Realisierung eines akustischen Neutrinoteleskops wird die Rate des
korrelierten Untergrunds sein, das hei"st nat"urlicher oder
anthropogener Schallquellen, die bipolare Pulse mit langer
Korrelationsl"ange erzeugen, die f"alschlicherweise als Neutrinos
identifiziert werden k"onnten. Da diese Rate v"ollig unbekannt ist,
plant die Erlanger ANTARES Gruppe ein Experiment mit dem diese Rate
innerhalb der experimentellen Infrastruktur von ANTARES gemessen
werden kann.

Wir konnten zeigen, dass ein akustisches Unterwasser-Neutrinoteleskop
mit einem instrumentierten Volumen von 1500\,km$^3$, wobei lediglich
$200 \, \mathrm{AM} / \mathrm{km}^3$ ben"otigt w"urden, in der Lage
w"are viele kosmologische Modelle, die einen Fluss
ultrahochenergetischer Neutrinos vorhersagen, innerhalb von f"unf
Jahren Messzeit zu testen
(Abb.~\ref{fig:flux_limit_rand200_1500km3_05mPa_5y} und
\ref{fig:flux_limit_rand200_1500km3_05mPa_models}). Ein solches
Experiment w"urde einen komplement"aren Ansatz zu den verschiedenen
Radio \v{C}erenkov Experimenten darstellen, die zur Zeit entwickelt
werden.

\selectlanguage{british}


\appendix
\chapter{Derivation of the thermoacoustic model}
\label{chap:hydrodynamics}

In this chapter the basic wave equation describing the connection
between energy deposited in a fluid and the resulting acoustic
pressure field is derived.

We consider an arbitrary compressible fluid, which can be described by
five fields: the pressure $p (\vec{r}, t)$, the density $\rho
(\vec{r}, t)$, and the velocity $\vec{v} (\vec{r}, t)$. It is
important to note, that all fields describe the properties of the
fluid at a fixed point $(\vec{r}, t)$ in space and time, and {\em not}
the properties of a moving volume element of the fluid.

First we note, that in the processes observed no fluid is created or
destroyed, i.e.~we have an equation of continuity for the mass of the
fluid. Considering an arbitrary volume $V$ with border $\partial V$ we
have:

\begin{equation}
  \label{eq:continuity_integral}
  \frac{\mathrm{d}}{\mathrm{d} t} \, \int_V \mathrm{d}^3 r \, \rho = -
  \oint_{\partial V} \vec{\mathrm{d} \sigma} \cdot \left( \rho
    \vec{v} \right)
\end{equation}

\noindent The left hand side of this equation represents the increase
of mass with time inside the volume $V$, which has to be equal to the
mass flow into the element through its surface, which is written on
the right hand side ($\vec{\mathrm{d} \sigma}$ is defined to point to
the outside of the volume, thus the minus sign). Using Gauss's
integral theorem $\oint_{\partial V} \vec{\mathrm{d} \sigma} \cdot
\left( \rho \vec{v} \right) = \int_V \mathrm{d}^3 r \, \vec{\nabla}
\cdot \left( \rho \vec{v} \right)$, and noting that
(\ref{eq:continuity_integral}) is valid for all volumes $V$, we can
write it in the differential form:

\begin{equation}
  \label{eq:continuity}
  \frac{\partial \rho}{\partial t} + \vec{\nabla} \cdot \left( \rho
    \vec{v} \right) = 0
\end{equation}

\smallskip Next, we will derive Euler's law by noting, that the force
$\vec{F}$ applied by the whole fluid on the volume element $V$ is:

\begin{equation}
  \vec{F} = - \oint_{\partial V} \vec{\mathrm{d} \sigma} \, p = -
  \int_V \mathrm{d}^3 r \, \vec{\nabla} p
\end{equation}

\noindent This allows us to write down the equation of motion for a
single volume element $\mathrm{d} V$ with mass $\rho \mathrm{d} V$:

\begin{equation}
  \label{eq:F_ma}
  \left( \rho \mathrm{d} V \right) \, \frac{\mathrm{d}
    \vec{v}}{\mathrm{d} t} = - \mathrm{d}V \, \vec{\nabla} p
\end{equation}

\noindent We can write the total derivative of $\vec{v}$ as a sum of
partial derivatives:

\begin{displaymath}
  \mathrm{d} \vec{v} = \frac{\partial \vec{v}}{\partial t} \,
  \mathrm{d} t + \sum_{j \in \{x, y, z\}} \, \frac{\partial
    \vec{v}}{\partial j} \, \mathrm{d} j = \frac{\partial
    \vec{v}}{\partial t} \, \mathrm{d} t + \left( \vec{\mathrm{d} r}
    \cdot \vec{\nabla} \right) \vec{v}
\end{displaymath}

\noindent or

\begin{displaymath}
  \frac{\mathrm{d} \vec{v}}{\mathrm{d} t} = \frac{\partial
    \vec{v}}{\partial t} + \left( \vec{v} \cdot \vec{\nabla} \right)
  \vec{v}
\end{displaymath}

\noindent Inserting this into (\ref{eq:F_ma}) leads to Euler's law:

\begin{equation}
  \label{eq:eulers_law}
  \frac{\partial \vec{v}}{\partial t} + \left( \vec{v} \cdot
    \vec{\nabla} \right) \vec{v} = - \frac{\vec{\nabla} p}{\rho}
\end{equation}

\smallskip For the derivation of the thermoacoustic model we consider
a fluid at rest in a static state. The acoustic waves are considered
to be small deviations from the static quantities:

\begin{equation}
  \begin{array}{rcccll}
    p_\mathrm{tot} (\vec{r}, t) & = & p_0 & + & p (\vec{r}, t) & p \ll
    p_0 \, \forall (\vec{r}, t) \\
    \rho_\mathrm{tot} (\vec{r}, t) & = & \rho_0 & + & \rho (\vec{r},
    t) & \rho \ll \rho_0 \, \forall (\vec{r}, t) \\
    \vec{v}_\mathrm{tot} (\vec{r}, t) & = & \vec{0} & + &  \vec{v}
    (\vec{r}, t)
  \end{array}
\end{equation}

We further assume that these deviations vary slowly enough in space
and time, that their derivatives can also be considered as small
quantities. Substituting these terms in the equation of continuity
(\ref{eq:continuity}), and neglecting terms which are the product of
two or more small quantities, we get:

\begin{equation}
  \label{eq:continuity_lin}
  \frac{\partial \rho}{\partial t} + \rho_0 \left( \vec{\nabla} \cdot
    \vec{v} \right) = 0
\end{equation}

\noindent Euler's law (\ref{eq:eulers_law}) is reduced to:

\begin{equation}
  \label{eq:eulers_law_lin}
  \frac{\partial \vec{v}}{\partial t} =  - \frac{\vec{\nabla}
    p}{\rho_0}
\end{equation}

We now take the partial time derivative of (\ref{eq:continuity_lin}),
exchange the spatial and temporal derivatives, and substitute
(\ref{eq:eulers_law_lin}):

\begin{displaymath}
  0 = \frac{\partial^2 \rho}{\partial t^2} + \rho_0 \left(
    \vec{\nabla} \cdot \frac{\partial \vec{v}}{\partial t} \right) =
  \frac{\partial^2 \rho}{\partial t^2} - \vec{\nabla} \cdot \left(
    \vec{\nabla} p \right)
\end{displaymath}

\noindent or

\begin{equation}
  \label{eq:wave_eq}
  \frac{\partial^2 \rho}{\partial t^2} - \Delta p = 0
\end{equation}

Let us now, for simplicity, consider first the adiabatic case, where
the density is a function of the pressure only: $\rho_\mathrm{tot} =
\rho_\mathrm{tot} (p_\mathrm{tot})$. We expand $\rho_\mathrm{tot}
(p_\mathrm{tot})$ in a Taylor series around $p_0$:

\begin{displaymath}
  \rho = \rho_\mathrm{tot} (p_0 + p) - \rho_\mathrm{tot}(p_0) \approx
  \frac{\mathrm{d} \rho}{\mathrm{d} p} (p_0) \, p =: \frac{1}{c^2} \,
  p
\end{displaymath}

\noindent Therefore, for the adiabatic case we get the following
simple wave equation, where $c$ is the speed of sound:

\begin{equation}
  \frac{1}{c^2} \frac{\partial^2 p}{\partial t^2} - \Delta p = 0
\end{equation}

If we consider the case important for particle detection, we have to
investigate the process where energy is deposited with an energy
density $\varepsilon (\vec{r}, t)$. This process is not adiabatic, and
leads to local heating and expansion of the fluid. Thus, the local
density is not a function of the pressure only any more, but of the
pressure {\em and} the deposited energy. We get:

\begin{displaymath}
  \mathrm{d} \rho = \left( \frac{\partial \rho}{\partial p} \right)_E
  \mathrm{d} p + \left( \frac{\partial \rho}{\partial E} \right)_p
  \mathrm{d} E
\end{displaymath}

\noindent where the derivative $\partial \rho / \partial p$ at
constant energy $E$ results in the speed of sound $1 / c^2$ again. For
the derivative $\partial \rho / \partial E$ at constant pressure we
look at an infinitesimal volume element $V$ with mass $m$:

\begin{displaymath}
  \left( \frac{\partial \rho}{\partial E} \right)_p = \left(
    \frac{\partial}{\partial E} \frac{m}{V} \right)_p = -
  \frac{m}{V^2} \left( \frac{\partial V}{\partial E} \right)_p \approx
  - \frac{\rho}{V} \left( \frac{\Delta V}{\Delta E} \right)
\end{displaymath}

\noindent The change of the volume $\Delta V$ can be calculated from
specific heat capacity at constant pressure $C_p$, and the thermal
bulk expansion coefficient $\alpha$ as

\begin{displaymath}
  \Delta V = \frac{\alpha}{\rho C_p} \, \Delta E
\end{displaymath}

\noindent We finally get:

\begin{displaymath}
  \mathrm{d} \rho = \frac{1}{c^2} \, \mathrm{d} p - \frac{\alpha}{C_p}
  \, \mathrm{d} \varepsilon
\end{displaymath}

\noindent where $\mathrm{d} \varepsilon = \mathrm{d} E / V$ is the
energy density in the infinitesimal volume element $V$. Substituting
this into (\ref{eq:wave_eq}) we arrive at the thermoacoustic model
which describes the pressure field produced by energy with an energy
density distribution $\varepsilon (\vec{r}, t)$ deposited in a fluid.

\begin{equation}
  \label{eq:thermoacoustic_model_app}
  \frac{1}{c^2} \frac{\partial^2 p}{\partial t^2} - \Delta p =
  \frac{\alpha}{C_p} \frac{\partial^2 \varepsilon}{\partial t^2}
\end{equation}


\chapter{On the calculation of flux limits}
\label{chap:limits}

No existing experiment has detected any extra-galactic neutrinos so
far, so the only published results are upper limits on the neutrino
flux. On the other hand, new experiments publish limits on the lowest
neutrino flux they will be able to detect in a given lifetime

When comparing limits on the neutrino flux presented by different
experiments one has to be aware that there are different methods to
calculate these limits.  In this section, these methods are derived,
and the possibility to compare results obtained with different methods
are discussed.

We assume a constant and isotropic flux of neutrinos $\Phi (E)$, which
depends only on the neutrino energy $E$. Then the mean number of
events $\langle N_\mathrm{ev} \rangle$ expected to be detected by an
experiment is given by

\begin{equation}
  \langle N_\mathrm{ev} \rangle = \int \mathrm{d} E \int \mathrm{d} t
  \int \mathrm{d} \Omega \, \Phi (E) A_\mathrm{eff} (E, \Omega)
\end{equation}

The above equation can also be read as a definition for the effective
area $A_\mathrm{eff}$ of the experiment. The effective area depends on
the neutrino cross section, and thus on its energy, and it depends
further on the detection mechanism and the geometrical acceptance of
the detector, i.e. on the direction of incidence $\Omega$.
$A_\mathrm{eff}$ is either calculated analytically or by Monte Carlo
simulations including all aspects of a given detector.

For an isotropic neutrino flux one can define the aperture
$\mathcal{A} (E)$ of the detector:

\begin{equation}
  \label{eq:aperture}
  \mathcal{A} (E) = \int \mathrm{d} \Omega \, A_\mathrm{eff} (E,
  \Omega)
\end{equation}

\noindent and thus

\begin{equation}
  \langle N_\mathrm{ev} \rangle = \int \mathrm{d} E \int \mathrm{d} t
  \, \Phi (E) \mathcal{A} (E)
\end{equation}

If further the flux is constant in time, the lifetime $T$ of the
experiment can be introduced:

\begin{equation}
  \langle N_\mathrm{ev} \rangle = T \int \mathrm{d} E \, \Phi (E)
  \mathcal{A} (E)
\end{equation}

Let us assume that the number of events $N_\mathrm{ev}$ detected by a
given experiment during its lifetime $T$ follows a Poisson
distribution with mean value $\langle N_\mathrm{ev} \rangle$, and that
the mean number of background events expected to be detected during
the time $T$ is zero (which is true for most ultra high energy
neutrino experiments). Then, if the experiment has detected no events
($N_\mathrm{ev} = 0$), it can be derived \cite{Feldman:1997qc} that
the true mean value $\langle N_\mathrm{ev} \rangle$ is inside the
interval [0,~2.44] with a confidence of 90\% ([0,~3.09] with a
confidence of 95\%).

Thus, setting $\langle N_\mathrm{ev} \rangle = 2.44$ results in an
upper limit $\Phi_{90} (E)$ on the real neutrino flux with a
confidence level of 90\%:

\begin{equation}
  \label{eq:flux_90_cl}
  \int \mathrm{d} E \, \Phi (E) \mathcal{A} (E) \le \, \frac{2.44}{T}
  \quad (90\% \, \mathrm{C.L.})
\end{equation}

\noindent or

\begin{equation}
  \int \mathrm{d} E \, \Phi_{90} (E) \mathcal{A} (E) = \,
  \frac{2.44}{T}
\end{equation}

To solve this equation further to get $\Phi_{90} (E)$, one has to make
certain assumptions on the flux. There are three assumptions commonly
used in the literature:

\paragraph{Limit on a given source model.} It is assumed that the
neutrino spectrum to be measured follows a given functional form
determined by a certain theoretical source model:

\begin{equation}
  \Phi_{90} (E) = \mathcal{N} \Phi_\mathrm{model} (E)
\end{equation}

\noindent where $\mathcal{N}$ is a normalisation constant for the
neutrino flux, which can then be calculated from:

\begin{equation}
  \label{eq:model_dep_limit}
  \mathcal{N} = \, \frac{2.44}{T \int \mathrm{d} E \,
    \Phi_\mathrm{model} (E) \mathcal{A} (E)}
\end{equation}

\noindent which can be solved analytically or numerically, since the
aperture $\mathcal{A} (E)$ is a known quantity.

Often power-law spectra $\Phi (E) \propto E^{- \alpha}$ are used,
which are typical hadronic source spectra expected from many
cosmological accelerator models. Especially a spectral index of
$\alpha = 2$ is expected from first order Fermi acceleration.

\begin{equation}
  \Phi_{90} (E) = \mathcal{N} E^{-2}
\end{equation}

Limits based on power-law spectra can be recognised as straight lines
in double-logarithmic flux plots, where $\alpha$ determines the
slope. Especially the important case of $\alpha =$ 2 reduces to a
constant in the $E^2 \Phi$ plots used in this work.

\paragraph{Non-detection in a given energy-interval.} There are two
possibilities of calculating model independent limits. The above model
dependent limit was derived from the non-detection of any neutrinos
over the whole energy range covered by the model. For an experiment
with an energy resolution better than this range, a limit can be set
on every energy interval compatible with the energy resolution.

Typical energy resolutions $\Delta E / E$ for neutrino telescopes are
of the order of 3 to 10. So a common value for the non-detection
energy interval found in the literature is one decade in energy, i.e.
an energy range $[E, 10 E]$.

If further the aperture can be assumed constant around some mean
energy $\bar{E}$ of the energy interval with length $\Delta E$, or
some mean aperture $\bar{\mathcal{A}} (\bar{E})$ over the interval can
be used, the equation for the flux limit can be linearised to

\begin{equation}
  \Delta E \, \Phi_{90} (\bar{E}) \bar{\mathcal{A}} (\bar{E}) = \,
  \frac{2.44}{T}
\end{equation}

\noindent and thus

\begin{equation}
  \Phi_{90} (\bar{E}) = \frac{2.44}{T \, \Delta E \, \bar{\mathcal{A}}
    (\bar{E})}
\end{equation}

\paragraph{Model independent estimate.} The FORTE collaboration
suggests a method to derive a flux limit which is independent of a
given functional form of the source spectrum, but only assumes that
the source spectrum is smooth (i.e. the source spectrum can be written
as linear combination of a set of functions to be defined below)
\cite{Lehtinen:2003xv}.

We assume a source model flux, that depends on an additional
parameter or set of parameters $P$:

\begin{equation}
  \Phi (E, P) = \mathcal{N} \Phi_\mathrm{model} (E, P)
\end{equation}

\noindent Then, using the model dependent limit
(\ref{eq:model_dep_limit}), we come to the following relation:

\begin{equation}
  \Phi (E, P) \le \, \frac{2.44 \, \Phi_\mathrm{model} (E, P)}{T \int
    \mathrm{d} E^\prime \, \Phi_\mathrm{model} (E^\prime, P)
    \mathcal{A} (E^\prime)}
\end{equation}

We get a conservative, but always valid upper limit which is
independent of the parameter $P$ if we take the maximum value of the
right hand side for every energy:

\begin{equation}
  \Phi (E) \le \max_P \, \frac{2.44 \, \Phi_\mathrm{model} (E, P)}{T
    \int \mathrm{d} E^\prime \, \Phi_\mathrm{model} (E^\prime, P)
    \mathcal{A} (E^\prime)}
\end{equation}

In the next step we will prove that the above relation is valid for
{\em any} linear combination of flux models, i.e.:

\begin{equation}
  \label{eq:flux_lin_comb}
  \Phi (E) = \int \mathrm{d} P \, \mathcal{N} (P) \Phi_\mathrm{model}
  (E, P)
\end{equation}

The proof assumes the opposite of the above relation, i.e. we assume
that {\em for all} $P$:

\begin{displaymath}
  \Phi (E) > \, \frac{2.44 \, \Phi_\mathrm{model} (E, P)}{T \int
    \mathrm{d} E^\prime \, \Phi_\mathrm{model} (E^\prime, P)
    \mathcal{A} (E^\prime)}
\end{displaymath}

\noindent or

\begin{equation}
  \int \mathrm{d} E^\prime \, \Phi_\mathrm{model} (E^\prime, P)
  \mathcal{A} (E^\prime) > \, \frac{2.44 \, \Phi_\mathrm{model} (E,
    P)}{T \, \Phi (E)}
\end{equation}

\noindent we multiply both sides by $\mathcal{N} (P)$ and integrate
over all $P$

\begin{displaymath}
  \int \mathrm{d} E^\prime \, \left( \int \mathrm{d} P \, \mathcal{N}
    (P) \Phi_\mathrm{model} (E^\prime, P) \right) \mathcal{A}
  (E^\prime) > \, \frac{2.44}{T \, \Phi (E)} \, \int \mathrm{d} P \,
  \mathcal{N} (P) \Phi_\mathrm{model} (E^\prime, P)
\end{displaymath}

\noindent using definition (\ref{eq:flux_lin_comb}) of $\Phi (E)$
we get

\begin{displaymath}
  \int \mathrm{d} E^\prime \, \Phi (E^\prime) \mathcal{A} (E^\prime) >
  \, \frac{2.44}{T}
\end{displaymath}

\noindent which is in contradiction to our initial assumption
(\ref{eq:flux_90_cl}) about the flux at 90\% C.L.

\smallskip So for any flux which can be represented as an arbitrary
linear combination of functions $\Phi_\mathrm{model} (E, P)$, the
upper limit at 90\% confidence level can be written as:

\begin{equation}
  \label{eq:independent_limit_long}
  \Phi_{90} (E) = \max_P \, \frac{2.44 \, \Phi_\mathrm{model} (E,
    P)}{T \int \mathrm{d} E^\prime \, \Phi_\mathrm{model} (E^\prime,
    P) \mathcal{A} (E^\prime)}
\end{equation}

This can be simplified very much if we restrain ourselves to a set of
convex curves $\Phi_C (E, E_0)$, which have their maximum at $E_0$
each, and have a width of $\approx E_0$. The curves shall be
normalised, so that $\int \mathrm{d} E \, \Phi_C (E, E_0) = 1$. Then
we have at the maximum $\max_E \Phi_C (E, E_0) \lesssim 1 / E_0$. If further
the aperture $\mathcal{A} (E)$ is smooth enough that it can be assumed
constant over the validity of a certain $\Phi_C$, so that $\int
\mathrm{d} E \, \Phi_C (E, E_0) \mathcal{A} (E) \approx \mathcal{A} (E_0)$,
equation (\ref{eq:independent_limit_long}) is reduced to:

\begin{equation}
  \label{eq:independent_limit}
  \Phi_{90} (E) = \, \frac{2.44}{T E \mathcal{A} (E)}
\end{equation}

This result deviates from the one of the non-detection in a given
interval by a constant factor if energy intervals of constant width in
$\log \, E$ are used, which usually is the case. Further, this factor
is of the order of unity if the energy $E$ in the latter model
compares to the mean value $\bar{E}$ of a decade in energy of the
second model presented.



\backmatter
\bibliographystyle{utphys}
\bibliography{bibliography}

\selectlanguage{ngerman}
\chapter*{Danksagung}
\pagestyle{plain}

Viele Menschen haben zum Gelingen dieser Arbeit beigetragen, so dass ich mich zuerst bei denjenigen entschuldigen m\"ochte, die hier keine namentliche Erw\"ahnung finden.

\smallskip Mein besonderer Dank gilt Frau Prof.~Dr.~Gisela Anton, die mir beim Anfertigen dieser Arbeit alle Freiheiten lie{\ss}, und mir dennoch stets mit Rat und Tat weiterhelfen konnte.

\smallskip Den Mitarbeitern der Arbeitsgruppen Astroteilchenphysik und Medizinphysik, mittlerweile mehr, als dass ich sie hier alle einzeln nennen k\"onnte, danke ich f\"ur die angenehme Arbeitsatmosph\"are und fortw\"ahrende Diskussionsbereitschaft, ohne die diese Arbeit niemals vollendet worden w\"are. Besonders m\"ochte ich bei meinem langj\"ahrigen Zimmerkollegen Norman Uhlmann bedanken, der mir geholfen hat dem st\"andigen Rechner-Wahnsinn Herr zu werden.

\smallskip Den Mitgliedern der ANTARES Kollaboration schulde ich Dank f\"ur die freundliche Aufnahme und die steten Bem\"uhungen mein Franz\"osisch zu verbessern. Insbesondere sei hier Valentin Niess genannt, mit dem ich viele fruchtbare Diskussionen \"uber akustische Detektion f\"uhren durfte.

\smallskip Ich danke Felix, J\"urgen, Kay und Klaus f\"ur das Korrekturlesen der Arbeit, sowie Herrn Prof.~Dr.~Karl-Heinz Kampert f\"ur die bereitwillige \"Ubernahme des Zweitgutachtens.

\chapter*{Lebenslauf}

\begin{center}
  \begin{tabular}{ll}
    \multicolumn{2}{l}{\bf Pers"onliche Daten:} \\[1mm]
    & Timo Thomas Karg \\
    & Hartmannstra\ss e 89 \\
    & 91052 Erlangen \\
    Geburtsdatum: & 30. September 1976 \\
    Geburtsort: & N"urnberg \\
    \\
    \multicolumn{2}{l}{\bf Hochschulausbildung:} \\[1mm]
    08/2002 & Beginn des Promotionsstudiums \\
    & Physikalisches Institut, Abt. 4 \\
    & Universit"at Erlangen-N"urnberg \\
    10/1997 -- 07/2002 & Grund- und Hauptstudium (Physik, Diplom) \\
    & Universit"at Erlangen-N"urnberg \\
    06/2001 -- 06/2002 & Anfertigung der Diplomarbeit \\
    & Physikalisches Institut, Abt. 4 \\
    & Thema: \glqq Entwicklung eines Bildrekonstruktions- \\
    & algorithmus f"ur die Compton-Kamera\grqq \\
    10/1999 -- 03/2000 & ERASMUS Stipendiat \\
    & Imperial College, London \\
    07/1999 & Diplom-Vorpr"ufung \\
    \\
    \multicolumn{2}{l}{\bf Zivildienst:} \\[1mm]
    07/1996 -- 07/1997 & Klinikum S"ud der Stadt N"urnberg \\
    \\
    \multicolumn{2}{l}{\bf Schulausbildung:} \\[1mm]
    1987 -- 1996 & Martin-Behaim-Gymnasium, N"urnberg \\
    1983 -- 1987 & Grundschule Zugspitzstra\ss e, N"urnberg \\
    \\
    \multicolumn{2}{l}{\bf Berufserfahrung:} \\[1mm]
    seit 08/2002 & Wissenschaftlicher Angestellter \\
    & Physikalisches Institut, Abt. 4 \\
    & Universit"at Erlangen-N"urnberg
  \end{tabular}
\end{center}

\newpage \strut


\end{document}